\documentclass[aps, twocolumn, superscriptaddress, longbibliography]{revtex4-2}

\usepackage{graphicx}

\usepackage{amsmath}
\usepackage{amssymb}
\usepackage{graphicx}
\usepackage{afterpage}
\usepackage{amsfonts}
\usepackage{amssymb}
\usepackage{graphicx}
\usepackage{braket}
\usepackage{textcomp}
\usepackage{bm}
\usepackage{color}
\usepackage{multirow}

\begin{document}
\title{{\normalsize\textrm{ Accepted version of Optica 9(5), 522-531 (2022)\\
https://doi.org/10.1364/OPTICA.448124}}
\rule{\textwidth}{1pt}
Resonant and phonon-assisted ultrafast coherent control of a single hBN color center}

\author{Johann A. Preu{\ss}}
\thanks{These two authors contributed equally}
\affiliation{Institute of Physics and Center for Nanotechnology, University of M\"unster, 48149 M\"unster, Germany}

\author{Daniel Groll}
\thanks{These two authors contributed equally}
\affiliation{Institute of Solid State Theory, University of M\"unster, 48149 M\"unster, Germany}

\author{Robert Schmidt}
\affiliation{Institute of Physics and Center for Nanotechnology, University of M\"unster, 48149 M\"unster, Germany}

\author{Thilo Hahn}
\affiliation{Institute of Solid State Theory, University of M\"unster, 48149 M\"unster, Germany}

\author{Pawe\l{} Machnikowski}
\affiliation{Department of Theoretical Physics, Wroc\l{}aw University of Science and Technology, 50-370~Wroc\l{}aw, Poland}

\author{Rudolf Bratschitsch}
\affiliation{Institute of Physics and Center for Nanotechnology, University of M\"unster, 48149 M\"unster, Germany}

\author{Tilmann Kuhn}
\affiliation{Institute of Solid State Theory, University of M\"unster, 48149 M\"unster, Germany}

\author{Steffen Michaelis de Vasconcellos}
\email{michaelis@uni-muenster.de}
\affiliation{Institute of Physics and Center for Nanotechnology, University of M\"unster, 48149 M\"unster, Germany}

\author{Daniel~Wigger}
\email{daniel.wigger@pwr.edu.pl}
\affiliation{Department of Theoretical Physics, Wroc\l{}aw University of Science and Technology, 50-370~Wroc\l{}aw, Poland}


\begin{abstract}
Single-photon emitters in solid-state systems are important building blocks for scalable quantum technologies. Recently, quantum light emitters have been discovered in the wide-gap van der Waals insulator hBN. These color centers have attracted considerable attention due to their quantum performance at elevated temperatures and wide range of transition energies. Here, we demonstrate coherent state manipulation of a single hBN color center with ultrafast laser pulses and investigate in our joint experiment-theory study the coupling between the electronic system and phonons. We demonstrate that coherent control can not only be performed resonantly on the optical transition giving access to the decoherence but also phonon-assisted, which reveals the internal phonon quantum dynamics. In the case of optical phonons we measure their decoherence, stemming in part from their anharmonic decay. Dephasing induced by the creation of acoustic phonons manifests as a rapid decrease of the coherent control signal when traveling phonon wave packets are emitted. Furthermore, we demonstrate that the quantum superposition between a phonon-assisted process and the resonant excitation causes ultrafast oscillations of the coherent control signal. Our results pave the way for ultrafast phonon quantum state control on the nanoscale and open up a new promising perspective for hybrid quantum technologies.
\end{abstract}


\date{\today}

\maketitle

\section{Introduction}
The ultrafast optical coherent manipulation of individual quantum systems is a key challenge in the development of quantum technologies~\cite{dowling2003quantum}. Over the last decades several different solid-state systems, such as color centers in diamond~\cite{aharonovich2011diamond}, perovskite~\cite{utzat2019coherent}, colloidal~\cite{klimov2017nanocrystal}, and self-assembled quantum dots~\cite{michler2003single}, have been realized, improved, and utilized~\cite{bayer2019bridging}. Recently, the family of quantum emitters has been joined by single-photon sources in layered van der Waals materials~\cite{toth2019single, michaelis2022single}. Compared to bulk materials these anisotropic crystals with their two-dimensional character are particularly promising due to their structural flexibility~\cite{liu2014elastic}. Emitters have not only been found in semiconducting transition metal dichalcogenides~\cite{tonndorf2015single, 2015_NatNano_Chakraborty, 2015_NatNano_He, 2015_NatNano_Koperski, 2015_NatNano_Srivastava} but also in insulating hexagonal boron nitride (hBN) in the form of color centers~\cite{tran2016quantum}. The latter have attracted massive attention because of their efficient single-photon emission at room temperature, emission over a wide spectral range spanning the near-infrared~\cite{camphausen2020observation}, visible~\cite{tran2016robust} and ultraviolet~\cite{bourrellier2016bright}, and spectral tunability via external electric fields~\cite{xia2019room} or strain~\cite{mendelson2020strain}. Furthermore, their characteristic coupling to phonons is remarkable. Due to the polar character, the color centers couple to both, acoustic and optical phonon modes, giving rise to strong phonon sidebands at room temperature~\cite{vuong2016phonon, wigger2019phonon}. The bright emitters are not only found in deliberately grown samples~\cite{mendelson2021grain} but also in commercially available nanocrystals, which makes their processability particularly easy~\cite{preuss2021assembly}. Although some progress has been made regarding resonant optical excitation by demonstrating Rabi oscillations in $g^{(2)}$-measurements~\cite{konthasinghe2019rabi} or the detection of lifetime-limited linewidths \cite{dietrich2020solid}, the microscopic impact of phonons still requires deeper understanding. Studies indicate that there might be a connection between the phonon coupling and the emission linewidth of the emitter~\cite{white2021phonon} but a consistent study in the spectral and the temporal domain, taking also phonon-assisted processes into account, has been pending. Here, we present the ultrafast coherent optical control of a single hBN color center. We demonstrate the role of different spectral jitter environments and reveal the impact of different phononic excitations. By combining experiment and theory, we explicitly quantify the emitter's pure and phonon-assisted coherence and at the same time access the impact of spectral noise. By performing a systematic study and thorough simulations, we reach a consistent picture of the emitter's coherence properties both in the spectral and the temporal domain and unravel the role of phonons in this context. To achieve an excellent agreement between the measurements and the simulation, we need to go beyond the calculation of correlation functions in the independent boson model~\cite{wigger2019phonon} and also take into account the exact optical excitation and detection conditions, as well as phonon decay dynamics.

\section{Results}
\subsection{Optical emission}

\begin{figure*}[t]
	\centering
	\includegraphics[width=0.9\textwidth]{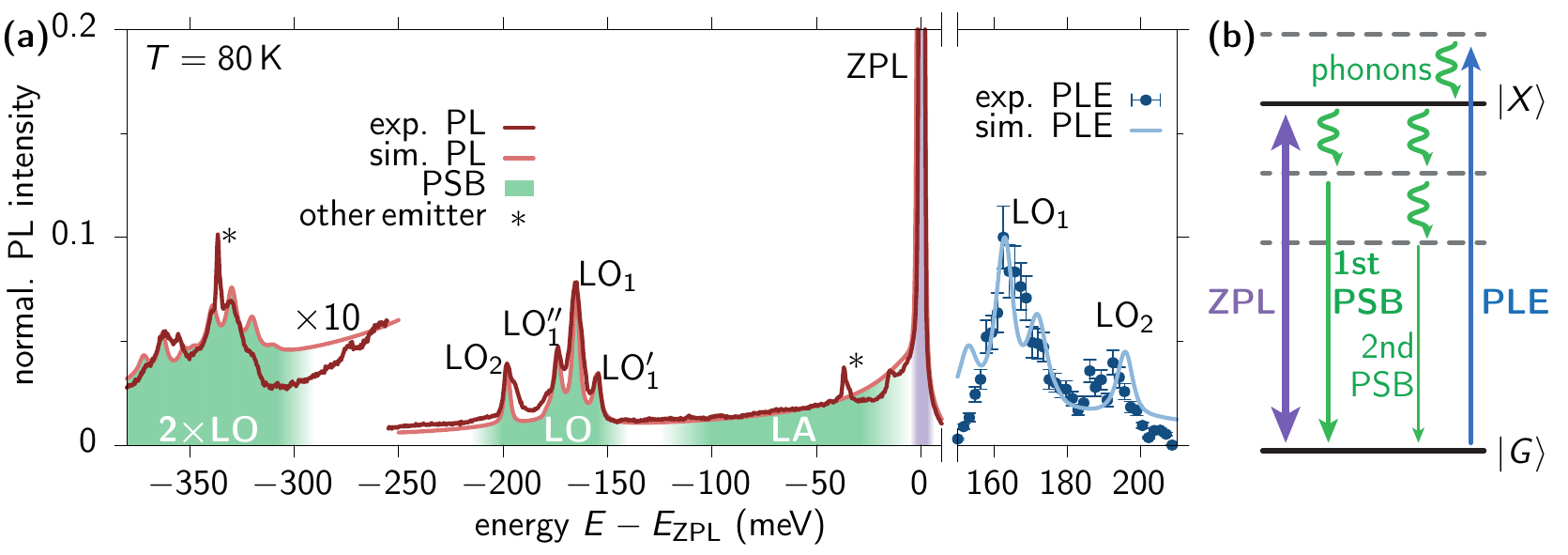}
	\caption{\textbf{Light emission from an individual hBN color center.} (a) Photoluminescence (PL) and photoluminescence excitation (PLE) spectra in red and blue, respectively. Measurement (dark) and simulation (bright) with green areas marking the different phonon sidebands (PSBs). The zero-phonon line (ZPL) energy is $E_{\rm ZPL}=2.036$~eV. (b) Schematic drawing of the two-level structure (solid lines) including ZPL (violet), phonon-assisted processes via virtual excited states (dashed lines) as PSBs (green), and PLE (blue).}
		\label{fig:spec}
\end{figure*}

For our experiments, we use commercially available hBN nanopowder crystals, which host bright and stable single-photon emitters~{\cite{wigger2019phonon, boll2020photophysics, preuss2021assembly}}. Details about sample preparation, the optical setup, and data acquisition are given in the Appendix and the Supplementary Material (SM), Sec. S1. Figure~\ref{fig:spec}(a) displays the light emission from an individual color center. Single-photon emission is confirmed by a photon correlation measurement using a Hanbury Brown and Twiss setup presented in the  SM, Sec. S1.C.  We measure (dark lines) and calculate (bright lines) photoluminescence (PL, red) and PL excitation spectra (PLE, blue). The PL spectrum (dark red) is measured at a temperature of $T = 80$~K under pulsed excitation at 2.20~eV with a pulse duration of 230~fs. We find the characteristic sharp zero-phonon line (ZPL) at $E_{\rm ZPL}=2.036$~eV, i.e., 160~meV below the excitation energy. The ZPL is accompanied by the typical peaked longitudinal optical (LO) phonon sidebands (PSBs) between 150~meV and 200~meV below the ZPL and a broad, continuous PSB stemming from longitudinal acoustic (LA) modes, exceeding $E-E_{\rm ZPL}=-100$~meV. We also detect multi-phonon sidebands for the sharp LO PSBs in the range of twice the LO phonon energies below the ZPL, i.e., between $-300$~meV and $-400$~meV. To measure the PLE spectrum, we scan the excitation laser energy above the ZPL and detect the PL integrated over ZPL and PSBs. The respective data in dark blue shows prominent maxima between $E-E_{\rm ZPL}=150$~meV and 200~meV, which excellently agrees with the LO phonon energies. This demonstrates that the emitter is efficiently excited via a LO-phonon--assisted process~\cite{wigger2019phonon, grosso2020low, hoese2020mechanical}. Such a phonon-assisted excitation was for example combined with stimulated emission depletion experiments which, when compared to PLE measurements, revealed a red-shift of one LO mode in the excited state~\cite{malein2021stimulated}. While at room temperature two prominent optical PSBs LO$_1$ at 165~meV and LO$_2$ at 200~meV are typically observed~\cite{vuong2016phonon, wigger2019phonon}, at cryogenic temperatures the LO$_1$ emission exhibits two additional lines (see e.g. Fig.~6 in Ref.~\cite{wigger2019phonon}), marked as LO$_1'$ and LO$_1''$ in Fig.~\ref{fig:spec}(a). We assume that the resonances LO$_1$ and LO$_2$ stem from the phonon dispersion maxima at the $\Gamma$ point, and the additional resonances LO$_1'$ and LO$_1''$ from Van Hove singularities at non-vanishing lattice momenta~\cite{kern1999ab, serrano2007vibrational}. The 2nd LO-PSBs between $E-E_\text{ZPL}=-300$ and $-400$~meV can be derived from these four lines LO$_1$, LO$_1'$, LO$_1''$, and LO$_2$ by summing all ten possible combinations of them (Fig.~\ref{fig:spec}(a)).

The model, schematically displayed in Fig.~\ref{fig:spec}(b), includes an optically driven two-level system (TLS) with the ground $\left|G\right>$ and the excited state $\left|X\right>$, the spectrum of the laser pulse, the detection process, and the coupling to phonons giving rise to the characteristic sidebands (see Eqs.~(\ref{eq:PL}) and \eqref{eq:PLE} for details). The coupling to phonons is considered via the independent boson model~\cite{mahan2013many} taking into account a continuous distribution of LA phonons via a standard super-ohmic spectral density with Lorentz-Drude cutoff and four discrete LO modes via the Einstein model~\cite{note1}. Anharmonic decay of the LO modes into lower energy phonons and pure dephasing of the emitter, both leading to a finite width of the spectral lines, are included via Lindblad dissipators~\cite{groll2021controlling} (see Eq.~(\ref{eq:H}) and SM, Sec. S2.A). The phonon parameters remain unchanged throughout this work and are summarized together with all other parameters used in the model in the SM, Sec. S3.E. The simulated spectrum of the 2nd LO-PSBs agrees well with the measured one, which indicates that all PSBs marked by green areas in Fig.~\ref{fig:spec}(a) belong to the same ZPL. The narrow peaks at $E-E_\text{ZPL}\approx -40$~meV and $-340$~meV (marked by $*$) do not belong to the investigated emitter (see SM, Sec. S1.E) and are therefore not considered in the following.

Overall, we find an excellent agreement between the measured spectroscopy data and the theoretical model considering one single bright state coupled to phonons (Fig.~\ref{fig:spec}(b)). All aspects of the measured time-integrated optical spectra of the emitter, namely all peak positions, their intensities and widths in both PL and PLE are reproduced by our model. The applicability of the model is further confirmed by measurements and simulations at a temperature of $T=8$~K, which are shown in the SM, Sec. S3.C. In the following, we demonstrate that the ultrafast coherent control of the emitter is consistently reproduced and explained in the considered model.

\subsection{Coherent control experiment}
\begin{figure}[t!]
\centering
	\includegraphics[width=0.95\columnwidth]{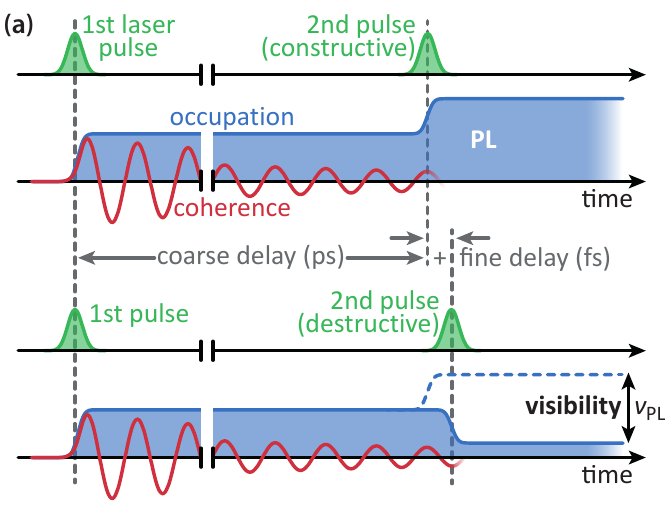}\\[2mm]
	\includegraphics[width=\columnwidth]{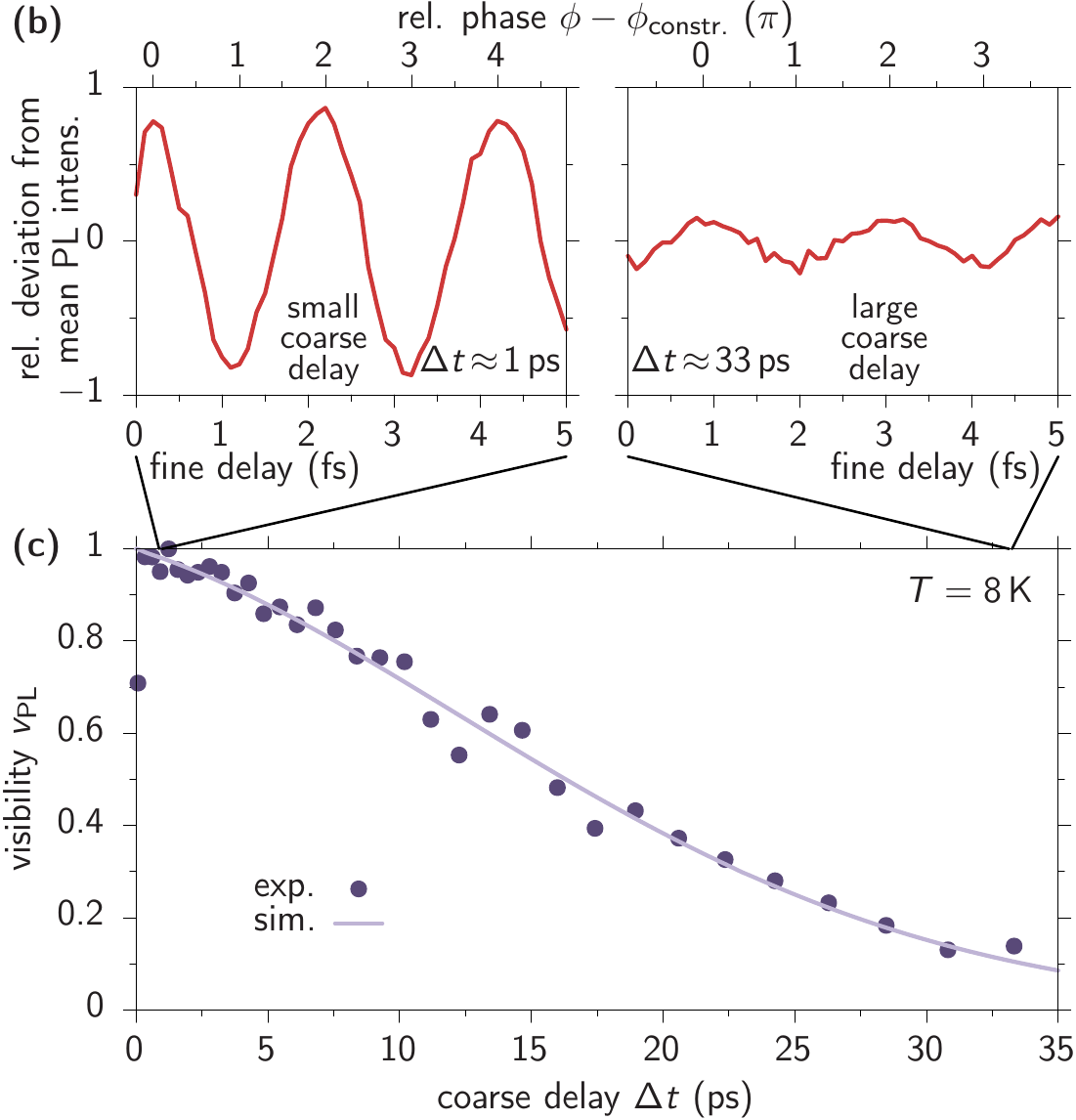}
	\caption{\textbf{Ultrafast coherent control experiment on a single-photon emitter in hBN.} (a)~Schematic of the coherent control experiment. Top: Constructive timing of the second pulse, increasing the emitter occupation. Bottom: Destructive timing decreasing the occupation. (b)~Measured variations of the PL intensity for a small coarse delay ($\Delta t= 1$~ps, left) and large coarse delay ($\Delta t= 33$~ps, right) for a resonant excitation on the ZPL at $T=8$~K. The upper x-axis marks the relative phase between emitter and second pulse. (c) PL visibility dynamics for resonant excitation of the ZPL. Experiment as dark dots and simulation as bright line.}
		\label{fig:ZPL}
\end{figure}

The time-integrated PL spectrum of a quantum emitter provides, however, only a very limited insight into its coherence properties and the impact of the specific environments such as spectral noise or lattice vibrations. To unravel these aspects, we perform ultrafast coherent control experiments on a single quantum emitter. We apply a double-pulse sequence, where the two laser pulses have equal intensities and a fixed phase relation, as schematically shown in Fig.~\ref{fig:ZPL}(a)~\cite{heberle1995ultrafast, heberle1996coherent, bonadeo1998coherent, htoon2002interplay, stufler2005quantum, kolodka2007inversion, ramsay2010review}. To gain a general understanding of the coherent control signal, it is instructive to first consider the TLS without any phonon coupling. The first laser pulse creates a coherent superposition  $\left|\psi\right> = \alpha_G\left|G\right> + \alpha_X\left|X\right>$ of ground and excited state of the emitter, characterized by its occupation $f=\left<\left|X\right>\left<X\right|\right>=|\alpha_X|^2$ and coherence $p=\left<\left|G\right>\left<X\right|\right>=|\alpha_G^*\alpha_X^{ }|e^{-i\phi_0}$, as depicted as blue and red curves in Fig.~\ref{fig:ZPL}(a), respectively. The occupation typically decays within a few nanoseconds~\cite{wigger2019phonon} and can therefore be assumed constant during the considered time scales in the picosecond range. The initial phase $\phi_0$ of the quantum coherence $\phi$ is determined by the phase of the first laser pulse. During the following dynamics it rotates with the optical transition frequency of the emitter $\phi(t)=\phi_0+\omega_\text{ZPL} t$, with $\omega_{\rm ZPL}=E_{\rm ZPL}/\hbar =2\pi/T_{\rm ZPL} \approx 3\ {\rm fs}^{-1}$ and its amplitude $|p|$ is damped with the pure dephasing rate $\gamma_{\rm pd}$. After a delay time $\Delta t$, the second laser pulse interacts with the quantum emitter. The final occupation of the quantum system depends on the relative phase between the second pulse and the quantum coherence. Because the applied pulse pair has a fixed phase relation, the phase difference between pulse and quantum state can be varied by tuning the time delay between the pulses $\Delta t$. This is visualized in Fig.~\ref{fig:ZPL}(a) by the two distinct timings for the second pulse (green Gaussians). When the second pulse arrives with a delay $\Delta t_{\rm constr.}$ such that the emitter's occupation gets efficiently enhanced (Fig.~\ref{fig:ZPL}(a), top), we call this ``constructive". In contrast, as shown in the bottom of Fig.~\ref{fig:ZPL}(a) for a delay of half a coherence period later $ \Delta t_{\rm constr.}+T_{\rm ZPL}/2=\Delta t_{\rm destr.}$ the impact of the second pulse is ``destructive" and the occupation is decreased. The occupation after the second pulse will therefore oscillate with the frequency $\omega_{\rm ZPL}$ as a function of the pulse delay $\Delta t$. If the coherence has decayed entirely, the impact of the second pulse does not depend on the delay anymore, because the system is in a statistical mixture of $\left|G\right>$ and $\left|X\right>$ with $|p|=0$.

After the second laser pulse, the occupation of the emitter decays fully into the ground state within its nanosecond radiative lifetime by emitting a photon. The next pulse pair hits the emitter at the repetition time of the laser (ns), which is long enough to guarantee the full decay of the emitter. Note, that a potential decay into a long-lived intermediate state with a lifetime longer than the repetition time does not disturb the coherent control experiment. Such a state would not be affected by the next pulse pair and therefore not contribute to the detected signal. The measured PL intensity is proportional to the occupation of the emitter, and we can compare the PL between constructive and destructive action of the second pulse within one period $T_{\rm ZPL}$. With this, we define the visibility of the coherent control experiment via 
\begin{equation}
v_{\rm PL} = \frac{{\rm max}(I_{\rm PL})-{\rm min}(I_{\rm PL})}{{\rm max}(I_{\rm PL})+{\rm min}(I_{\rm PL})}\,,
\end{equation}
which is a measure for the quantum coherence of the color center. In the experiment the visibility is retrieved from the data by a more involved procedure to reduce the impact of noise, as discussed in the SM, Sec. S1.D. Details on the simulation of the visibility are given in the SM, Sec. S2.D.3.

In the experiment, we use laser pulses (duration $1/\alpha_{\rm l}= 230$~fs) provided by a femtosecond Er:fiber laser system with nonlinear conversion stages (see Appendix). The visibility of the coherent control signal is detected by varying the delay on a femtosecond time scale (rotation of the phase), while the typical decay of the signal due to decoherence is on a picosecond time scale. Therefore, we use a Mach-Zehnder interferometer to create two pulses with a coarse delay varying on a picosecond time scale and a fine delay scanning over a few femtoseconds (see Fig.~\ref{fig:ZPL}(a) and SM, Sec. S1.B). While changing the fine delay $\Delta t_{\rm f}$, we detect the PL of the emitter via the PSBs and observe the oscillating signal depicted in Fig.~\ref{fig:ZPL}(b). The dynamics cover several oscillations of the quantum coherence as marked by the relative phase between the emitter and the second pulse on the upper x-axis. For small coarse delay times ($\Delta t= 1$~ps), the oscillation amplitude is close to unity (Fig.~\ref{fig:ZPL}(b), left panel) due to the almost perfect constructive and destructive interaction for short delays, implying $v_{\rm PL}\approx1$ as discussed in the SM, Sec. S1.D. As mentioned earlier, for longer delays ($\Delta t= 33$~ps in the right panel) the decoherence of the system before the second pulse results in a reduced oscillation amplitude of the detected PL intensity and consequently a smaller coherent control visibility.

\subsection{Resonant Excitation}\label{sec:res}
\begin{figure*}[t]
\centering
	\includegraphics[width=0.9\textwidth]{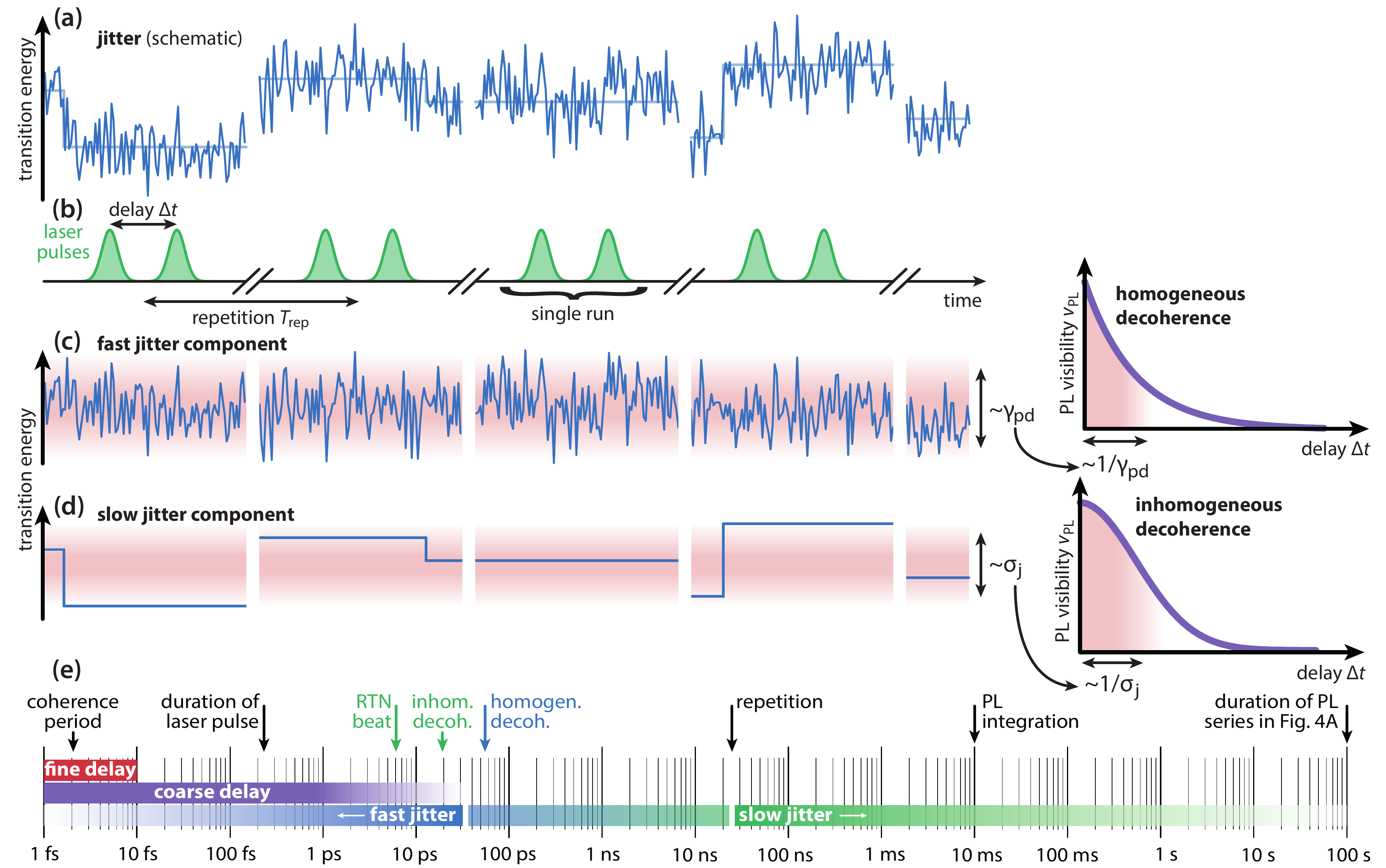}
	\caption{\textbf{Jitter dynamics and time scales for coherent control.} (a) Illustration of the spectral jitter changing the transition energy in time. (b)~Illustration of the repetition of the coherent control experiment. (c, d) Fast and slow jitter component with corresponding homogeneous and inhomogeneous decay of the PL visibility. (e) Illustration of time scales in the experiment. Quantities marked in blue (homogeneous decoherence) originate from fast and those in green (random telegraph noise (RTN), inhomogeneous decoherence) from slow jitters.}
		\label{fig:times}
\end{figure*}

Figure~\ref{fig:ZPL}(c) displays measurements (dark dots) and corresponding simulations (bright line) of the coherent control PL visibility for delays between $\Delta t=0$ and 35~ps at a temperature of $T=8$~K. The optical excitation is in resonance with the ZPL and results in a steady decay of the visibility over time. The first crucial observation is that this decay cannot be fitted by a single mono-exponential or Gaussian function (see SM, Sec. S3.D). Instead, the measured decoherence dynamics are excellently reproduced by the theory when considering a combination of an exponential (homogeneous/pure) and a Gaussian (inhomogeneous) dephasing contribution via (see Eq.~\eqref{eq:I_0} and discussion after Eq.~\eqref{eq:v})
\begin{gather}
v_\text{PL} (\Delta t) = \exp\left[ -\frac{\gamma_\text{pd}}{2}\Delta t- \frac{\sigma_{\rm j}^2(\Delta t)^2}{2}\right]\,, \label{eq:exp_gauss}\\ 
 \text{with}\quad 2/\gamma_{\rm pd}= 55\ {\rm ps}\quad {\rm and}\quad 1/\sigma_{\rm j}= 19\ {\rm ps}\,. \notag
\end{gather}
As illustrated in Fig.~\ref{fig:times}, these two dephasing contributions can be identified with spectral jitter components (also often referred to as spectral noise or spectral diffusion)~\cite{white2021phonon}, which act on different characteristic time scales. The critical value for this time scale is the repetition of the two-pulse coherent control experiment, which is $T_\text{rep}=25$~ns in our case (Fig.~\ref{fig:times}(b)). If the considered spectral jitter component is faster than the delay between the pulses, the transition energy of the emitter is likely to change between the two pulses in a single run of the experiment (Fig.~\ref{fig:times}(c)). This disturbs the quantum phase and results in an exponential loss of the coherence~\cite{weiss2012quantum}, which is usually referred to as homogeneous dephasing. If the jitter is slower than the repetition time of the experiment, it is more likely that the transition energy of the emitter changes between single runs of the experiment (Fig.~\ref{fig:times}(d))~\cite{spokoyny2020effect}. In this situation the  successive repetition of the coherent control experiment can be treated as an ensemble average over the spectral distribution representing the jitter (see SM, Sec. S2.D.5 for details). In our case, the Gaussian dephasing dynamics correspond to a Gaussian spectral distribution of the jitter-induced energy shifts, which is usually called inhomogeneous broadening. The combined spectral jitter is schematically shown in Fig.~\ref{fig:times}(a). Figure~\ref{fig:times}(e) and Table~\ref{tab:1} provide an overview of the various time scales including fast and slow spectral jitter. In addition, the table lists the determined dephasing times and their corresponding widths in optical spectra and in which figures the respective time scales are found.

From homogeneous and inhomogeneous dephasing, we estimate an effective emitter coherence time, where Eq.~\eqref{eq:exp_gauss} has dropped to $1/e$ of the initial amplitude, to
\begin{equation}
T_{\rm eff}=\frac{\gamma_{\rm pd}}{2\sigma_{\rm j}^2}\left(\sqrt{8\sigma_{\rm j}^2/\gamma_{\rm pd}^2 + 1} - 1\right) \approx 21\ {\rm ps}\,.
\end{equation}
This value is on the same order of magnitude as previously reported coherence times of around 80~ps in a hBN emitter~\cite{sontheimer2017photodynamics}. $T_\text{eff}\approx 21$~ps indicates that the investigated emitter experiences a strong spectral jitter on different time scales. The jitter might stem from laser induced charge fluctuations in hBN or the substrate~\cite{empedocles1997quantum}, which influence the transition energy of the quantum emitter, e.g., via the Stark effect.

\begin{table}[h]
\caption{\label{tab:1} List of the different time scales in the experiment. ``Fast" and ``slow" spectral jitter are defined with respect to the repetition time of the experiment. The decoherence times are given for a temperature of $T=8$~K. The listed times are also illustrated in Fig.~\ref{fig:times}(e).}
\begin{tabular}{c|c|c|c|c}
~ & Fig. & time scale & jitter & \begin{tabular}{@{}c@{}}spectral \\  width \end{tabular}  \\
\hline
\hline
\begin{tabular}{@{}c@{}}coherence \\ period $T_\text{ZPL}$\end{tabular}  	& \ref{fig:ZPL}(b) & 2.1 fs &  &  \\ \cline{1-3}
fine delay 	$\Delta t_\text{f}$		& \ref{fig:ZPL}(b) & 0\,--\,10 fs &   & \\ \cline{1-3}
\begin{tabular}{@{}c@{}}duration of \\  laser pulse $1/\alpha_\text{l}$ \end{tabular}& \ref{fig:LO}(a) & 230 fs & fast &  \\ \cline{1-3}
coarse delay $\Delta t$			& \ref{fig:ZPL}(c), \ref{fig:jitter}(c) & 0\,--\,35 ps  &$\uparrow$ & \\ \cline{1-4}
\begin{tabular}{@{}c@{}}repetition of \\ experiment $T_\text{rep}$ \end{tabular}  	& - & 25 ns & $\downarrow$ &  \\ \cline{1-3}
PL integration 					& \ref{fig:jitter}(a) & 10 ms & slow & \\ \cline{1-3}
duration of PL series 			& \ref{fig:jitter}(a) & 100 s &  & \\ \cline{1-3}
\hline
\hline
telegraph noise beat			 	& \ref{fig:jitter}(b, c) & 6 ps & (slow) & 0.7 meV  \\
inhom. decoh. $1/\sigma_\text{j}$ 	& \ref{fig:ZPL}(c) & 19 ps & (slow) & 35 \textmu eV \\
hom. decoh. $2/\gamma_\text{pd}$ 	& \ref{fig:ZPL}(c) & 55 ps & (fast) & 12 \textmu eV 
\end{tabular}
\end{table}

\subsection{Telegraph noise}\label{sec:RTN}
\begin{figure}[h]
\centering
	\includegraphics[width=\columnwidth]{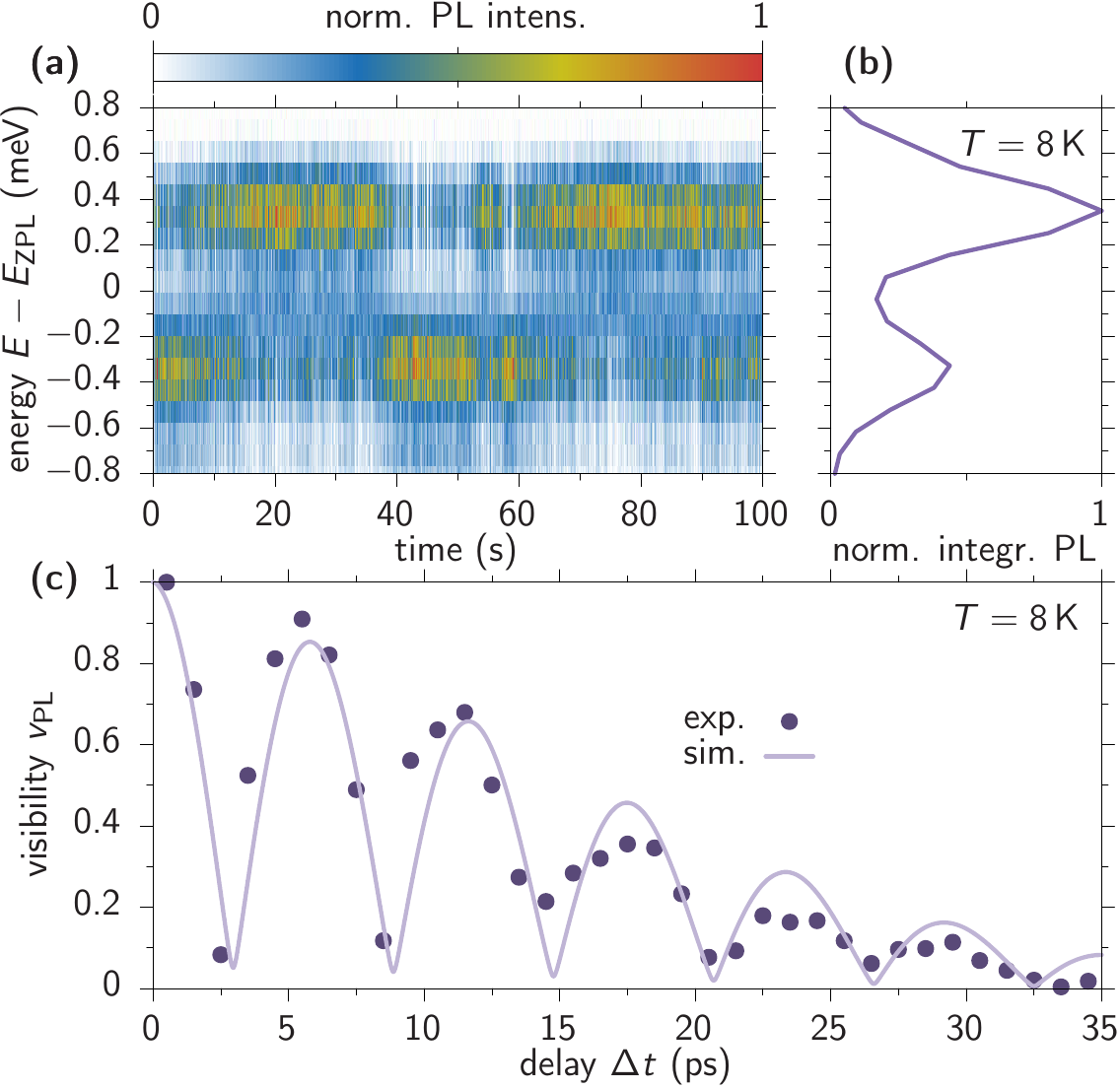}
	\caption{\textbf{Coherent control experiment with telegraph noise.} (a) PL spectra as a function of detection time covering 100 seconds showing spectral jumps between two energies. (b) Time-integrated PL spectrum retrieved from (a). The ZPL energy of $E_{\rm ZPL}=2.036$~eV is defined as the mean value of the two peak positions. (c) Coherent control visibility dynamics of the double-peaked ZPL. Experiment as dark dots and simulation as bright line.}
		\label{fig:jitter}
\end{figure}
Interestingly, in the course of the experiments after cooling from room temperature to cryogenic temperatures the same emitter sometimes performed random jumps between two distinct emission energies, i.e., random telegraph noise (RTN), on a millisecond time scale as shown in Fig.~\ref{fig:jitter}(a)~\cite{tran2019suppression}. When assuming that the spectral jitter is induced by variations of the local surrounding of the emitter, heating and cooling cycles can likely result in varying noise situations. For each time step in this measurement the PL was integrated over 10~ms. As explained in the previous section (see also Table~\ref{tab:1}) this spectral jitter is slow and can therefore be treated as a spectral ensemble. The RTN character is evident, because one of the emission lines is dim when the other is bright. We also find slight variations in the peak positions hinting towards an additional fluctuation of the peak energies with a smaller amplitude on the order of the spectral linewidth. The amplitude of this jitter is in the range of a few tens of \textmu eV (see SM, Sec. S1.F for details), which is in reasonable agreement with the value determined from the inhomogeneous decoherence $\hbar\sigma_{\rm j}=35$~\textmu eV in the coherent control experiment in the previous section. After time integration, the RTN results in the double-peak structure of the ZPL's PL spectrum in Fig.~\ref{fig:jitter}(b). The corresponding coherence dynamics in Fig.~\ref{fig:jitter}(c) exhibit a characteristic beat, given by the energy splitting of the RTN of $\Delta E_{\rm ZPL}=0.7$~meV. Also, this behavior is reproduced by our model when additionally considering the RTN as an ensemble average (see discussion after Eq.~(\ref{eq:v}) in the Appendix). These results demonstrate that even in the presence of different sources of spectral jitter, i.e., the fast jitter leading to the homogeneous decoherence, the slow Gaussian noise resulting in the inhomogeneous decoherence, and the RTN introducing a beat, coherent control of the quantum state of an individual hBN color center is possible. The respective measurements of the coherence dynamics allow us to determine the present noise characteristics with high precision.

\subsection{Optical phonons}
\begin{figure}[t]
	\centering
	\includegraphics[width=\columnwidth]{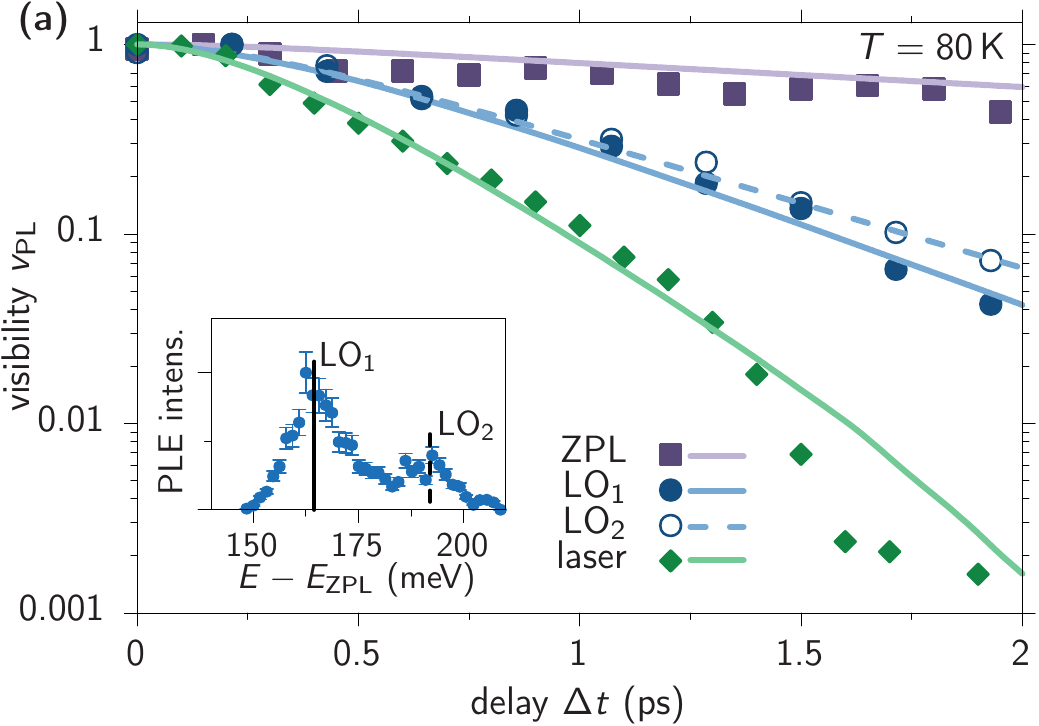}\\
	\includegraphics[width=\columnwidth]{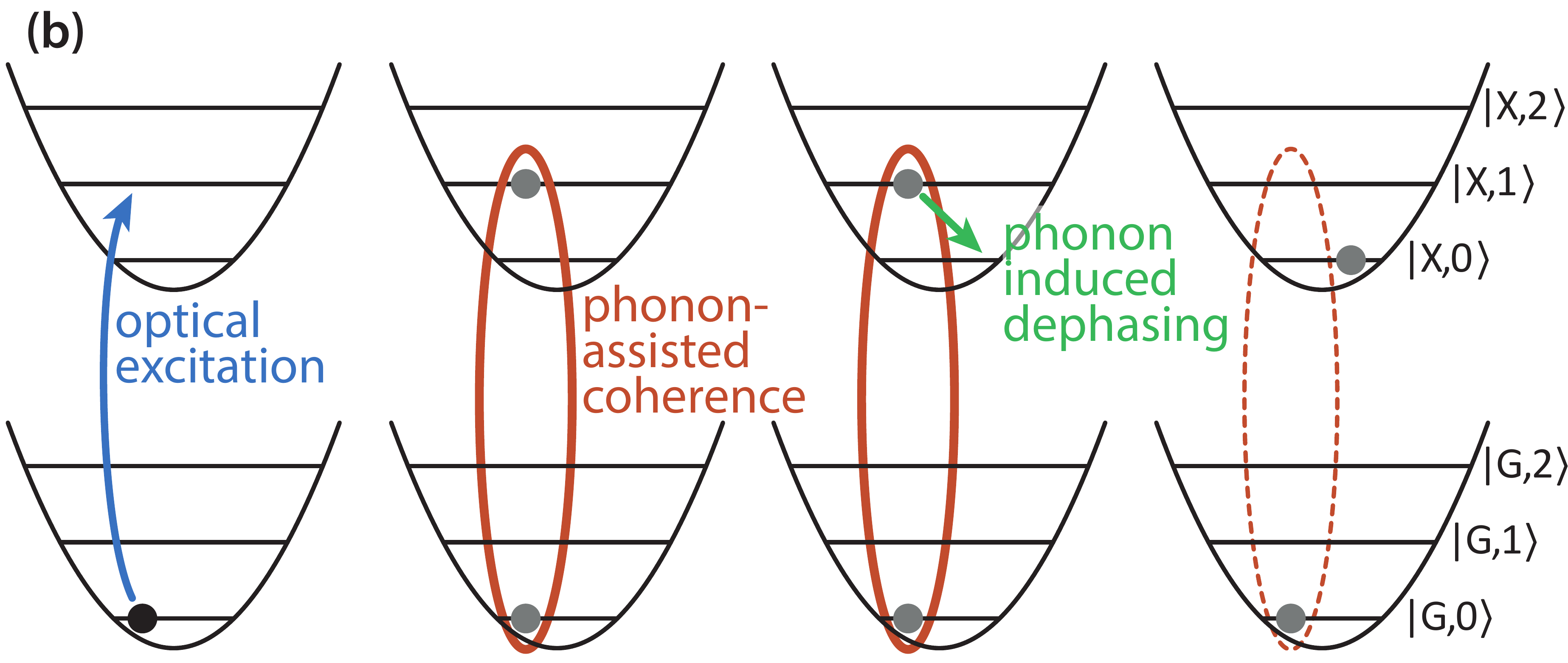}
	\caption{\textbf{Coherent control of optical phonon sidebands (PSBs).} (a) Coherent control visibility dynamics for LO-phonon--assisted excitation in blue. Solid curve for the LO$_1$ and dashed curve for the LO$_2$ mode, experiment as dark symbols and simulation as bright lines. In addition the visibility for resonant excitation is shown in purple and the laser autocorrelation in green. The respective PLE spectrum of the LO PSBs is shown in the inset. (b) Illustration of the phonon-assisted coherence decay. The harmonic potentials represent phonon excitations in ground $\left|G\right>$ and excited $\left|X\right>$ state. The coherence is indicated by the orange ellipse. Phonon induced dephasing (green arrow) results in the loss of coherence (dashed ellipse).}
		\label{fig:LO}
\end{figure}
Phonon-assisted excitation and emission have a remarkable strength in hBN color centers~\cite{vuong2016phonon, wigger2019phonon, feldman2019phonon, grosso2020low, khatri2020optical}, with PSBs reaching peak intensities of almost 10\% of the ZPL at $T=80$~K, as discussed for the PL and PLE spectra in Fig.~\ref{fig:spec}. We now detune the carrier frequency of the applied laser pulses to study the emitter's coherence when phonon-assisted excitations are addressed. As discussed throughout the SM and in the Appendix, the experimental results crucially depend on the spectral widths of the laser and the emitter. To reduce the impact of slow spectral jitters, we increase the temperature to $T=80$~K, which results in a broadening of the spectral features, especially the ZPL. This generates a more robust situation as the slow jitters have a reduced relative impact on the broadening. Note, that these experiments were performed when the emitter did not show any RTN, i.e., only a single ZPL was present. The violet data in Fig.~\ref{fig:LO}(a) (dark squares: experiment, bright line: simulation) shows the coherent control visibility for a resonant ZPL excitation which exhibits a dephasing time of $2/\gamma_{\rm pd} =3.5$~ps (see Eq.~\eqref{eq:I_0}). This more than ten-fold increase of the decoherence rate compared to $T=8$~K confirms the temperature-dependent broadening of the ZPL (see Eq.~\eqref{eq:PL} in Appendix). We now tune the pulsed optical excitation to the LO PSBs, LO$_1$ (filled circles, solid line) and LO$_2$ (open circles, dashed line) as depicted in the inset of Fig.~\ref{fig:LO}(a). Both resonances of the PSBs in the PLE spectrum stem from flat parts of the phonon dispersion with a vanishing group velocity~\cite{kern1999ab}. Consequently, the created phonons remain localized in the region of the color center's wave function. In Fig.~\ref{fig:LO}(a) we plot the measured (dark blue dots) and simulated (bright blue lines) visibility of the coherent control signals, i.e., the coherence dynamics, for optical excitations on the LO$_1$ and LO$_2$ PSB above the ZPL. As a reference, we additionally display the autocorrelation signal of the laser (green, dark diamonds: experiment, bright line: simulation, see SM, Sec. S3.B). It is measured by directly interfering the two laser pulses without interacting with the sample and calculated via Eq.~\eqref{eq:auto_laser}. We find that for the LO-assisted excitations, the coherence decays significantly faster than for the resonant excitation of the ZPL (violet), but decays slower than the autocorrelation of the laser pulses.

As schematically shown in Fig.~\ref{fig:LO}(b), the first laser pulse (blue arrow) creates a phonon-assisted coherence between the states $\left|G,0\right>$ and $\left|X,1\right>$, where the second entry describes the number of excited phonons (orange ellipse). Any phonon-induced dephasing (PID) then leads to the destruction of the quantum coherence between the optical phonon, the ground state, and the excited state of the quantum emitter (dashed ellipse). This process is naturally faster than the decoherence without phonons. One obvious source of PID is the finite lifetime of optical phonons, which typically decay anharmonically into lower energy modes~\cite{cusco2018isotopic}. This process is given by the transition $\left|X,1\right>\to\left|X,0\right>$. As shown in the SM, Sec. S2.E.3 a finite LO lifetime has a similar impact on the coherent control signal as a non-vanishing dispersion of the optical modes, which is another possible source of decoherence. To reproduce the measured visibility dynamics in Fig.~\ref{fig:LO}(a) we find phonon decay times of $1/\gamma_{\rm LO_1}=0.3$~ps and  $1/\gamma_{\rm LO_2}=0.4$~ps within our model. LO phonon lifetimes in hBN can also be measured by Raman scattering and are on the order of a few ps~\cite{cusco2018isotopic}, which is longer than the values found here. Therefore, additional decoherence mechanisms apart from their decay, e.g., a non-vanishing dispersion, are likely.

\subsection{Acoustic phonons}
\begin{figure}[h!]
	\centering
	\includegraphics[width=\columnwidth]{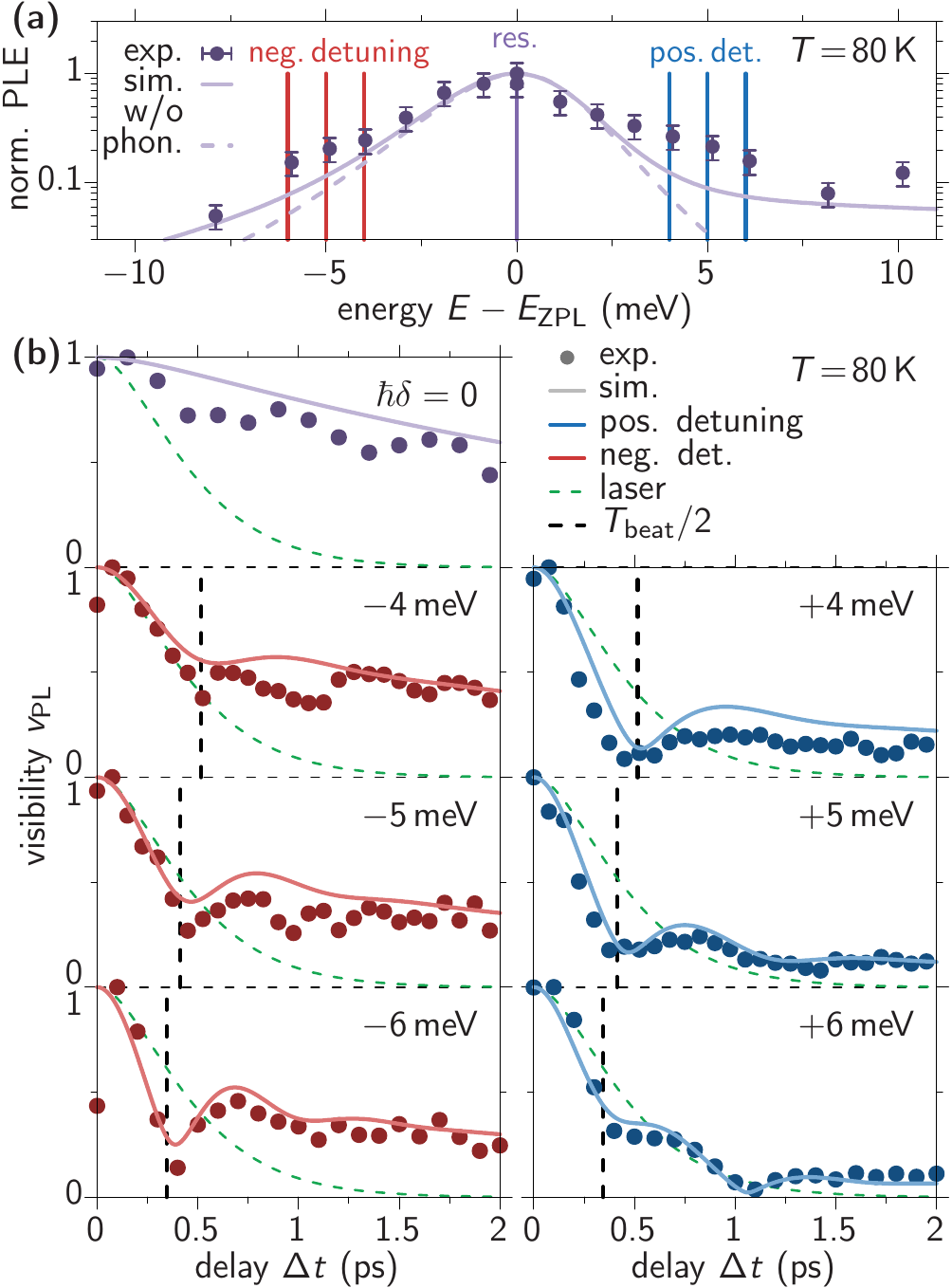}\\
	\includegraphics[width=\columnwidth]{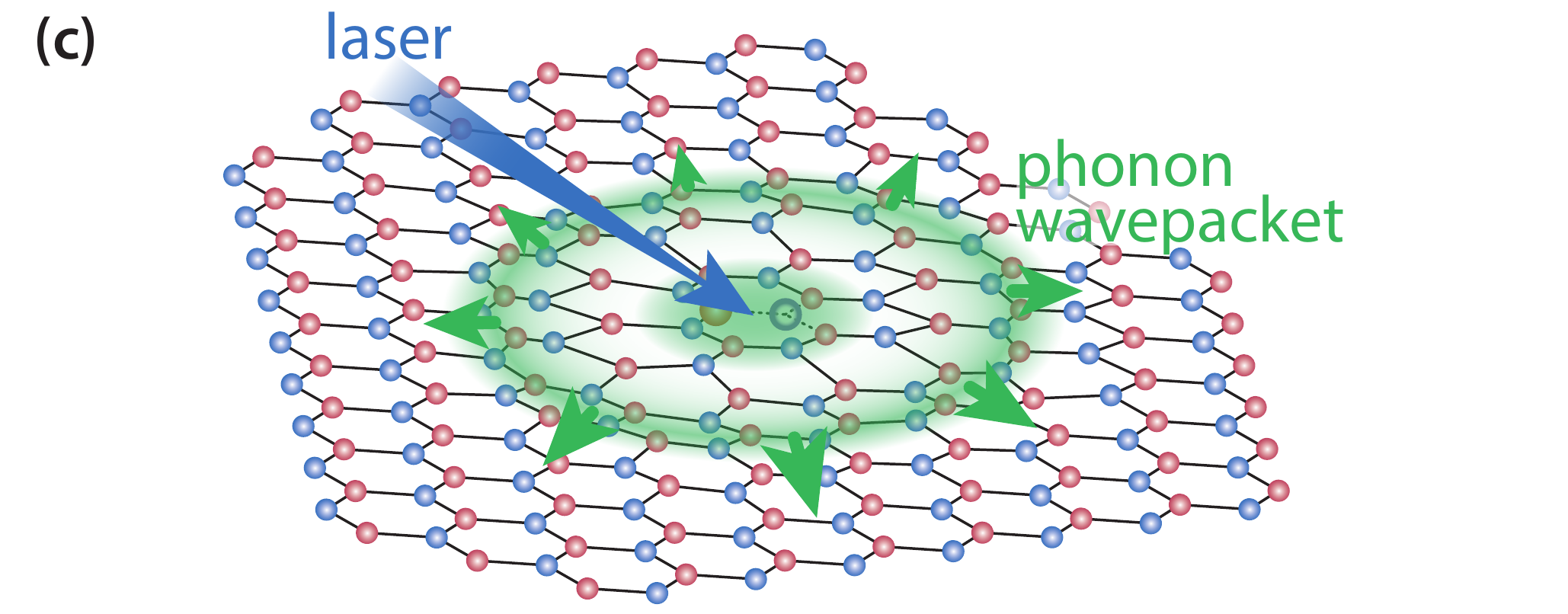}
	\caption{\textbf{Coherent control of acoustic PSBs.} (a) Measured and simulated PLE data as dark dots and bright curve, respectively. The colored vertical lines mark the detunings considered in (b). (b) Visibility dynamics for increasing detunings from top to bottom with negative detunings left and positive right. Experiment as dark dots and simulation as bright lines. Laser autocorrelation in green. Vertical dashed lines mark half a period corresponding to the frequency detuning between laser and ZPL. (c) Illustration of the laser pulse induced generation of a phonon wave packet.}
		\label{fig:PID}
\end{figure}%
Next, we reveal the impact of LA phonons on the coherence of the hBN color center. To confirm that LA-phonon--assisted excitations are feasible~\cite{khatri2020optical, hoese2020mechanical}, we perform a PLE scan at $T=80$~K on a meV range around the ZPL as depicted in Fig.~\ref{fig:PID}(a), where the experiment is shown as dark dots and the simulation as bright line. We find that the PLE spectrum is asymmetric with a flatter slope on the low energy side and a plateau-like region on the high energy side. This asymmetry has two reasons: (i) at  $T=80$~K the LA phonon emission is more likely than LA phonon absorption, which results in a stronger LA PSB at positive detuning  forming the plateau. To verify this, the dashed line in Fig.~\ref{fig:PID}(a) depicts the calculated PLE spectrum without phonon coupling. (ii) The measured PLE spectrum is a convolution of the intrinsic absorption spectrum of the emitter and the laser spectrum. The asymmetry of the laser spectrum has a non-trivial impact on the final PLE spectrum, as seen in the dashed curve and discussed in detail in the SM, Sec. S4.B.

The coherent control visibility with acoustic-phonon--assisted optical excitation is shown in Fig.~\ref{fig:PID}(b) for $T=80$~K. The experiment is depicted as dark dots, the simulation as bright lines, and the detuning between the laser energy and the transition energy $\hbar\delta = E_\text{laser}-E_{\rm ZPL}$ increases from top to bottom, with negative detunings (red) in the left and positive (blue) in the right column. The resonant case (violet) is the same as in Fig.~\ref{fig:LO}(a) and the considered detunings ($\hbar\delta=\pm4$, $\pm5$, and $\pm6$~meV) are additionally marked as red (negative) and blue (positive) vertical lines in the PLE spectrum in Fig.~\ref{fig:PID}(a). As a reference, we again plot the laser autocorrelation as green dashed lines in Fig.~\ref{fig:PID}(b). For detuned (non-resonant) excitation, all coherent control signals exhibit a rapid initial visibility drop within 0.5~ps, which is followed by a slower decoherence or, interestingly, an intermediate recovery of the coherence. We find a remarkable agreement between the experimental data and the theoretical simulations. Notably, the asymmetric spectral shape of the femtosecond laser pulses has to be accurately considered in the theoretical model to achieve this almost perfect consistent match between the coherent control signal dynamics and the PL spectrum in Fig.~\ref{fig:spec}(a) (see SM, Sec. S4.A for details). According to our model, the observed coherence dynamics for delays below $\approx 1$~ps have two contributions (see Eq.~(\ref{eq:v_short}) and SM, Sec. S2.D.4): (i) An initial loss of coherence resulting in the rapid drop of the coherent control visibility, followed by a quasi-stationary coherence for delays larger than $\approx 1$~ps; (ii) a quantum beat resulting in an oscillation for delays below $\approx 1$~ps.

Finding (i) is specific for the coupling to acoustic phonons and well known from self-assembled semiconductor quantum dots as a PID effect~\cite{jakubczyk2016impact, wigger2020acoustic}. Its physical origin is the formation of a polaron, i.e., a deformed lattice configuration in the emitter region, which is created on the time scale of the optical excitation. As illustrated in Fig.~\ref{fig:PID}(c), this rapid lattice deformation after the first laser pulse is accompanied by the emission of a phonon wave packet due to the linear dispersion of the LA phonons. As this phonon pulse is emitted from the color center with the speed of sound (several nm/ps in hBN), it carries part of the coherence with it, which is lost for the quantum state of the emitter. This loss of coherence is then probed by the second laser pulse in the coherent control experiment. 

The amplitude of the visibility at $\Delta t= 2$~ps reflects the coherence of the polaron state, which is generated by the first pulse. In each line of Fig.~\ref{fig:PID}(b) positive (blue, right) and negative (red, left) detuning with the same absolute value are directly compared. From the visibility values at $\Delta t= 2$~ps we find that the PID is always more pronounced for positive detunings caused by the asymmetric laser spectrum and a mismatch between phonon emission and absorption due to the thermal occupation of phonons $n_{\rm ph}$ at $T = 80$~K ($n_{\rm ph}\approx 1$ for modes with $E_{\rm ph}\approx5$~meV). For positive detuning, phonons must be generated to occupy the polaron, while for negative detunings phonons need to be absorbed. The two processes scale with $(n_{\rm ph}+1)\approx 2$ and $n_{\rm ph}\approx 1$, respectively. This mismatch results in the more pronounced PID drop of the visibility for positive detunings. The quasi-stationary signal at $\Delta t\approx2$~ps is a measure for the overlap of the laser spectrum with the ZPL. The visibility is generally smaller for larger absolute detuning, e.g., $v_{\rm PL}(\Delta t=2\,{\rm ps})\approx 0.5$ at $\hbar\delta=+4$~meV compared to $\approx 0.25$ at $\hbar\delta=+6$~meV, as the polaron is excited less directly and more via phonon-assisted processes. This leads to a stronger loss of coherence induced by the creation of the phonon wave packet when the absolute detuning $|\delta|$ is larger.

The additional oscillation of the signal, i.e., effect (ii), results in dynamics which are even faster than the laser autocorrelation (green curves in Fig.~\ref{fig:PID}(b)). We can understand these oscillations when considering the theoretical expression for the coherent control signal for short delays in Eq.~(\ref{eq:v_short}) in the Appendix. We find that the detuning of the laser and the remaining overlap with the ZPL effectively select two spectral components: the pure emitter coherence and the phonon-assisted coherence at the carrier frequency of the laser pulse. These two contributions result in a quantum beat with the frequency $\omega_\text{beat}=|\delta|$ that shows up as an oscillation in the coherent control signal. This directly explains why the oscillations get faster for larger detunings. The expected minimum from such a beat at $T_{\rm beat}/2=\pi/|\delta|$ is marked by vertical dashed lines in Fig.~\ref{fig:PID}(b) and matches well with the measured minima in the visibility dynamics. This beat is a clear indication of the coherent coupling between the acoustic phonon modes and the emitter.

\section{Conclusions}
By performing optical coherent control experiments with ultrafast laser pulses we have revealed the coherence dynamics of a single hBN color center. We found an excellent agreement with simulations, being consistent for spectrally resolved time-integrated PL and the time-resolved coherent control signal. We highlighted the impact of different sources of spectral jitter: Fast jitter on or below the picosecond time scale resulted in homogenous/exponential decoherence dynamics, while slow jitter on a nanosecond and longer time scale led to inhomogeneous/Gaussian decoherence dynamics. The presence of random telegraph noise was revealed by a characteristic beat of the coherent control signal. Furthermore, we demonstrated the impact of phonons on the emitter coherence by using phonon-assisted excitation. In the case of optical phonons we found that the increased decoherence rates are due to dephasing processes of the phonon state, which partially stem from its anharmonic decay. When the phonon-assisted excitation involved acoustic phonon modes, we found that (i) the generation of the phonon wave packet led to a characteristic drop of the coherent control signal and (ii) the simultaneous resonant and phonon-assisted excitation resulted in the selection of two energetically detuned contributions, which manifested in beat dynamics in the signal.

Hence, the coherence of hBN color centers is limited by phonon induced decoherence and spectral jitter. While the phonon impact could be controlled by deliberately shaped laser pulses~\cite{wigger2014energy, grange2017reducing, kaldewey2017demonstrating}, control over the crystal environment could reduce spectral jitter. An overall increased coherence time will enable advanced coherent control schemes, e.g., employing coherent optoelectronics~\cite{Michaelis2010coherent}.

Our demonstration of phonon-assisted coherent control of individual hBN color centers is a key step towards hybrid quantum technologies, which combine electronic and phononic excitations~\cite{groll2021controlling}. We envision that the impact of the phonons can further be controlled by external acoustic driving~\cite{wigger2021remote}, which will allow to deterministically address the photon emission properties.

\section*{Appendix A: Methods}
\subsection*{Experiment}
Commercially available hBN multilayer nanopowder (Sigma-Aldrich, diameter $<150$~nm) is annealed at 850$^\circ$\,C to increase the number of single-photon emitters and randomly scattered on a clean quartz substrate. The sample is then placed in a variable temperature helium bath cryostat with a low-temperature transmission microscope inside the sample chamber. The microscope consists of two objective lenses with a numerical aperture NA of 0.82, enabling efficient photon collection. The sample, as well as the microscope, is cooled by evaporating the coolant in the sample chamber. The cryostat is operated using liquid helium or nitrogen, resulting in sample temperatures of 8K and 80K respectively. Spectrally tunable laser pulses with durations of $230$~fs are provided by a frequency-doubled supercontinuum from a femtosecond Er:fiber laser. To obtain two identical pulses, a single laser pulse is split at a 50:50 plate beamsplitter. The path of the split pulses can be controlled by a combination of a 204~mm long translation stage and a 100~\textmu m closed-loop piezo scanner, resulting in a coarse and fine delay of the pulses in the range of ps and sub-fs respectively. The beam path of the two pulses is rejoined by an identical 50:50 plate beamsplitter and then focused on the single-photon emitter. The photoluminescence is detected either by a spectrometer and cryogenically-cooled CCD for spectroscopy and the dynamics measurements or a single-mode fiber-coupled Hanbury Brown and Twiss setup for photon correlation measurements. Further details can be found in the SM, Sec. S1.

\subsection*{Theory}
In the SM in Sec. S2 we derive all required equations to model the depicted optical signals in detail. Here, we summarize the most important final results for each measurement. We model the emitter as a two-level system (TLS) and consider the coupling to phonons in the independent boson model and the coupling to light semi-classically leading to the Hamiltonian
\begin{align}\label{eq:H}
	H &= \hbar\omega_{\mathrm{TLS}} X^{\dagger}X + \frac{1}{2}\left[\mathcal{E}(t)X^{\dagger}+\mathcal{E}(t)^*X\right]\notag\\
	&+\sum_n \hbar\omega_n b_n^{\dagger}b_n^{}+X^{\dagger}X\sum_n \hbar\left(g_n^{}b_n^{\dagger}+g_n^*b_n^{}\right)\notag\\
	& + H_{\mathrm{pd}} + H_{\mathrm{anh}}\,.
\end{align}
Here, $\hbar\omega_{\rm TLS}$ is the transition energy of the emitter in the absence of phonon coupling. The zero phonon line energy additionally includes the polaron shift
$\hbar\omega_{\rm ZPL} = \hbar\omega_{\rm TLS}-\hbar\sum_n |g_n|^2/\omega_n$, depending on the phonon frequencies $\omega_n$ and their coupling strengths $g_n$. $X^{(\dagger)}$ and $b_n^{(\dagger)}$ are the TLS and phonon annihilation (creation) operators, respectively, and $\mathcal E(t)$ describes the laser pulses, being the projection of the electric field of the laser onto the emitter's dipole moment. We also include $H_{\mathrm{pd}} $ and $ H_{\mathrm{anh}}$ to describe the pure dephasing (pd) of the TLS and the anharmonic decay (anh) of the LO phonons, which are treated in terms of Lindblad dissipators in the model.

The PL spectrum of the emitter is calculated via
\begin{align}\label{eq:PL}
	I_{\rm PL}(\omega) &= 
	\int\limits_{-\infty}^{\infty}{\rm d} \Omega R(\omega,\Omega) \int\limits_{-\infty}^{\infty}{\rm d} t\,G_{-+}(t)e^{-i(\Omega-\omega_{\rm ZPL}) t}e^{-\gamma_{\rm pd}|t|/2}\,,
\end{align}
where $R$ is the spectrometer response and $\gamma_{\rm pd}$ is the pure dephasing rate. For the PL presented in Fig.~\ref{fig:spec}(a) the spectrometer response is given by a normal distribution with mean $\omega$ and standard deviation $\omega \Delta \lambda/\lambda(\omega)$, where $\lambda(\omega)$ is the corresponding wavelength and $\Delta\lambda=0.3$~nm is the resolution of the spectrometer. We emphasize that the spectrometer response is crucial for accurately modeling the PL spectra as the resolution influences the relative height of ZPL and LA-PSB.  For some phonon parameters, this implies that they cannot be accurately determined from the PL spectra, unless the spectrometer response is well-known, as discussed in the SM, Sec. S3.C.

 $G_{-+}(\tau)$ in Eq.~\eqref{eq:PL} is the phonon correlation function, which captures the information on the emitter-phonon coupling. It is given by
\begin{align}\label{eq:G}
G_{-+}(\tau)&=\exp[\phi(\tau)-\phi(0)]\\
\phi(\tau)&=\int\limits_{0}^{\infty}{\rm d}\omega\,\frac{J_{\rm{LA}}(\omega)}{\omega^2}\left\lbrace\left[n(\omega)+1\right]e^{-i\omega\tau}+n(\omega)e^{i\omega\tau}\right\rbrace\notag\\
&+\int\limits_{0}^{\infty}{\rm d}\omega\,\frac{J_{\rm{LO}}(\omega)}{\omega^2}\exp\left[-i\omega\tau-\frac{\gamma(\omega)}{2}|\tau|\right]\,.\label{eq:phi}
\end{align}
Here, $J_{\rm LA}(\omega)=\sum_{n\in\lbrace{\rm LA}\rbrace}|g_n|^2\delta(\omega_n-\omega)$ is the LA spectral density, where the sum runs over all LA modes. The LO spectral density is defined in an analogous way. $n(\omega)$ is the thermal occupation of the phonon mode with frequency $\omega$. In the case of the LO modes we assume $n(\omega)=0$ since they have energies $>150$~meV and we consider temperatures well below 300~K. $\gamma(\omega)$ is the decay rate of the LO mode with frequency $\omega$. Equation~\eqref{eq:phi} shows that the central quantity determining the phonon correlation function is the spectral density. All details on the structure of the couplings $g_n$, i.e., its dependence on the quantum number $n$, is lost in this effective description and we can in principle choose a spectral density for our modeling that fits best with the experimental observation. For the LAs we choose a super-ohmic spectral density with a Lorentz-Drude cutoff. The LO spectral density is modeled as a sum of four $\delta$-functions, i.e., it describes four dispersionless modes in the Einstein model.

The PLE spectrum and the visibility dynamics of the coherent control experiment can be calculated from the same expression
\begin{align}\label{eq:I_0}
I_0(\omega_l,\Delta t)&=\int\limits_{-\infty}^{\infty}{\rm d}t\,\int\limits_{-\infty}^{\infty}{\rm d}\tau\,\tilde{\mathcal{E}}^*(t)\tilde{\mathcal{E}}(t+\tau)e^{-i\omega_l\tau} \\
	&\times G_{-+}(\Delta t-\tau)e^{-i\omega_{\rm ZPL}(\Delta t-\tau)}e^{-\gamma_{\rm pd}|\Delta t-\tau|/2}\,,\notag
\end{align}
where $\tilde{\mathcal{E}}(t)$ is the envelope of a single laser pulse and $\omega_l$ its carrier frequency. In this paper we define the carrier frequency as the maximum of the detected laser spectrum, as this can be conveniently identified in experiment. The temporal shape of the laser pulse is derived from an asymmetric spectrum, modeled by a Pearson-4 distribution (see SM, Sec. S3.B for details). The PLE is then given by
\begin{equation}\label{eq:PLE}
I_{\rm PLE}(\omega_l)=I_0(\omega_l,\Delta t=0)
\end{equation}
and the visibility for a fixed carrier frequency, depending on the delay $\Delta t$ between the pulses, is
\begin{equation}\label{eq:v}
v(\Delta t)=\frac{|I_0(\omega_l,\Delta t)|}{I_0(\omega_l,\Delta t=0)}\,,
\end{equation}
i.e., it is the absolute value of $I_0$, normalized by the PLE signal, which satisfies $v(\Delta t)\leq1$ with equality for $\Delta t=0$.

Slow spectral jitter, i.e., jumps of the ZPL frequency on a time scale longer than a single repetition of the experiment, can be incorporated via an ensemble average. This is done by replacing $\exp[-i\omega_{\rm ZPL}(\Delta t-\tau)]$ under the integral in Eq.~(\ref{eq:I_0}) by the weighted average over a probability distribution $p(\omega_{\rm ZPL})$. We show in the SM in Sec. S2.D.5 that Gaussian jitter on a spectral scale well below the width of the laser spectrum simply leads to a prefactor $\exp[-\sigma_{\rm j}^2(\Delta t)^2/2]$ in Eq.~(\ref{eq:I_0}) with $1/\sigma_{\rm j}\approx 19$~ps throughout the paper. Telegraph noise jitter, as seen in Fig.~\ref{fig:ZPL}, is modeled by two $\delta$-functions in $p(\omega_{\rm ZPL})$.

As discussed in the SM in Sec. S2.D.7, the visibility of the direct laser interference, measured in the same coherent control setup, is given by the autocorrelation of the electric field of a single pulse. Here, we deal with linearly polarized light, such that the amplitude of the electric field is proportional to $\mathcal{E}(t)$. Therefore the visibility of the laser interference is modeled by normalizing
\begin{equation}\label{eq:auto_laser}
\int\limits_{-\infty}^{\infty}{\rm d}t\,\tilde{\mathcal{E}}^*(t)\tilde{\mathcal{E}}(t+\Delta t)e^{-i\omega_l\Delta t}\,.
\end{equation}

In the SM in Sec. S2.D.4 we discuss an approximation for the visibility dynamics in Eq.~\eqref{eq:v} when exciting close to the ZPL and considering short time scales where the pure dephasing can be neglected. In this case we arrive at the expression
\begin{equation}\label{eq:v_short}
v(\Delta t)\approx \frac{|a(\Delta t)e^{-i\delta\Delta t}+b|}{a(\Delta t=0)+b}\,,
\end{equation}
where $\delta=\omega_l-\omega_{\rm ZPL}$ is the detuning between the laser and the ZPL. The function $a(\Delta t)$ is given by
\begin{equation}
a(\Delta t)=I_{\rm abs, PSB}(\omega_l)\int\limits_{-\infty}^{\infty}{\rm d}t\,\tilde{\mathcal{E}}^*(t)\tilde{\mathcal{E}}(t+\Delta t)\,,
\end{equation}
i.e., it is the envelope of the pulse autocorrelation, weighted by the LA-PSB contribution to the absorption spectrum at the position of the carrier frequency $\omega_l$, as defined in the SM in Sec. S2.D.4. The constant $b$ is determined by
\begin{equation}
b=B^2 I_{\rm laser}^{\rm ideal}(\omega_{\rm ZPL})\,,
\end{equation}
where $B^2=\exp[-\phi(0)]$ with $\phi(\tau)$ from Eq.~\eqref{eq:phi} is a measure of the phonon coupling strength. For strong phonon coupling or large temperatures, $B^2$ tends towards zero. In the PL spectra, $B^2$ is also the prefactor of the pure ZPL contribution. For weak phonon coupling, $B^2$ tends towards unity, i.e., the spectrum consists only of a ZPL and contains no PSBs. $I_{\rm laser}^{\rm ideal}(\omega_{\rm ZPL})$ is the ideal (perfect resolution) laser spectrum at the position of the ZPL, as defined in the SM, Sec. S2.D.2. Thus, $b$ measures the overlap of the laser pulse spectrum with the ZPL.\\
$a(\Delta t)$ measures the strength of the LA-PSB at the carrier frequency for $\Delta t=0$ and decays along with the pulse autocorrelation. Thus we expect dynamics that decay on the time scale of the laser autocorrelation towards a final value of $b/(a(0)+b)$, which is a measure of how efficiently we excite the system resonantly on the ZPL compared to off-resonant excitation via PSBs. The visibility dynamics exhibits a beat with the detuning frequency $|\delta|$, additionally to the decay, which leads to dynamics faster than the pulse autocorrelation in Fig.~\ref{fig:PID}, as discussed in the main text.\\
As a final note on the visibility dynamics on short time scales we want to emphasize that the derivation was made for the case $\gamma_{\rm pd}=0$, i.e., no pure dephasing. This leads to a $\delta$-shaped ZPL and a strong dependence of the parameter $b$ on the ZPL frequency $\omega_{\rm ZPL}$ via the overlap with the laser spectrum. Therefore in this regime ($\gamma_{\rm pd}\rightarrow 0$), spectral jitter can have a significant impact and makes the experimental identification of the discussed phenomenon challenging. Thus, rather counter-intuitive, it is more instructive to perform the coherent control experiment with detuned excitation at more elevated temperatures. In that case $\gamma_{\rm pd}$ is larger, i.e., the system dephases more quickly, leading to a broader ZPL and therefore less sensitivity on spectral jitter.

\section*{Funding}
Narodowa Agencja Wymiany Akademickiej (PPN/ULM/2019/1/00064).

\section*{Disclosures} The authors declare no conflicts of interest.

\section*{Data availability} Data underlying the results presented in this paper are not publicly available at this time but may be obtained from the authors upon reasonable request.


\section*{Supplemental document}
This document contains details on
\begin{itemize}
\item[S1] Experiment
\item[S2] Theory
\item[S3] Impact of the system parameters
\item[S4] Simulations for different pulse shapes
\end{itemize}



\begin{thebibliography}{57}%
\makeatletter
\providecommand \@ifxundefined [1]{%
 \@ifx{#1\undefined}
}%
\providecommand \@ifnum [1]{%
 \ifnum #1\expandafter \@firstoftwo
 \else \expandafter \@secondoftwo
 \fi
}%
\providecommand \@ifx [1]{%
 \ifx #1\expandafter \@firstoftwo
 \else \expandafter \@secondoftwo
 \fi
}%
\providecommand \natexlab [1]{#1}%
\providecommand \enquote  [1]{``#1''}%
\providecommand \bibnamefont  [1]{#1}%
\providecommand \bibfnamefont [1]{#1}%
\providecommand \citenamefont [1]{#1}%
\providecommand \href@noop [0]{\@secondoftwo}%
\providecommand \href [0]{\begingroup \@sanitize@url \@href}%
\providecommand \@href[1]{\@@startlink{#1}\@@href}%
\providecommand \@@href[1]{\endgroup#1\@@endlink}%
\providecommand \@sanitize@url [0]{\catcode `\\12\catcode `\$12\catcode
  `\&12\catcode `\#12\catcode `\^12\catcode `\_12\catcode `\%12\relax}%
\providecommand \@@startlink[1]{}%
\providecommand \@@endlink[0]{}%
\providecommand \url  [0]{\begingroup\@sanitize@url \@url }%
\providecommand \@url [1]{\endgroup\@href {#1}{\urlprefix }}%
\providecommand \urlprefix  [0]{URL }%
\providecommand \Eprint [0]{\href }%
\providecommand \doibase [0]{https://doi.org/}%
\providecommand \selectlanguage [0]{\@gobble}%
\providecommand \bibinfo  [0]{\@secondoftwo}%
\providecommand \bibfield  [0]{\@secondoftwo}%
\providecommand \translation [1]{[#1]}%
\providecommand \BibitemOpen [0]{}%
\providecommand \bibitemStop [0]{}%
\providecommand \bibitemNoStop [0]{.\EOS\space}%
\providecommand \EOS [0]{\spacefactor3000\relax}%
\providecommand \BibitemShut  [1]{\csname bibitem#1\endcsname}%
\let\auto@bib@innerbib\@empty
\bibitem [{\citenamefont {Dowling}\ and\ \citenamefont
  {Milburn}(2003)}]{dowling2003quantum}%
  \BibitemOpen
  \bibfield  {author} {\bibinfo {author} {\bibfnamefont {J.~P.}\ \bibnamefont
  {Dowling}}\ and\ \bibinfo {author} {\bibfnamefont {G.~J.}\ \bibnamefont
  {Milburn}},\ }\bibfield  {title} {\bibinfo {title} {Quantum technology: the
  second quantum revolution},\ }\href@noop {} {\bibfield  {journal} {\bibinfo
  {journal} {Philos. Trans. R. Soc. A}\ }\textbf {\bibinfo {volume} {361}},\
  \bibinfo {pages} {1655} (\bibinfo {year} {2003})}\BibitemShut {NoStop}%
\bibitem [{\citenamefont {Aharonovich}\ \emph {et~al.}(2011)\citenamefont
  {Aharonovich}, \citenamefont {Castelletto}, \citenamefont {Simpson},
  \citenamefont {Su}, \citenamefont {Greentree},\ and\ \citenamefont
  {Prawer}}]{aharonovich2011diamond}%
  \BibitemOpen
  \bibfield  {author} {\bibinfo {author} {\bibfnamefont {I.}~\bibnamefont
  {Aharonovich}}, \bibinfo {author} {\bibfnamefont {S.}~\bibnamefont
  {Castelletto}}, \bibinfo {author} {\bibfnamefont {D.}~\bibnamefont
  {Simpson}}, \bibinfo {author} {\bibfnamefont {C.}~\bibnamefont {Su}},
  \bibinfo {author} {\bibfnamefont {A.}~\bibnamefont {Greentree}},\ and\
  \bibinfo {author} {\bibfnamefont {S.}~\bibnamefont {Prawer}},\ }\bibfield
  {title} {\bibinfo {title} {Diamond-based single-photon emitters},\
  }\href@noop {} {\bibfield  {journal} {\bibinfo  {journal} {Rep. Prog. Phys.}\
  }\textbf {\bibinfo {volume} {74}},\ \bibinfo {pages} {076501} (\bibinfo
  {year} {2011})}\BibitemShut {NoStop}%
\bibitem [{\citenamefont {Utzat}\ \emph {et~al.}(2019)\citenamefont {Utzat},
  \citenamefont {Sun}, \citenamefont {Kaplan}, \citenamefont {Krieg},
  \citenamefont {Ginterseder}, \citenamefont {Spokoyny}, \citenamefont {Klein},
  \citenamefont {Shulenberger}, \citenamefont {Perkinson}, \citenamefont
  {Kovalenko} \emph {et~al.}}]{utzat2019coherent}%
  \BibitemOpen
  \bibfield  {author} {\bibinfo {author} {\bibfnamefont {H.}~\bibnamefont
  {Utzat}}, \bibinfo {author} {\bibfnamefont {W.}~\bibnamefont {Sun}}, \bibinfo
  {author} {\bibfnamefont {A.~E.}\ \bibnamefont {Kaplan}}, \bibinfo {author}
  {\bibfnamefont {F.}~\bibnamefont {Krieg}}, \bibinfo {author} {\bibfnamefont
  {M.}~\bibnamefont {Ginterseder}}, \bibinfo {author} {\bibfnamefont
  {B.}~\bibnamefont {Spokoyny}}, \bibinfo {author} {\bibfnamefont {N.~D.}\
  \bibnamefont {Klein}}, \bibinfo {author} {\bibfnamefont {K.~E.}\ \bibnamefont
  {Shulenberger}}, \bibinfo {author} {\bibfnamefont {C.~F.}\ \bibnamefont
  {Perkinson}}, \bibinfo {author} {\bibfnamefont {M.~V.}\ \bibnamefont
  {Kovalenko}}, \emph {et~al.},\ }\bibfield  {title} {\bibinfo {title}
  {Coherent single-photon emission from colloidal lead halide perovskite
  quantum dots},\ }\href@noop {} {\bibfield  {journal} {\bibinfo  {journal}
  {Science}\ }\textbf {\bibinfo {volume} {363}},\ \bibinfo {pages} {1068}
  (\bibinfo {year} {2019})}\BibitemShut {NoStop}%
\bibitem [{\citenamefont {Klimov}(2017)}]{klimov2017nanocrystal}%
  \BibitemOpen
  \bibfield  {author} {\bibinfo {author} {\bibfnamefont {V.~I.}\ \bibnamefont
  {Klimov}},\ }\href@noop {} {\emph {\bibinfo {title} {Nanocrystal quantum
  dots}}}\ (\bibinfo  {publisher} {CRC Press},\ \bibinfo {year}
  {2017})\BibitemShut {NoStop}%
\bibitem [{\citenamefont {Michler}(2003)}]{michler2003single}%
  \BibitemOpen
  \bibfield  {author} {\bibinfo {author} {\bibfnamefont {P.}~\bibnamefont
  {Michler}},\ }\href@noop {} {\emph {\bibinfo {title} {Single quantum dots:
  Fundamentals, applications and new concepts}}}\ (\bibinfo  {publisher}
  {Springer Science \& Business Media},\ \bibinfo {year} {2003})\BibitemShut
  {NoStop}%
\bibitem [{\citenamefont {Bayer}(2019)}]{bayer2019bridging}%
  \BibitemOpen
  \bibfield  {author} {\bibinfo {author} {\bibfnamefont {M.}~\bibnamefont
  {Bayer}},\ }\bibfield  {title} {\bibinfo {title} {Bridging two worlds:
  colloidal versus epitaxial quantum dots},\ }\href@noop {} {\bibfield
  {journal} {\bibinfo  {journal} {Ann. Phys.}\ }\textbf {\bibinfo {volume}
  {531}},\ \bibinfo {pages} {1900039} (\bibinfo {year} {2019})}\BibitemShut
  {NoStop}%
\bibitem [{\citenamefont {Toth}\ and\ \citenamefont
  {Aharonovich}(2019)}]{toth2019single}%
  \BibitemOpen
  \bibfield  {author} {\bibinfo {author} {\bibfnamefont {M.}~\bibnamefont
  {Toth}}\ and\ \bibinfo {author} {\bibfnamefont {I.}~\bibnamefont
  {Aharonovich}},\ }\bibfield  {title} {\bibinfo {title} {Single photon sources
  in atomically thin materials},\ }\href@noop {} {\bibfield  {journal}
  {\bibinfo  {journal} {Annu. Rev. Phys. Chem.}\ }\textbf {\bibinfo {volume}
  {70}},\ \bibinfo {pages} {123} (\bibinfo {year} {2019})}\BibitemShut
  {NoStop}%
\bibitem [{\citenamefont {Michaelis~de Vasconcellos}\ \emph
  {et~al.}(2022)\citenamefont {Michaelis~de Vasconcellos}, \citenamefont
  {Wigger}, \citenamefont {Wurstbauer}, \citenamefont {Holleitner},
  \citenamefont {Bratschitsch},\ and\ \citenamefont
  {Kuhn}}]{michaelis2022single}%
  \BibitemOpen
  \bibfield  {author} {\bibinfo {author} {\bibfnamefont {S.}~\bibnamefont
  {Michaelis~de Vasconcellos}}, \bibinfo {author} {\bibfnamefont
  {D.}~\bibnamefont {Wigger}}, \bibinfo {author} {\bibfnamefont
  {U.}~\bibnamefont {Wurstbauer}}, \bibinfo {author} {\bibfnamefont {A.~W.}\
  \bibnamefont {Holleitner}}, \bibinfo {author} {\bibfnamefont
  {R.}~\bibnamefont {Bratschitsch}},\ and\ \bibinfo {author} {\bibfnamefont
  {T.}~\bibnamefont {Kuhn}},\ }\bibfield  {title} {\bibinfo {title}
  {{Single-photon emitters in layered van der Waals materials}},\ }\href@noop
  {} {\bibfield  {journal} {\bibinfo  {journal} {Phys. Status Solidi B}\ }
  (\bibinfo {year} {2022})}\BibitemShut {NoStop}%
\bibitem [{\citenamefont {Liu}\ \emph {et~al.}(2014)\citenamefont {Liu},
  \citenamefont {Yan}, \citenamefont {Chen}, \citenamefont {Fan}, \citenamefont
  {Sun}, \citenamefont {Suh}, \citenamefont {Fu}, \citenamefont {Lee},
  \citenamefont {Zhou}, \citenamefont {Tongay} \emph
  {et~al.}}]{liu2014elastic}%
  \BibitemOpen
  \bibfield  {author} {\bibinfo {author} {\bibfnamefont {K.}~\bibnamefont
  {Liu}}, \bibinfo {author} {\bibfnamefont {Q.}~\bibnamefont {Yan}}, \bibinfo
  {author} {\bibfnamefont {M.}~\bibnamefont {Chen}}, \bibinfo {author}
  {\bibfnamefont {W.}~\bibnamefont {Fan}}, \bibinfo {author} {\bibfnamefont
  {Y.}~\bibnamefont {Sun}}, \bibinfo {author} {\bibfnamefont {J.}~\bibnamefont
  {Suh}}, \bibinfo {author} {\bibfnamefont {D.}~\bibnamefont {Fu}}, \bibinfo
  {author} {\bibfnamefont {S.}~\bibnamefont {Lee}}, \bibinfo {author}
  {\bibfnamefont {J.}~\bibnamefont {Zhou}}, \bibinfo {author} {\bibfnamefont
  {S.}~\bibnamefont {Tongay}}, \emph {et~al.},\ }\bibfield  {title} {\bibinfo
  {title} {{Elastic properties of chemical-vapor-deposited monolayer MoS$_2$,
  WS$_2$, and their bilayer heterostructures}},\ }\href@noop {} {\bibfield
  {journal} {\bibinfo  {journal} {Nano Lett.}\ }\textbf {\bibinfo {volume}
  {14}},\ \bibinfo {pages} {5097} (\bibinfo {year} {2014})}\BibitemShut
  {NoStop}%
\bibitem [{\citenamefont {Tonndorf}\ \emph {et~al.}(2015)\citenamefont
  {Tonndorf}, \citenamefont {Schmidt}, \citenamefont {Schneider}, \citenamefont
  {Kern}, \citenamefont {Buscema}, \citenamefont {Steele}, \citenamefont
  {Castellanos-Gomez}, \citenamefont {van~der Zant}, \citenamefont
  {de~Vasconcellos},\ and\ \citenamefont {Bratschitsch}}]{tonndorf2015single}%
  \BibitemOpen
  \bibfield  {author} {\bibinfo {author} {\bibfnamefont {P.}~\bibnamefont
  {Tonndorf}}, \bibinfo {author} {\bibfnamefont {R.}~\bibnamefont {Schmidt}},
  \bibinfo {author} {\bibfnamefont {R.}~\bibnamefont {Schneider}}, \bibinfo
  {author} {\bibfnamefont {J.}~\bibnamefont {Kern}}, \bibinfo {author}
  {\bibfnamefont {M.}~\bibnamefont {Buscema}}, \bibinfo {author} {\bibfnamefont
  {G.~A.}\ \bibnamefont {Steele}}, \bibinfo {author} {\bibfnamefont
  {A.}~\bibnamefont {Castellanos-Gomez}}, \bibinfo {author} {\bibfnamefont
  {H.~S.}\ \bibnamefont {van~der Zant}}, \bibinfo {author} {\bibfnamefont
  {S.~M.}\ \bibnamefont {de~Vasconcellos}},\ and\ \bibinfo {author}
  {\bibfnamefont {R.}~\bibnamefont {Bratschitsch}},\ }\bibfield  {title}
  {\bibinfo {title} {Single-photon emission from localized excitons in an
  atomically thin semiconductor},\ }\href@noop {} {\bibfield  {journal}
  {\bibinfo  {journal} {Optica}\ }\textbf {\bibinfo {volume} {2}},\ \bibinfo
  {pages} {347} (\bibinfo {year} {2015})}\BibitemShut {NoStop}%
\bibitem [{\citenamefont {Chakraborty}\ \emph {et~al.}(2015)\citenamefont
  {Chakraborty}, \citenamefont {Kinnischtzke}, \citenamefont {Goodfellow},
  \citenamefont {Beams},\ and\ \citenamefont
  {Vamivakas}}]{2015_NatNano_Chakraborty}%
  \BibitemOpen
  \bibfield  {author} {\bibinfo {author} {\bibfnamefont {C.}~\bibnamefont
  {Chakraborty}}, \bibinfo {author} {\bibfnamefont {L.}~\bibnamefont
  {Kinnischtzke}}, \bibinfo {author} {\bibfnamefont {K.~M.}\ \bibnamefont
  {Goodfellow}}, \bibinfo {author} {\bibfnamefont {R.}~\bibnamefont {Beams}},\
  and\ \bibinfo {author} {\bibfnamefont {A.~N.}\ \bibnamefont {Vamivakas}},\
  }\bibfield  {title} {\bibinfo {title} {Voltage-controlled quantum light from
  an atomically thin semiconductor},\ }\href@noop {} {\bibfield  {journal}
  {\bibinfo  {journal} {Nat. Nanotechnol.}\ }\textbf {\bibinfo {volume} {10}},\
  \bibinfo {pages} {507} (\bibinfo {year} {2015})}\BibitemShut {NoStop}%
\bibitem [{\citenamefont {He}\ \emph {et~al.}(2015)\citenamefont {He},
  \citenamefont {Clark}, \citenamefont {Schaibley}, \citenamefont {He},
  \citenamefont {Chen}, \citenamefont {Wei}, \citenamefont {Ding},
  \citenamefont {Zhang}, \citenamefont {Yao}, \citenamefont {Xu}, \citenamefont
  {Lu},\ and\ \citenamefont {Pan}}]{2015_NatNano_He}%
  \BibitemOpen
  \bibfield  {author} {\bibinfo {author} {\bibfnamefont {Y.-M.}\ \bibnamefont
  {He}}, \bibinfo {author} {\bibfnamefont {G.}~\bibnamefont {Clark}}, \bibinfo
  {author} {\bibfnamefont {J.~R.}\ \bibnamefont {Schaibley}}, \bibinfo {author}
  {\bibfnamefont {Y.}~\bibnamefont {He}}, \bibinfo {author} {\bibfnamefont
  {M.-C.}\ \bibnamefont {Chen}}, \bibinfo {author} {\bibfnamefont {Y.-J.}\
  \bibnamefont {Wei}}, \bibinfo {author} {\bibfnamefont {X.}~\bibnamefont
  {Ding}}, \bibinfo {author} {\bibfnamefont {Q.}~\bibnamefont {Zhang}},
  \bibinfo {author} {\bibfnamefont {W.}~\bibnamefont {Yao}}, \bibinfo {author}
  {\bibfnamefont {X.}~\bibnamefont {Xu}}, \bibinfo {author} {\bibfnamefont
  {C.-Y.}\ \bibnamefont {Lu}},\ and\ \bibinfo {author} {\bibfnamefont {J.-W.}\
  \bibnamefont {Pan}},\ }\bibfield  {title} {\bibinfo {title} {Single quantum
  emitters in monolayer semiconductors},\ }\href@noop {} {\bibfield  {journal}
  {\bibinfo  {journal} {Nat. Nanotechnol.}\ }\textbf {\bibinfo {volume} {10}},\
  \bibinfo {pages} {497} (\bibinfo {year} {2015})}\BibitemShut {NoStop}%
\bibitem [{\citenamefont {Koperski}\ \emph {et~al.}(2015)\citenamefont
  {Koperski}, \citenamefont {Nogajewski}, \citenamefont {Arora}, \citenamefont
  {Cherkez}, \citenamefont {Mallet}, \citenamefont {Veuillen}, \citenamefont
  {Marcus}, \citenamefont {Kossacki},\ and\ \citenamefont
  {Potemski}}]{2015_NatNano_Koperski}%
  \BibitemOpen
  \bibfield  {author} {\bibinfo {author} {\bibfnamefont {M.}~\bibnamefont
  {Koperski}}, \bibinfo {author} {\bibfnamefont {K.}~\bibnamefont
  {Nogajewski}}, \bibinfo {author} {\bibfnamefont {A.}~\bibnamefont {Arora}},
  \bibinfo {author} {\bibfnamefont {V.}~\bibnamefont {Cherkez}}, \bibinfo
  {author} {\bibfnamefont {P.}~\bibnamefont {Mallet}}, \bibinfo {author}
  {\bibfnamefont {J.-Y.}\ \bibnamefont {Veuillen}}, \bibinfo {author}
  {\bibfnamefont {J.}~\bibnamefont {Marcus}}, \bibinfo {author} {\bibfnamefont
  {P.}~\bibnamefont {Kossacki}},\ and\ \bibinfo {author} {\bibfnamefont
  {M.}~\bibnamefont {Potemski}},\ }\bibfield  {title} {\bibinfo {title}
  {{Single photon emitters in exfoliated WSe$_2$ structures}},\ }\href@noop {}
  {\bibfield  {journal} {\bibinfo  {journal} {Nat. Nanotechnol.}\ }\textbf
  {\bibinfo {volume} {10}},\ \bibinfo {pages} {503} (\bibinfo {year}
  {2015})}\BibitemShut {NoStop}%
\bibitem [{\citenamefont {Srivastava}\ \emph {et~al.}(2015)\citenamefont
  {Srivastava}, \citenamefont {Sidler}, \citenamefont {Allain}, \citenamefont
  {Lembke}, \citenamefont {Kis},\ and\ \citenamefont
  {Imamo{\u{g}}lu}}]{2015_NatNano_Srivastava}%
  \BibitemOpen
  \bibfield  {author} {\bibinfo {author} {\bibfnamefont {A.}~\bibnamefont
  {Srivastava}}, \bibinfo {author} {\bibfnamefont {M.}~\bibnamefont {Sidler}},
  \bibinfo {author} {\bibfnamefont {A.~V.}\ \bibnamefont {Allain}}, \bibinfo
  {author} {\bibfnamefont {D.~S.}\ \bibnamefont {Lembke}}, \bibinfo {author}
  {\bibfnamefont {A.}~\bibnamefont {Kis}},\ and\ \bibinfo {author}
  {\bibfnamefont {A.}~\bibnamefont {Imamo{\u{g}}lu}},\ }\bibfield  {title}
  {\bibinfo {title} {{Optically active quantum dots in monolayer WSe$_2$}},\
  }\href@noop {} {\bibfield  {journal} {\bibinfo  {journal} {Nat.
  Nanotechnol.}\ }\textbf {\bibinfo {volume} {10}},\ \bibinfo {pages} {491}
  (\bibinfo {year} {2015})}\BibitemShut {NoStop}%
\bibitem [{\citenamefont {Tran}\ \emph
  {et~al.}(2016{\natexlab{a}})\citenamefont {Tran}, \citenamefont {Bray},
  \citenamefont {Ford}, \citenamefont {Toth},\ and\ \citenamefont
  {Aharonovich}}]{tran2016quantum}%
  \BibitemOpen
  \bibfield  {author} {\bibinfo {author} {\bibfnamefont {T.~T.}\ \bibnamefont
  {Tran}}, \bibinfo {author} {\bibfnamefont {K.}~\bibnamefont {Bray}}, \bibinfo
  {author} {\bibfnamefont {M.~J.}\ \bibnamefont {Ford}}, \bibinfo {author}
  {\bibfnamefont {M.}~\bibnamefont {Toth}},\ and\ \bibinfo {author}
  {\bibfnamefont {I.}~\bibnamefont {Aharonovich}},\ }\bibfield  {title}
  {\bibinfo {title} {Quantum emission from hexagonal boron nitride
  monolayers},\ }\href@noop {} {\bibfield  {journal} {\bibinfo  {journal} {Nat.
  Nanotechnol.}\ }\textbf {\bibinfo {volume} {11}},\ \bibinfo {pages} {37}
  (\bibinfo {year} {2016}{\natexlab{a}})}\BibitemShut {NoStop}%
\bibitem [{\citenamefont {Camphausen}\ \emph {et~al.}(2020)\citenamefont
  {Camphausen}, \citenamefont {Marini}, \citenamefont {Tawfik}, \citenamefont
  {Tran}, \citenamefont {Ford},\ and\ \citenamefont
  {Palomba}}]{camphausen2020observation}%
  \BibitemOpen
  \bibfield  {author} {\bibinfo {author} {\bibfnamefont {R.}~\bibnamefont
  {Camphausen}}, \bibinfo {author} {\bibfnamefont {L.}~\bibnamefont {Marini}},
  \bibinfo {author} {\bibfnamefont {S.~A.}\ \bibnamefont {Tawfik}}, \bibinfo
  {author} {\bibfnamefont {T.~T.}\ \bibnamefont {Tran}}, \bibinfo {author}
  {\bibfnamefont {M.~J.}\ \bibnamefont {Ford}},\ and\ \bibinfo {author}
  {\bibfnamefont {S.}~\bibnamefont {Palomba}},\ }\bibfield  {title} {\bibinfo
  {title} {Observation of near-infrared sub-poissonian photon emission in
  hexagonal boron nitride at room temperature},\ }\href@noop {} {\bibfield
  {journal} {\bibinfo  {journal} {APL Photonics}\ }\textbf {\bibinfo {volume}
  {5}},\ \bibinfo {pages} {076103} (\bibinfo {year} {2020})}\BibitemShut
  {NoStop}%
\bibitem [{\citenamefont {Tran}\ \emph
  {et~al.}(2016{\natexlab{b}})\citenamefont {Tran}, \citenamefont {Elbadawi},
  \citenamefont {Totonjian}, \citenamefont {Lobo}, \citenamefont {Grosso},
  \citenamefont {Moon}, \citenamefont {Englund}, \citenamefont {Ford},
  \citenamefont {Aharonovich},\ and\ \citenamefont {Toth}}]{tran2016robust}%
  \BibitemOpen
  \bibfield  {author} {\bibinfo {author} {\bibfnamefont {T.~T.}\ \bibnamefont
  {Tran}}, \bibinfo {author} {\bibfnamefont {C.}~\bibnamefont {Elbadawi}},
  \bibinfo {author} {\bibfnamefont {D.}~\bibnamefont {Totonjian}}, \bibinfo
  {author} {\bibfnamefont {C.~J.}\ \bibnamefont {Lobo}}, \bibinfo {author}
  {\bibfnamefont {G.}~\bibnamefont {Grosso}}, \bibinfo {author} {\bibfnamefont
  {H.}~\bibnamefont {Moon}}, \bibinfo {author} {\bibfnamefont {D.~R.}\
  \bibnamefont {Englund}}, \bibinfo {author} {\bibfnamefont {M.~J.}\
  \bibnamefont {Ford}}, \bibinfo {author} {\bibfnamefont {I.}~\bibnamefont
  {Aharonovich}},\ and\ \bibinfo {author} {\bibfnamefont {M.}~\bibnamefont
  {Toth}},\ }\bibfield  {title} {\bibinfo {title} {Robust multicolor single
  photon emission from point defects in hexagonal boron nitride},\ }\href@noop
  {} {\bibfield  {journal} {\bibinfo  {journal} {ACS Nano}\ }\textbf {\bibinfo
  {volume} {10}},\ \bibinfo {pages} {7331} (\bibinfo {year}
  {2016}{\natexlab{b}})}\BibitemShut {NoStop}%
\bibitem [{\citenamefont {Bourrellier}\ \emph {et~al.}(2016)\citenamefont
  {Bourrellier}, \citenamefont {Meuret}, \citenamefont {Tararan}, \citenamefont
  {St{\'e}phan}, \citenamefont {Kociak}, \citenamefont {Tizei},\ and\
  \citenamefont {Zobelli}}]{bourrellier2016bright}%
  \BibitemOpen
  \bibfield  {author} {\bibinfo {author} {\bibfnamefont {R.}~\bibnamefont
  {Bourrellier}}, \bibinfo {author} {\bibfnamefont {S.}~\bibnamefont {Meuret}},
  \bibinfo {author} {\bibfnamefont {A.}~\bibnamefont {Tararan}}, \bibinfo
  {author} {\bibfnamefont {O.}~\bibnamefont {St{\'e}phan}}, \bibinfo {author}
  {\bibfnamefont {M.}~\bibnamefont {Kociak}}, \bibinfo {author} {\bibfnamefont
  {L.~H.}\ \bibnamefont {Tizei}},\ and\ \bibinfo {author} {\bibfnamefont
  {A.}~\bibnamefont {Zobelli}},\ }\bibfield  {title} {\bibinfo {title} {{Bright
  UV single photon emission at point defects in h-BN}},\ }\href@noop {}
  {\bibfield  {journal} {\bibinfo  {journal} {Nano Lett.}\ }\textbf {\bibinfo
  {volume} {16}},\ \bibinfo {pages} {4317} (\bibinfo {year}
  {2016})}\BibitemShut {NoStop}%
\bibitem [{\citenamefont {Xia}\ \emph {et~al.}(2019)\citenamefont {Xia},
  \citenamefont {Li}, \citenamefont {Kim}, \citenamefont {Bao}, \citenamefont
  {Gong}, \citenamefont {Yang}, \citenamefont {Wang},\ and\ \citenamefont
  {Zhang}}]{xia2019room}%
  \BibitemOpen
  \bibfield  {author} {\bibinfo {author} {\bibfnamefont {Y.}~\bibnamefont
  {Xia}}, \bibinfo {author} {\bibfnamefont {Q.}~\bibnamefont {Li}}, \bibinfo
  {author} {\bibfnamefont {J.}~\bibnamefont {Kim}}, \bibinfo {author}
  {\bibfnamefont {W.}~\bibnamefont {Bao}}, \bibinfo {author} {\bibfnamefont
  {C.}~\bibnamefont {Gong}}, \bibinfo {author} {\bibfnamefont {S.}~\bibnamefont
  {Yang}}, \bibinfo {author} {\bibfnamefont {Y.}~\bibnamefont {Wang}},\ and\
  \bibinfo {author} {\bibfnamefont {X.}~\bibnamefont {Zhang}},\ }\bibfield
  {title} {\bibinfo {title} {{Room-temperature giant Stark effect of single
  photon emitter in van der Waals material}},\ }\href@noop {} {\bibfield
  {journal} {\bibinfo  {journal} {Nano Lett.}\ }\textbf {\bibinfo {volume}
  {19}},\ \bibinfo {pages} {7100} (\bibinfo {year} {2019})}\BibitemShut
  {NoStop}%
\bibitem [{\citenamefont {Mendelson}\ \emph {et~al.}(2020)\citenamefont
  {Mendelson}, \citenamefont {Doherty}, \citenamefont {Toth}, \citenamefont
  {Aharonovich},\ and\ \citenamefont {Tran}}]{mendelson2020strain}%
  \BibitemOpen
  \bibfield  {author} {\bibinfo {author} {\bibfnamefont {N.}~\bibnamefont
  {Mendelson}}, \bibinfo {author} {\bibfnamefont {M.}~\bibnamefont {Doherty}},
  \bibinfo {author} {\bibfnamefont {M.}~\bibnamefont {Toth}}, \bibinfo {author}
  {\bibfnamefont {I.}~\bibnamefont {Aharonovich}},\ and\ \bibinfo {author}
  {\bibfnamefont {T.~T.}\ \bibnamefont {Tran}},\ }\bibfield  {title} {\bibinfo
  {title} {Strain-induced modification of the optical characteristics of
  quantum emitters in hexagonal boron nitride},\ }\href@noop {} {\bibfield
  {journal} {\bibinfo  {journal} {Adv. Mater.}\ }\textbf {\bibinfo {volume}
  {32}},\ \bibinfo {pages} {1908316} (\bibinfo {year} {2020})}\BibitemShut
  {NoStop}%
\bibitem [{\citenamefont {Vuong}\ \emph {et~al.}(2016)\citenamefont {Vuong},
  \citenamefont {Cassabois}, \citenamefont {Valvin}, \citenamefont {Ouerghi},
  \citenamefont {Chassagneux}, \citenamefont {Voisin},\ and\ \citenamefont
  {Gil}}]{vuong2016phonon}%
  \BibitemOpen
  \bibfield  {author} {\bibinfo {author} {\bibfnamefont {T.}~\bibnamefont
  {Vuong}}, \bibinfo {author} {\bibfnamefont {G.}~\bibnamefont {Cassabois}},
  \bibinfo {author} {\bibfnamefont {P.}~\bibnamefont {Valvin}}, \bibinfo
  {author} {\bibfnamefont {A.}~\bibnamefont {Ouerghi}}, \bibinfo {author}
  {\bibfnamefont {Y.}~\bibnamefont {Chassagneux}}, \bibinfo {author}
  {\bibfnamefont {C.}~\bibnamefont {Voisin}},\ and\ \bibinfo {author}
  {\bibfnamefont {B.}~\bibnamefont {Gil}},\ }\bibfield  {title} {\bibinfo
  {title} {Phonon-photon mapping in a color center in hexagonal boron
  nitride},\ }\href@noop {} {\bibfield  {journal} {\bibinfo  {journal} {Phys.
  Rev. Lett.}\ }\textbf {\bibinfo {volume} {117}},\ \bibinfo {pages} {097402}
  (\bibinfo {year} {2016})}\BibitemShut {NoStop}%
\bibitem [{\citenamefont {Wigger}\ \emph {et~al.}(2019)\citenamefont {Wigger},
  \citenamefont {Schmidt}, \citenamefont {Del Pozo-Zamudio}, \citenamefont
  {Preu{\ss}}, \citenamefont {Tonndorf}, \citenamefont {Schneider},
  \citenamefont {Steeger}, \citenamefont {Kern}, \citenamefont {Khodaei},
  \citenamefont {Sperling} \emph {et~al.}}]{wigger2019phonon}%
  \BibitemOpen
  \bibfield  {author} {\bibinfo {author} {\bibfnamefont {D.}~\bibnamefont
  {Wigger}}, \bibinfo {author} {\bibfnamefont {R.}~\bibnamefont {Schmidt}},
  \bibinfo {author} {\bibfnamefont {O.}~\bibnamefont {Del Pozo-Zamudio}},
  \bibinfo {author} {\bibfnamefont {J.~A.}\ \bibnamefont {Preu{\ss}}}, \bibinfo
  {author} {\bibfnamefont {P.}~\bibnamefont {Tonndorf}}, \bibinfo {author}
  {\bibfnamefont {R.}~\bibnamefont {Schneider}}, \bibinfo {author}
  {\bibfnamefont {P.}~\bibnamefont {Steeger}}, \bibinfo {author} {\bibfnamefont
  {J.}~\bibnamefont {Kern}}, \bibinfo {author} {\bibfnamefont {Y.}~\bibnamefont
  {Khodaei}}, \bibinfo {author} {\bibfnamefont {J.}~\bibnamefont {Sperling}},
  \emph {et~al.},\ }\bibfield  {title} {\bibinfo {title} {Phonon-assisted
  emission and absorption of individual color centers in hexagonal boron
  nitride},\ }\href@noop {} {\bibfield  {journal} {\bibinfo  {journal} {2D
  Mater.}\ }\textbf {\bibinfo {volume} {6}},\ \bibinfo {pages} {035006}
  (\bibinfo {year} {2019})}\BibitemShut {NoStop}%
\bibitem [{\citenamefont {Mendelson}\ \emph {et~al.}(2021)\citenamefont
  {Mendelson}, \citenamefont {Morales-Inostroza}, \citenamefont {Li},
  \citenamefont {Ritika}, \citenamefont {Nguyen}, \citenamefont
  {Loyola-Echeverria}, \citenamefont {Kim}, \citenamefont {G{\"o}tzinger},
  \citenamefont {Toth},\ and\ \citenamefont
  {Aharonovich}}]{mendelson2021grain}%
  \BibitemOpen
  \bibfield  {author} {\bibinfo {author} {\bibfnamefont {N.}~\bibnamefont
  {Mendelson}}, \bibinfo {author} {\bibfnamefont {L.}~\bibnamefont
  {Morales-Inostroza}}, \bibinfo {author} {\bibfnamefont {C.}~\bibnamefont
  {Li}}, \bibinfo {author} {\bibfnamefont {R.}~\bibnamefont {Ritika}}, \bibinfo
  {author} {\bibfnamefont {M.~A.~P.}\ \bibnamefont {Nguyen}}, \bibinfo {author}
  {\bibfnamefont {J.}~\bibnamefont {Loyola-Echeverria}}, \bibinfo {author}
  {\bibfnamefont {S.}~\bibnamefont {Kim}}, \bibinfo {author} {\bibfnamefont
  {S.}~\bibnamefont {G{\"o}tzinger}}, \bibinfo {author} {\bibfnamefont
  {M.}~\bibnamefont {Toth}},\ and\ \bibinfo {author} {\bibfnamefont
  {I.}~\bibnamefont {Aharonovich}},\ }\bibfield  {title} {\bibinfo {title}
  {Grain dependent growth of bright quantum emitters in hexagonal boron
  nitride},\ }\href@noop {} {\bibfield  {journal} {\bibinfo  {journal} {Adv.
  Opt. Mater.}\ }\textbf {\bibinfo {volume} {9}},\ \bibinfo {pages} {2001271}
  (\bibinfo {year} {2021})}\BibitemShut {NoStop}%
\bibitem [{\citenamefont {Preu{\ss}}\ \emph {et~al.}(2021)\citenamefont
  {Preu{\ss}}, \citenamefont {Rudi}, \citenamefont {Kern}, \citenamefont
  {Schmidt}, \citenamefont {Bratschitsch},\ and\ \citenamefont {Michaelis~de
  Vasconcellos}}]{preuss2021assembly}%
  \BibitemOpen
  \bibfield  {author} {\bibinfo {author} {\bibfnamefont {J.~A.}\ \bibnamefont
  {Preu{\ss}}}, \bibinfo {author} {\bibfnamefont {E.}~\bibnamefont {Rudi}},
  \bibinfo {author} {\bibfnamefont {J.}~\bibnamefont {Kern}}, \bibinfo {author}
  {\bibfnamefont {R.}~\bibnamefont {Schmidt}}, \bibinfo {author} {\bibfnamefont
  {R.}~\bibnamefont {Bratschitsch}},\ and\ \bibinfo {author} {\bibfnamefont
  {S.}~\bibnamefont {Michaelis~de Vasconcellos}},\ }\bibfield  {title}
  {\bibinfo {title} {{Assembly of large hBN nanocrystal arrays for quantum
  light emission}},\ }\href@noop {} {\bibfield  {journal} {\bibinfo  {journal}
  {2D Mater.}\ }\textbf {\bibinfo {volume} {8}},\ \bibinfo {pages} {035005}
  (\bibinfo {year} {2021})}\BibitemShut {NoStop}%
\bibitem [{\citenamefont {Konthasinghe}\ \emph {et~al.}(2019)\citenamefont
  {Konthasinghe}, \citenamefont {Chakraborty}, \citenamefont {Mathur},
  \citenamefont {Qiu}, \citenamefont {Mukherjee}, \citenamefont {Fuchs},\ and\
  \citenamefont {Vamivakas}}]{konthasinghe2019rabi}%
  \BibitemOpen
  \bibfield  {author} {\bibinfo {author} {\bibfnamefont {K.}~\bibnamefont
  {Konthasinghe}}, \bibinfo {author} {\bibfnamefont {C.}~\bibnamefont
  {Chakraborty}}, \bibinfo {author} {\bibfnamefont {N.}~\bibnamefont {Mathur}},
  \bibinfo {author} {\bibfnamefont {L.}~\bibnamefont {Qiu}}, \bibinfo {author}
  {\bibfnamefont {A.}~\bibnamefont {Mukherjee}}, \bibinfo {author}
  {\bibfnamefont {G.~D.}\ \bibnamefont {Fuchs}},\ and\ \bibinfo {author}
  {\bibfnamefont {A.~N.}\ \bibnamefont {Vamivakas}},\ }\bibfield  {title}
  {\bibinfo {title} {Rabi oscillations and resonance fluorescence from a single
  hexagonal boron nitride quantum emitter},\ }\href@noop {} {\bibfield
  {journal} {\bibinfo  {journal} {Optica}\ }\textbf {\bibinfo {volume} {6}},\
  \bibinfo {pages} {542} (\bibinfo {year} {2019})}\BibitemShut {NoStop}%
\bibitem [{\citenamefont {Dietrich}\ \emph {et~al.}(2020)\citenamefont
  {Dietrich}, \citenamefont {Doherty}, \citenamefont {Aharonovich},\ and\
  \citenamefont {Kubanek}}]{dietrich2020solid}%
  \BibitemOpen
  \bibfield  {author} {\bibinfo {author} {\bibfnamefont {A.}~\bibnamefont
  {Dietrich}}, \bibinfo {author} {\bibfnamefont {M.}~\bibnamefont {Doherty}},
  \bibinfo {author} {\bibfnamefont {I.}~\bibnamefont {Aharonovich}},\ and\
  \bibinfo {author} {\bibfnamefont {A.}~\bibnamefont {Kubanek}},\ }\bibfield
  {title} {\bibinfo {title} {{Solid-state single photon source with Fourier
  transform limited lines at room temperature}},\ }\href@noop {} {\bibfield
  {journal} {\bibinfo  {journal} {Phys. Rev. B}\ }\textbf {\bibinfo {volume}
  {101}},\ \bibinfo {pages} {081401} (\bibinfo {year} {2020})}\BibitemShut
  {NoStop}%
\bibitem [{\citenamefont {White}\ \emph {et~al.}(2021)\citenamefont {White},
  \citenamefont {Stewart}, \citenamefont {Solntsev}, \citenamefont {Li},
  \citenamefont {Toth}, \citenamefont {Kianinia},\ and\ \citenamefont
  {Aharonovich}}]{white2021phonon}%
  \BibitemOpen
  \bibfield  {author} {\bibinfo {author} {\bibfnamefont {S.}~\bibnamefont
  {White}}, \bibinfo {author} {\bibfnamefont {C.}~\bibnamefont {Stewart}},
  \bibinfo {author} {\bibfnamefont {A.~S.}\ \bibnamefont {Solntsev}}, \bibinfo
  {author} {\bibfnamefont {C.}~\bibnamefont {Li}}, \bibinfo {author}
  {\bibfnamefont {M.}~\bibnamefont {Toth}}, \bibinfo {author} {\bibfnamefont
  {M.}~\bibnamefont {Kianinia}},\ and\ \bibinfo {author} {\bibfnamefont
  {I.}~\bibnamefont {Aharonovich}},\ }\bibfield  {title} {\bibinfo {title}
  {Phonon dephasing and spectral diffusion of quantum emitters in hexagonal
  boron nitride},\ }\href@noop {} {\bibfield  {journal} {\bibinfo  {journal}
  {Optica}\ }\textbf {\bibinfo {volume} {8}},\ \bibinfo {pages} {1153}
  (\bibinfo {year} {2021})}\BibitemShut {NoStop}%
\bibitem [{\citenamefont {Boll}\ \emph {et~al.}(2020)\citenamefont {Boll},
  \citenamefont {Radko}, \citenamefont {Huck},\ and\ \citenamefont
  {Andersen}}]{boll2020photophysics}%
  \BibitemOpen
  \bibfield  {author} {\bibinfo {author} {\bibfnamefont {M.~K.}\ \bibnamefont
  {Boll}}, \bibinfo {author} {\bibfnamefont {I.~P.}\ \bibnamefont {Radko}},
  \bibinfo {author} {\bibfnamefont {A.}~\bibnamefont {Huck}},\ and\ \bibinfo
  {author} {\bibfnamefont {U.~L.}\ \bibnamefont {Andersen}},\ }\bibfield
  {title} {\bibinfo {title} {Photophysics of quantum emitters in hexagonal
  boron-nitride nano-flakes},\ }\href@noop {} {\bibfield  {journal} {\bibinfo
  {journal} {Optics Express}\ }\textbf {\bibinfo {volume} {28}},\ \bibinfo
  {pages} {7475} (\bibinfo {year} {2020})}\BibitemShut {NoStop}%
\bibitem [{\citenamefont {Grosso}\ \emph {et~al.}(2020)\citenamefont {Grosso},
  \citenamefont {Moon}, \citenamefont {Ciccarino}, \citenamefont {Flick},
  \citenamefont {Mendelson}, \citenamefont {Mennel}, \citenamefont {Toth},
  \citenamefont {Aharonovich}, \citenamefont {Narang},\ and\ \citenamefont
  {Englund}}]{grosso2020low}%
  \BibitemOpen
  \bibfield  {author} {\bibinfo {author} {\bibfnamefont {G.}~\bibnamefont
  {Grosso}}, \bibinfo {author} {\bibfnamefont {H.}~\bibnamefont {Moon}},
  \bibinfo {author} {\bibfnamefont {C.~J.}\ \bibnamefont {Ciccarino}}, \bibinfo
  {author} {\bibfnamefont {J.}~\bibnamefont {Flick}}, \bibinfo {author}
  {\bibfnamefont {N.}~\bibnamefont {Mendelson}}, \bibinfo {author}
  {\bibfnamefont {L.}~\bibnamefont {Mennel}}, \bibinfo {author} {\bibfnamefont
  {M.}~\bibnamefont {Toth}}, \bibinfo {author} {\bibfnamefont {I.}~\bibnamefont
  {Aharonovich}}, \bibinfo {author} {\bibfnamefont {P.}~\bibnamefont
  {Narang}},\ and\ \bibinfo {author} {\bibfnamefont {D.~R.}\ \bibnamefont
  {Englund}},\ }\bibfield  {title} {\bibinfo {title} {{Low-temperature
  electron--phonon interaction of quantum emitters in hexagonal boron
  nitride}},\ }\href@noop {} {\bibfield  {journal} {\bibinfo  {journal} {ACS
  Photonics}\ }\textbf {\bibinfo {volume} {7}},\ \bibinfo {pages} {1410}
  (\bibinfo {year} {2020})}\BibitemShut {NoStop}%
\bibitem [{\citenamefont {Hoese}\ \emph {et~al.}(2020)\citenamefont {Hoese},
  \citenamefont {Reddy}, \citenamefont {Dietrich}, \citenamefont {Koch},
  \citenamefont {Fehler}, \citenamefont {Doherty},\ and\ \citenamefont
  {Kubanek}}]{hoese2020mechanical}%
  \BibitemOpen
  \bibfield  {author} {\bibinfo {author} {\bibfnamefont {M.}~\bibnamefont
  {Hoese}}, \bibinfo {author} {\bibfnamefont {P.}~\bibnamefont {Reddy}},
  \bibinfo {author} {\bibfnamefont {A.}~\bibnamefont {Dietrich}}, \bibinfo
  {author} {\bibfnamefont {M.~K.}\ \bibnamefont {Koch}}, \bibinfo {author}
  {\bibfnamefont {K.~G.}\ \bibnamefont {Fehler}}, \bibinfo {author}
  {\bibfnamefont {M.~W.}\ \bibnamefont {Doherty}},\ and\ \bibinfo {author}
  {\bibfnamefont {A.}~\bibnamefont {Kubanek}},\ }\bibfield  {title} {\bibinfo
  {title} {Mechanical decoupling of quantum emitters in hexagonal boron nitride
  from low-energy phonon modes},\ }\href@noop {} {\bibfield  {journal}
  {\bibinfo  {journal} {Sci. Adv.}\ }\textbf {\bibinfo {volume} {6}},\ \bibinfo
  {pages} {eaba6038} (\bibinfo {year} {2020})}\BibitemShut {NoStop}%
\bibitem [{\citenamefont {Malein}\ \emph {et~al.}(2021)\citenamefont {Malein},
  \citenamefont {Khatri}, \citenamefont {Ramsay},\ and\ \citenamefont
  {Luxmoore}}]{malein2021stimulated}%
  \BibitemOpen
  \bibfield  {author} {\bibinfo {author} {\bibfnamefont {R.~N.~E.}\
  \bibnamefont {Malein}}, \bibinfo {author} {\bibfnamefont {P.}~\bibnamefont
  {Khatri}}, \bibinfo {author} {\bibfnamefont {A.~J.}\ \bibnamefont {Ramsay}},\
  and\ \bibinfo {author} {\bibfnamefont {I.~J.}\ \bibnamefont {Luxmoore}},\
  }\bibfield  {title} {\bibinfo {title} {Stimulated emission depletion
  spectroscopy of color centers in hexagonal boron nitride},\ }\href@noop {}
  {\bibfield  {journal} {\bibinfo  {journal} {ACS Photonics}\ }\textbf
  {\bibinfo {volume} {8}},\ \bibinfo {pages} {1007} (\bibinfo {year}
  {2021})}\BibitemShut {NoStop}%
\bibitem [{\citenamefont {Kern}\ \emph {et~al.}(1999)\citenamefont {Kern},
  \citenamefont {Kresse},\ and\ \citenamefont {Hafner}}]{kern1999ab}%
  \BibitemOpen
  \bibfield  {author} {\bibinfo {author} {\bibfnamefont {G.}~\bibnamefont
  {Kern}}, \bibinfo {author} {\bibfnamefont {G.}~\bibnamefont {Kresse}},\ and\
  \bibinfo {author} {\bibfnamefont {J.}~\bibnamefont {Hafner}},\ }\bibfield
  {title} {\bibinfo {title} {Ab initio calculation of the lattice dynamics and
  phase diagram of boron nitride},\ }\href@noop {} {\bibfield  {journal}
  {\bibinfo  {journal} {Phys. Rev. B}\ }\textbf {\bibinfo {volume} {59}},\
  \bibinfo {pages} {8551} (\bibinfo {year} {1999})}\BibitemShut {NoStop}%
\bibitem [{\citenamefont {Serrano}\ \emph {et~al.}(2007)\citenamefont
  {Serrano}, \citenamefont {Bosak}, \citenamefont {Arenal}, \citenamefont
  {Krisch}, \citenamefont {Watanabe}, \citenamefont {Taniguchi}, \citenamefont
  {Kanda}, \citenamefont {Rubio},\ and\ \citenamefont
  {Wirtz}}]{serrano2007vibrational}%
  \BibitemOpen
  \bibfield  {author} {\bibinfo {author} {\bibfnamefont {J.}~\bibnamefont
  {Serrano}}, \bibinfo {author} {\bibfnamefont {A.}~\bibnamefont {Bosak}},
  \bibinfo {author} {\bibfnamefont {R.}~\bibnamefont {Arenal}}, \bibinfo
  {author} {\bibfnamefont {M.}~\bibnamefont {Krisch}}, \bibinfo {author}
  {\bibfnamefont {K.}~\bibnamefont {Watanabe}}, \bibinfo {author}
  {\bibfnamefont {T.}~\bibnamefont {Taniguchi}}, \bibinfo {author}
  {\bibfnamefont {H.}~\bibnamefont {Kanda}}, \bibinfo {author} {\bibfnamefont
  {A.}~\bibnamefont {Rubio}},\ and\ \bibinfo {author} {\bibfnamefont
  {L.}~\bibnamefont {Wirtz}},\ }\bibfield  {title} {\bibinfo {title}
  {Vibrational properties of hexagonal boron nitride: inelastic x-ray
  scattering and ab initio calculations},\ }\href@noop {} {\bibfield  {journal}
  {\bibinfo  {journal} {Phys. Rev. Lett.}\ }\textbf {\bibinfo {volume} {98}},\
  \bibinfo {pages} {095503} (\bibinfo {year} {2007})}\BibitemShut {NoStop}%
\bibitem [{\citenamefont {Mahan}(2013)}]{mahan2013many}%
  \BibitemOpen
  \bibfield  {author} {\bibinfo {author} {\bibfnamefont {G.~D.}\ \bibnamefont
  {Mahan}},\ }\href@noop {} {\emph {\bibinfo {title} {Many-particle physics}}}\
  (\bibinfo  {publisher} {Springer Science \& Business Media, New York},\
  \bibinfo {year} {2013})\BibitemShut {NoStop}%
\bibitem [{\citenamefont {Groll}\ \emph {et~al.}(2021)\citenamefont {Groll},
  \citenamefont {Hahn}, \citenamefont {Machnikowski}, \citenamefont {Wigger},\
  and\ \citenamefont {Kuhn}}]{groll2021controlling}%
  \BibitemOpen
  \bibfield  {author} {\bibinfo {author} {\bibfnamefont {D.}~\bibnamefont
  {Groll}}, \bibinfo {author} {\bibfnamefont {T.}~\bibnamefont {Hahn}},
  \bibinfo {author} {\bibfnamefont {P.}~\bibnamefont {Machnikowski}}, \bibinfo
  {author} {\bibfnamefont {D.}~\bibnamefont {Wigger}},\ and\ \bibinfo {author}
  {\bibfnamefont {T.}~\bibnamefont {Kuhn}},\ }\bibfield  {title} {\bibinfo
  {title} {{Controlling photoluminescence spectra of hBN color centers by
  selective phonon-assisted excitation: a theoretical proposal}},\ }\href@noop
  {} {\bibfield  {journal} {\bibinfo  {journal} {Mater. Quantum Technol.}\
  }\textbf {\bibinfo {volume} {1}},\ \bibinfo {pages} {015004} (\bibinfo {year}
  {2021})}\BibitemShut {NoStop}%
\bibitem [{\citenamefont {Heberle}\ \emph {et~al.}(1995)\citenamefont
  {Heberle}, \citenamefont {Baumberg},\ and\ \citenamefont
  {K{\"o}hler}}]{heberle1995ultrafast}%
  \BibitemOpen
  \bibfield  {author} {\bibinfo {author} {\bibfnamefont {A.~P.}\ \bibnamefont
  {Heberle}}, \bibinfo {author} {\bibfnamefont {J.~J.}\ \bibnamefont
  {Baumberg}},\ and\ \bibinfo {author} {\bibfnamefont {K.}~\bibnamefont
  {K{\"o}hler}},\ }\bibfield  {title} {\bibinfo {title} {Ultrafast coherent
  control and destruction of excitons in quantum wells},\ }\href@noop {}
  {\bibfield  {journal} {\bibinfo  {journal} {Phys. Rev. Lett.}\ }\textbf
  {\bibinfo {volume} {75}},\ \bibinfo {pages} {2598} (\bibinfo {year}
  {1995})}\BibitemShut {NoStop}%
\bibitem [{\citenamefont {Heberle}\ \emph {et~al.}(1996)\citenamefont
  {Heberle}, \citenamefont {Baumberg}, \citenamefont {Binder}, \citenamefont
  {Kuhn}, \citenamefont {Kohler},\ and\ \citenamefont
  {Ploog}}]{heberle1996coherent}%
  \BibitemOpen
  \bibfield  {author} {\bibinfo {author} {\bibfnamefont {A.~P.}\ \bibnamefont
  {Heberle}}, \bibinfo {author} {\bibfnamefont {J.~J.}\ \bibnamefont
  {Baumberg}}, \bibinfo {author} {\bibfnamefont {E.}~\bibnamefont {Binder}},
  \bibinfo {author} {\bibfnamefont {T.}~\bibnamefont {Kuhn}}, \bibinfo {author}
  {\bibfnamefont {K.}~\bibnamefont {Kohler}},\ and\ \bibinfo {author}
  {\bibfnamefont {K.~H.}\ \bibnamefont {Ploog}},\ }\bibfield  {title} {\bibinfo
  {title} {Coherent control of exciton density and spin},\ }\href@noop {}
  {\bibfield  {journal} {\bibinfo  {journal} {IEEE J. Sel. Top. Quantum
  Electron.}\ }\textbf {\bibinfo {volume} {2}},\ \bibinfo {pages} {769}
  (\bibinfo {year} {1996})}\BibitemShut {NoStop}%
\bibitem [{\citenamefont {Bonadeo}\ \emph {et~al.}(1998)\citenamefont
  {Bonadeo}, \citenamefont {Erland}, \citenamefont {Gammon}, \citenamefont
  {Park}, \citenamefont {Katzer},\ and\ \citenamefont
  {Steel}}]{bonadeo1998coherent}%
  \BibitemOpen
  \bibfield  {author} {\bibinfo {author} {\bibfnamefont {N.~H.}\ \bibnamefont
  {Bonadeo}}, \bibinfo {author} {\bibfnamefont {J.}~\bibnamefont {Erland}},
  \bibinfo {author} {\bibfnamefont {D.}~\bibnamefont {Gammon}}, \bibinfo
  {author} {\bibfnamefont {D.}~\bibnamefont {Park}}, \bibinfo {author}
  {\bibfnamefont {D.~S.}\ \bibnamefont {Katzer}},\ and\ \bibinfo {author}
  {\bibfnamefont {D.~G.}\ \bibnamefont {Steel}},\ }\bibfield  {title} {\bibinfo
  {title} {Coherent optical control of the quantum state of a single quantum
  dot},\ }\href@noop {} {\bibfield  {journal} {\bibinfo  {journal} {Science}\
  }\textbf {\bibinfo {volume} {282}},\ \bibinfo {pages} {1473} (\bibinfo {year}
  {1998})}\BibitemShut {NoStop}%
\bibitem [{\citenamefont {Htoon}\ \emph {et~al.}(2002)\citenamefont {Htoon},
  \citenamefont {Takagahara}, \citenamefont {Kulik}, \citenamefont {Baklenov},
  \citenamefont {Holmes~Jr},\ and\ \citenamefont {Shih}}]{htoon2002interplay}%
  \BibitemOpen
  \bibfield  {author} {\bibinfo {author} {\bibfnamefont {H.}~\bibnamefont
  {Htoon}}, \bibinfo {author} {\bibfnamefont {T.}~\bibnamefont {Takagahara}},
  \bibinfo {author} {\bibfnamefont {D.}~\bibnamefont {Kulik}}, \bibinfo
  {author} {\bibfnamefont {O.}~\bibnamefont {Baklenov}}, \bibinfo {author}
  {\bibfnamefont {A.~L.}\ \bibnamefont {Holmes~Jr}},\ and\ \bibinfo {author}
  {\bibfnamefont {C.~K.}\ \bibnamefont {Shih}},\ }\bibfield  {title} {\bibinfo
  {title} {{Interplay of Rabi oscillations and quantum interference in
  semiconductor quantum dots}},\ }\href@noop {} {\bibfield  {journal} {\bibinfo
   {journal} {Phys. Rev. Lett.}\ }\textbf {\bibinfo {volume} {88}},\ \bibinfo
  {pages} {087401} (\bibinfo {year} {2002})}\BibitemShut {NoStop}%
\bibitem [{\citenamefont {Stufler}\ \emph {et~al.}(2005)\citenamefont
  {Stufler}, \citenamefont {Ester}, \citenamefont {Zrenner},\ and\
  \citenamefont {Bichler}}]{stufler2005quantum}%
  \BibitemOpen
  \bibfield  {author} {\bibinfo {author} {\bibfnamefont {S.}~\bibnamefont
  {Stufler}}, \bibinfo {author} {\bibfnamefont {P.}~\bibnamefont {Ester}},
  \bibinfo {author} {\bibfnamefont {A.}~\bibnamefont {Zrenner}},\ and\ \bibinfo
  {author} {\bibfnamefont {M.}~\bibnamefont {Bichler}},\ }\bibfield  {title}
  {\bibinfo {title} {{Quantum optical properties of a single In$_x$Ga$_{1-
  x}$As-GaAs quantum dot two-level system}},\ }\href@noop {} {\bibfield
  {journal} {\bibinfo  {journal} {Phys. Rev. B}\ }\textbf {\bibinfo {volume}
  {72}},\ \bibinfo {pages} {121301} (\bibinfo {year} {2005})}\BibitemShut
  {NoStop}%
\bibitem [{\citenamefont {Kolodka}\ \emph {et~al.}(2007)\citenamefont
  {Kolodka}, \citenamefont {Ramsay}, \citenamefont {Skiba-Szymanska},
  \citenamefont {Fry}, \citenamefont {Liu}, \citenamefont {Fox},\ and\
  \citenamefont {Skolnick}}]{kolodka2007inversion}%
  \BibitemOpen
  \bibfield  {author} {\bibinfo {author} {\bibfnamefont {R.~S.}\ \bibnamefont
  {Kolodka}}, \bibinfo {author} {\bibfnamefont {A.~J.}\ \bibnamefont {Ramsay}},
  \bibinfo {author} {\bibfnamefont {J.}~\bibnamefont {Skiba-Szymanska}},
  \bibinfo {author} {\bibfnamefont {P.~W.}\ \bibnamefont {Fry}}, \bibinfo
  {author} {\bibfnamefont {H.~Y.}\ \bibnamefont {Liu}}, \bibinfo {author}
  {\bibfnamefont {A.~M.}\ \bibnamefont {Fox}},\ and\ \bibinfo {author}
  {\bibfnamefont {M.~S.}\ \bibnamefont {Skolnick}},\ }\bibfield  {title}
  {\bibinfo {title} {Inversion recovery of single quantum-dot exciton based
  qubit},\ }\href@noop {} {\bibfield  {journal} {\bibinfo  {journal} {Phys.
  Rev. B}\ }\textbf {\bibinfo {volume} {75}},\ \bibinfo {pages} {193306}
  (\bibinfo {year} {2007})}\BibitemShut {NoStop}%
\bibitem [{\citenamefont {Ramsay}(2010)}]{ramsay2010review}%
  \BibitemOpen
  \bibfield  {author} {\bibinfo {author} {\bibfnamefont {A.~J.}\ \bibnamefont
  {Ramsay}},\ }\bibfield  {title} {\bibinfo {title} {A review of the coherent
  optical control of the exciton and spin states of semiconductor quantum
  dots},\ }\href@noop {} {\bibfield  {journal} {\bibinfo  {journal} {Semicond.
  Sci. Technol.}\ }\textbf {\bibinfo {volume} {25}},\ \bibinfo {pages} {103001}
  (\bibinfo {year} {2010})}\BibitemShut {NoStop}%
\bibitem [{\citenamefont {Weiss}(2012)}]{weiss2012quantum}%
  \BibitemOpen
  \bibfield  {author} {\bibinfo {author} {\bibfnamefont {U.}~\bibnamefont
  {Weiss}},\ }\href@noop {} {\emph {\bibinfo {title} {Quantum dissipative
  systems}}}\ (\bibinfo  {publisher} {World scientific},\ \bibinfo {year}
  {2012})\BibitemShut {NoStop}%
\bibitem [{\citenamefont {Spokoyny}\ \emph {et~al.}(2020)\citenamefont
  {Spokoyny}, \citenamefont {Utzat}, \citenamefont {Moon}, \citenamefont
  {Grosso}, \citenamefont {Englund},\ and\ \citenamefont
  {Bawendi}}]{spokoyny2020effect}%
  \BibitemOpen
  \bibfield  {author} {\bibinfo {author} {\bibfnamefont {B.}~\bibnamefont
  {Spokoyny}}, \bibinfo {author} {\bibfnamefont {H.}~\bibnamefont {Utzat}},
  \bibinfo {author} {\bibfnamefont {H.}~\bibnamefont {Moon}}, \bibinfo {author}
  {\bibfnamefont {G.}~\bibnamefont {Grosso}}, \bibinfo {author} {\bibfnamefont
  {D.}~\bibnamefont {Englund}},\ and\ \bibinfo {author} {\bibfnamefont {M.~G.}\
  \bibnamefont {Bawendi}},\ }\bibfield  {title} {\bibinfo {title} {Effect of
  spectral diffusion on the coherence properties of a single quantum emitter in
  hexagonal boron nitride},\ }\href@noop {} {\bibfield  {journal} {\bibinfo
  {journal} {J. Phys. Chem. Lett.}\ }\textbf {\bibinfo {volume} {11}},\
  \bibinfo {pages} {1330} (\bibinfo {year} {2020})}\BibitemShut {NoStop}%
\bibitem [{\citenamefont {Sontheimer}\ \emph {et~al.}(2017)\citenamefont
  {Sontheimer}, \citenamefont {Braun}, \citenamefont {Nikolay}, \citenamefont
  {Sadzak}, \citenamefont {Aharonovich},\ and\ \citenamefont
  {Benson}}]{sontheimer2017photodynamics}%
  \BibitemOpen
  \bibfield  {author} {\bibinfo {author} {\bibfnamefont {B.}~\bibnamefont
  {Sontheimer}}, \bibinfo {author} {\bibfnamefont {M.}~\bibnamefont {Braun}},
  \bibinfo {author} {\bibfnamefont {N.}~\bibnamefont {Nikolay}}, \bibinfo
  {author} {\bibfnamefont {N.}~\bibnamefont {Sadzak}}, \bibinfo {author}
  {\bibfnamefont {I.}~\bibnamefont {Aharonovich}},\ and\ \bibinfo {author}
  {\bibfnamefont {O.}~\bibnamefont {Benson}},\ }\bibfield  {title} {\bibinfo
  {title} {Photodynamics of quantum emitters in hexagonal boron nitride
  revealed by low-temperature spectroscopy},\ }\href@noop {} {\bibfield
  {journal} {\bibinfo  {journal} {Phys. Rev. B}\ }\textbf {\bibinfo {volume}
  {96}},\ \bibinfo {pages} {121202} (\bibinfo {year} {2017})}\BibitemShut
  {NoStop}%
\bibitem [{\citenamefont {Empedocles}\ and\ \citenamefont
  {Bawendi}(1997)}]{empedocles1997quantum}%
  \BibitemOpen
  \bibfield  {author} {\bibinfo {author} {\bibfnamefont {S.~A.}\ \bibnamefont
  {Empedocles}}\ and\ \bibinfo {author} {\bibfnamefont {M.~G.}\ \bibnamefont
  {Bawendi}},\ }\bibfield  {title} {\bibinfo {title} {{Quantum-confined stark
  effect in single CdSe nanocrystallite quantum dots}},\ }\href@noop {}
  {\bibfield  {journal} {\bibinfo  {journal} {Science}\ }\textbf {\bibinfo
  {volume} {278}},\ \bibinfo {pages} {2114} (\bibinfo {year}
  {1997})}\BibitemShut {NoStop}%
\bibitem [{\citenamefont {Tran}\ \emph {et~al.}(2019)\citenamefont {Tran},
  \citenamefont {Bradac}, \citenamefont {Solntsev}, \citenamefont {Toth},\ and\
  \citenamefont {Aharonovich}}]{tran2019suppression}%
  \BibitemOpen
  \bibfield  {author} {\bibinfo {author} {\bibfnamefont {T.~T.}\ \bibnamefont
  {Tran}}, \bibinfo {author} {\bibfnamefont {C.}~\bibnamefont {Bradac}},
  \bibinfo {author} {\bibfnamefont {A.~S.}\ \bibnamefont {Solntsev}}, \bibinfo
  {author} {\bibfnamefont {M.}~\bibnamefont {Toth}},\ and\ \bibinfo {author}
  {\bibfnamefont {I.}~\bibnamefont {Aharonovich}},\ }\bibfield  {title}
  {\bibinfo {title} {{Suppression of spectral diffusion by anti-Stokes
  excitation of quantum emitters in hexagonal boron nitride}},\ }\href@noop {}
  {\bibfield  {journal} {\bibinfo  {journal} {Appl. Phys. Lett.}\ }\textbf
  {\bibinfo {volume} {115}},\ \bibinfo {pages} {071102} (\bibinfo {year}
  {2019})}\BibitemShut {NoStop}%
\bibitem [{\citenamefont {Feldman}\ \emph {et~al.}(2019)\citenamefont
  {Feldman}, \citenamefont {Puretzky}, \citenamefont {Lindsay}, \citenamefont
  {Tucker}, \citenamefont {Briggs}, \citenamefont {Evans}, \citenamefont
  {Haglund},\ and\ \citenamefont {Lawrie}}]{feldman2019phonon}%
  \BibitemOpen
  \bibfield  {author} {\bibinfo {author} {\bibfnamefont {M.~A.}\ \bibnamefont
  {Feldman}}, \bibinfo {author} {\bibfnamefont {A.}~\bibnamefont {Puretzky}},
  \bibinfo {author} {\bibfnamefont {L.}~\bibnamefont {Lindsay}}, \bibinfo
  {author} {\bibfnamefont {E.}~\bibnamefont {Tucker}}, \bibinfo {author}
  {\bibfnamefont {D.~P.}\ \bibnamefont {Briggs}}, \bibinfo {author}
  {\bibfnamefont {P.~G.}\ \bibnamefont {Evans}}, \bibinfo {author}
  {\bibfnamefont {R.~F.}\ \bibnamefont {Haglund}},\ and\ \bibinfo {author}
  {\bibfnamefont {B.~J.}\ \bibnamefont {Lawrie}},\ }\bibfield  {title}
  {\bibinfo {title} {{Phonon-induced multicolor correlations in hBN
  single-photon emitters}},\ }\href@noop {} {\bibfield  {journal} {\bibinfo
  {journal} {Phys. Rev. B}\ }\textbf {\bibinfo {volume} {99}},\ \bibinfo
  {pages} {020101} (\bibinfo {year} {2019})}\BibitemShut {NoStop}%
\bibitem [{\citenamefont {Khatri}\ \emph {et~al.}(2020)\citenamefont {Khatri},
  \citenamefont {Ramsay}, \citenamefont {Malein}, \citenamefont {Chong},\ and\
  \citenamefont {Luxmoore}}]{khatri2020optical}%
  \BibitemOpen
  \bibfield  {author} {\bibinfo {author} {\bibfnamefont {P.}~\bibnamefont
  {Khatri}}, \bibinfo {author} {\bibfnamefont {A.~J.}\ \bibnamefont {Ramsay}},
  \bibinfo {author} {\bibfnamefont {R.~N.~E.}\ \bibnamefont {Malein}}, \bibinfo
  {author} {\bibfnamefont {H.~M.}\ \bibnamefont {Chong}},\ and\ \bibinfo
  {author} {\bibfnamefont {I.~J.}\ \bibnamefont {Luxmoore}},\ }\bibfield
  {title} {\bibinfo {title} {Optical gating of photoluminescence from color
  centers in hexagonal boron nitride},\ }\href@noop {} {\bibfield  {journal}
  {\bibinfo  {journal} {Nano Lett.}\ }\textbf {\bibinfo {volume} {20}},\
  \bibinfo {pages} {4256} (\bibinfo {year} {2020})}\BibitemShut {NoStop}%
\bibitem [{\citenamefont {Cusc{\'o}}\ \emph {et~al.}(2018)\citenamefont
  {Cusc{\'o}}, \citenamefont {Art{\'u}s}, \citenamefont {Edgar}, \citenamefont
  {Liu}, \citenamefont {Cassabois},\ and\ \citenamefont
  {Gil}}]{cusco2018isotopic}%
  \BibitemOpen
  \bibfield  {author} {\bibinfo {author} {\bibfnamefont {R.}~\bibnamefont
  {Cusc{\'o}}}, \bibinfo {author} {\bibfnamefont {L.}~\bibnamefont
  {Art{\'u}s}}, \bibinfo {author} {\bibfnamefont {J.~H.}\ \bibnamefont
  {Edgar}}, \bibinfo {author} {\bibfnamefont {S.}~\bibnamefont {Liu}}, \bibinfo
  {author} {\bibfnamefont {G.}~\bibnamefont {Cassabois}},\ and\ \bibinfo
  {author} {\bibfnamefont {B.}~\bibnamefont {Gil}},\ }\bibfield  {title}
  {\bibinfo {title} {Isotopic effects on phonon anharmonicity in layered van
  der waals crystals: Isotopically pure hexagonal boron nitride},\ }\href@noop
  {} {\bibfield  {journal} {\bibinfo  {journal} {Phys. Rev. B}\ }\textbf
  {\bibinfo {volume} {97}},\ \bibinfo {pages} {155435} (\bibinfo {year}
  {2018})}\BibitemShut {NoStop}%
\bibitem [{\citenamefont {Jakubczyk}\ \emph {et~al.}(2016)\citenamefont
  {Jakubczyk}, \citenamefont {Delmonte}, \citenamefont {Fischbach},
  \citenamefont {Wigger}, \citenamefont {Reiter}, \citenamefont {Mermillod},
  \citenamefont {Schnauber}, \citenamefont {Kaganskiy}, \citenamefont
  {Schulze}, \citenamefont {Strittmatter} \emph
  {et~al.}}]{jakubczyk2016impact}%
  \BibitemOpen
  \bibfield  {author} {\bibinfo {author} {\bibfnamefont {T.}~\bibnamefont
  {Jakubczyk}}, \bibinfo {author} {\bibfnamefont {V.}~\bibnamefont {Delmonte}},
  \bibinfo {author} {\bibfnamefont {S.}~\bibnamefont {Fischbach}}, \bibinfo
  {author} {\bibfnamefont {D.}~\bibnamefont {Wigger}}, \bibinfo {author}
  {\bibfnamefont {D.~E.}\ \bibnamefont {Reiter}}, \bibinfo {author}
  {\bibfnamefont {Q.}~\bibnamefont {Mermillod}}, \bibinfo {author}
  {\bibfnamefont {P.}~\bibnamefont {Schnauber}}, \bibinfo {author}
  {\bibfnamefont {A.}~\bibnamefont {Kaganskiy}}, \bibinfo {author}
  {\bibfnamefont {J.-H.}\ \bibnamefont {Schulze}}, \bibinfo {author}
  {\bibfnamefont {A.}~\bibnamefont {Strittmatter}}, \emph {et~al.},\ }\bibfield
   {title} {\bibinfo {title} {Impact of phonons on dephasing of individual
  excitons in deterministic quantum dot microlenses},\ }\href@noop {}
  {\bibfield  {journal} {\bibinfo  {journal} {ACS Photonics}\ }\textbf
  {\bibinfo {volume} {3}},\ \bibinfo {pages} {2461} (\bibinfo {year}
  {2016})}\BibitemShut {NoStop}%
\bibitem [{\citenamefont {Wigger}\ \emph {et~al.}(2020)\citenamefont {Wigger},
  \citenamefont {Karakhanyan}, \citenamefont {Schneider}, \citenamefont {Kamp},
  \citenamefont {H{\"o}fling}, \citenamefont {Machnikowski}, \citenamefont
  {Kuhn},\ and\ \citenamefont {Kasprzak}}]{wigger2020acoustic}%
  \BibitemOpen
  \bibfield  {author} {\bibinfo {author} {\bibfnamefont {D.}~\bibnamefont
  {Wigger}}, \bibinfo {author} {\bibfnamefont {V.}~\bibnamefont {Karakhanyan}},
  \bibinfo {author} {\bibfnamefont {C.}~\bibnamefont {Schneider}}, \bibinfo
  {author} {\bibfnamefont {M.}~\bibnamefont {Kamp}}, \bibinfo {author}
  {\bibfnamefont {S.}~\bibnamefont {H{\"o}fling}}, \bibinfo {author}
  {\bibfnamefont {P.}~\bibnamefont {Machnikowski}}, \bibinfo {author}
  {\bibfnamefont {T.}~\bibnamefont {Kuhn}},\ and\ \bibinfo {author}
  {\bibfnamefont {J.}~\bibnamefont {Kasprzak}},\ }\bibfield  {title} {\bibinfo
  {title} {Acoustic phonon sideband dynamics during polaron formation in a
  single quantum dot},\ }\href@noop {} {\bibfield  {journal} {\bibinfo
  {journal} {Opt. Lett.}\ }\textbf {\bibinfo {volume} {45}},\ \bibinfo {pages}
  {919} (\bibinfo {year} {2020})}\BibitemShut {NoStop}%
\bibitem [{\citenamefont {Wigger}\ \emph {et~al.}(2014)\citenamefont {Wigger},
  \citenamefont {L{\"u}ker}, \citenamefont {Reiter}, \citenamefont {Axt},
  \citenamefont {Machnikowski},\ and\ \citenamefont {Kuhn}}]{wigger2014energy}%
  \BibitemOpen
  \bibfield  {author} {\bibinfo {author} {\bibfnamefont {D.}~\bibnamefont
  {Wigger}}, \bibinfo {author} {\bibfnamefont {S.}~\bibnamefont {L{\"u}ker}},
  \bibinfo {author} {\bibfnamefont {D.}~\bibnamefont {Reiter}}, \bibinfo
  {author} {\bibfnamefont {V.~M.}\ \bibnamefont {Axt}}, \bibinfo {author}
  {\bibfnamefont {P.}~\bibnamefont {Machnikowski}},\ and\ \bibinfo {author}
  {\bibfnamefont {T.}~\bibnamefont {Kuhn}},\ }\bibfield  {title} {\bibinfo
  {title} {Energy transport and coherence properties of acoustic phonons
  generated by optical excitation of a quantum dot},\ }\href@noop {} {\bibfield
   {journal} {\bibinfo  {journal} {J. Phys. Condens. Matter}\ }\textbf
  {\bibinfo {volume} {26}},\ \bibinfo {pages} {355802} (\bibinfo {year}
  {2014})}\BibitemShut {NoStop}%
\bibitem [{\citenamefont {Grange}\ \emph {et~al.}(2017)\citenamefont {Grange},
  \citenamefont {Somaschi}, \citenamefont {Ant{\'o}n}, \citenamefont
  {De~Santis}, \citenamefont {Coppola}, \citenamefont {Giesz}, \citenamefont
  {Lema{\^\i}tre}, \citenamefont {Sagnes}, \citenamefont {Auff{\`e}ves},\ and\
  \citenamefont {Senellart}}]{grange2017reducing}%
  \BibitemOpen
  \bibfield  {author} {\bibinfo {author} {\bibfnamefont {T.}~\bibnamefont
  {Grange}}, \bibinfo {author} {\bibfnamefont {N.}~\bibnamefont {Somaschi}},
  \bibinfo {author} {\bibfnamefont {C.}~\bibnamefont {Ant{\'o}n}}, \bibinfo
  {author} {\bibfnamefont {L.}~\bibnamefont {De~Santis}}, \bibinfo {author}
  {\bibfnamefont {G.}~\bibnamefont {Coppola}}, \bibinfo {author} {\bibfnamefont
  {V.}~\bibnamefont {Giesz}}, \bibinfo {author} {\bibfnamefont
  {A.}~\bibnamefont {Lema{\^\i}tre}}, \bibinfo {author} {\bibfnamefont
  {I.}~\bibnamefont {Sagnes}}, \bibinfo {author} {\bibfnamefont
  {A.}~\bibnamefont {Auff{\`e}ves}},\ and\ \bibinfo {author} {\bibfnamefont
  {P.}~\bibnamefont {Senellart}},\ }\bibfield  {title} {\bibinfo {title}
  {Reducing phonon-induced decoherence in solid-state single-photon sources
  with cavity quantum electrodynamics},\ }\href@noop {} {\bibfield  {journal}
  {\bibinfo  {journal} {Phys. Rev. Lett.}\ }\textbf {\bibinfo {volume} {118}},\
  \bibinfo {pages} {253602} (\bibinfo {year} {2017})}\BibitemShut {NoStop}%
\bibitem [{\citenamefont {Kaldewey}\ \emph {et~al.}(2017)\citenamefont
  {Kaldewey}, \citenamefont {L{\"u}ker}, \citenamefont {Kuhlmann},
  \citenamefont {Valentin}, \citenamefont {Chauveau}, \citenamefont {Ludwig},
  \citenamefont {Wieck}, \citenamefont {Reiter}, \citenamefont {Kuhn},\ and\
  \citenamefont {Warburton}}]{kaldewey2017demonstrating}%
  \BibitemOpen
  \bibfield  {author} {\bibinfo {author} {\bibfnamefont {T.}~\bibnamefont
  {Kaldewey}}, \bibinfo {author} {\bibfnamefont {S.}~\bibnamefont {L{\"u}ker}},
  \bibinfo {author} {\bibfnamefont {A.~V.}\ \bibnamefont {Kuhlmann}}, \bibinfo
  {author} {\bibfnamefont {S.~R.}\ \bibnamefont {Valentin}}, \bibinfo {author}
  {\bibfnamefont {J.-M.}\ \bibnamefont {Chauveau}}, \bibinfo {author}
  {\bibfnamefont {A.}~\bibnamefont {Ludwig}}, \bibinfo {author} {\bibfnamefont
  {A.~D.}\ \bibnamefont {Wieck}}, \bibinfo {author} {\bibfnamefont {D.~E.}\
  \bibnamefont {Reiter}}, \bibinfo {author} {\bibfnamefont {T.}~\bibnamefont
  {Kuhn}},\ and\ \bibinfo {author} {\bibfnamefont {R.~J.}\ \bibnamefont
  {Warburton}},\ }\bibfield  {title} {\bibinfo {title} {Demonstrating the
  decoupling regime of the electron-phonon interaction in a quantum dot using
  chirped optical excitation},\ }\href@noop {} {\bibfield  {journal} {\bibinfo
  {journal} {Phys. Rev. B}\ }\textbf {\bibinfo {volume} {95}},\ \bibinfo
  {pages} {241306} (\bibinfo {year} {2017})}\BibitemShut {NoStop}%
\bibitem [{\citenamefont {Michaelis~de Vasconcellos}\ \emph
  {et~al.}(2010)\citenamefont {Michaelis~de Vasconcellos}, \citenamefont
  {Gordon}, \citenamefont {Bichler}, \citenamefont {Meier},\ and\ \citenamefont
  {Zrenner}}]{Michaelis2010coherent}%
  \BibitemOpen
  \bibfield  {author} {\bibinfo {author} {\bibfnamefont {S.}~\bibnamefont
  {Michaelis~de Vasconcellos}}, \bibinfo {author} {\bibfnamefont
  {S.}~\bibnamefont {Gordon}}, \bibinfo {author} {\bibfnamefont
  {M.}~\bibnamefont {Bichler}}, \bibinfo {author} {\bibfnamefont
  {T.}~\bibnamefont {Meier}},\ and\ \bibinfo {author} {\bibfnamefont
  {A.}~\bibnamefont {Zrenner}},\ }\bibfield  {title} {\bibinfo {title}
  {Coherent control of a single exciton qubit by optoelectronic manipulation},\
  }\href@noop {} {\bibfield  {journal} {\bibinfo  {journal} {Nat. Photonics}\
  }\textbf {\bibinfo {volume} {4}},\ \bibinfo {pages} {545} (\bibinfo {year}
  {2010})}\BibitemShut {NoStop}%
\bibitem [{\citenamefont {Wigger}\ \emph {et~al.}(2021)\citenamefont {Wigger},
  \citenamefont {Gawarecki},\ and\ \citenamefont
  {Machnikowski}}]{wigger2021remote}%
  \BibitemOpen
  \bibfield  {author} {\bibinfo {author} {\bibfnamefont {D.}~\bibnamefont
  {Wigger}}, \bibinfo {author} {\bibfnamefont {K.}~\bibnamefont {Gawarecki}},\
  and\ \bibinfo {author} {\bibfnamefont {P.}~\bibnamefont {Machnikowski}},\
  }\bibfield  {title} {\bibinfo {title} {Remote phonon control of quantum dots
  and other artificial atoms},\ }\href@noop {} {\bibfield  {journal} {\bibinfo
  {journal} {Adv. Quantum Technol.}\ }\textbf {\bibinfo {volume} {4}},\
  \bibinfo {pages} {2000128} (\bibinfo {year} {2021})}\BibitemShut {NoStop}%
\end{thebibliography}

%

\end{document}


\title{
Resonant and phonon-assisted ultrafast coherent control of a single hBN color center\\
(\textit{Supplementary Material})}

\author{Johann A. Preu{\ss}}
\thanks{These two authors contributed equally}
\affiliation{Institute of Physics and Center for Nanotechnology, University of M\"unster, 48149 M\"unster, Germany}

\author{Daniel Groll}
\thanks{These two authors contributed equally}
\affiliation{Institute of Solid State Theory, University of M\"unster, 48149 M\"unster, Germany}

\author{Robert Schmidt}
\affiliation{Institute of Physics and Center for Nanotechnology, University of M\"unster, 48149 M\"unster, Germany}

\author{Thilo Hahn}
\affiliation{Institute of Solid State Theory, University of M\"unster, 48149 M\"unster, Germany}

\author{Pawe\l{} Machnikowski}
\affiliation{Department of Theoretical Physics, Wroc\l{}aw University of Science and Technology, 50-370~Wroc\l{}aw, Poland}

\author{Rudolf Bratschitsch}
\affiliation{Institute of Physics and Center for Nanotechnology, University of M\"unster, 48149 M\"unster, Germany}

\author{Tilmann Kuhn}
\affiliation{Institute of Solid State Theory, University of M\"unster, 48149 M\"unster, Germany}

\author{Steffen Michaelis de Vasconcellos}
\email{michaelis@uni-muenster.de}
\affiliation{Institute of Physics and Center for Nanotechnology, University of M\"unster, 48149 M\"unster, Germany}

\author{Daniel~Wigger}
\email{daniel.wigger@pwr.edu.pl}
\affiliation{Department of Theoretical Physics, Wroc\l{}aw University of Science and Technology, 50-370~Wroc\l{}aw, Poland}


\begin{abstract}
~
\end{abstract}


\date{\today}

\maketitle

This document contains:\\[-6mm] 
\begin{itemize}
\item[\textbf{S1}] \textbf{Experiment}\\[-6mm]
\begin{itemize}
	\item[\textbf{S1.A}] \textbf{Sample preparation}\\[-6mm]
	\item[\textbf{S1.B}] \textbf{Experimental setup}\\[-6mm]
	\item[\textbf{S1.C}] \textbf{Single-photon emission characteristics}\\[-6mm]
	\item[\textbf{S1.D}] \textbf{Data acquisition and analysis}\\[-6mm]
	\item[\textbf{S1.E}] \textbf{Ruling out additional zero-phonon lines}\\[-6mm]
	\item[\textbf{S1.F}] \textbf{Characterization of the Gaussian spectral jitter}\\[-6mm]
\end{itemize}
\item[\textbf{S2}] \textbf{Theory}\\[-6mm]
\begin{itemize}
	\item[\textbf{S2.A}] \textbf{Model of the color center}\\[-6mm]
	\item[\textbf{S2.B}] \textbf{Lindblad equation in the polaron frame}\\[-6mm]
	\item[\textbf{S2.C}] \textbf{Spectra}\\[-6mm]
	\item[\textbf{S2.D}] \textbf{Modeling the coherent control experiment}\\[-6mm]
	\item[\textbf{S2.E}] \textbf{Correlation functions}\\[-6mm]
	\item[\textbf{S2.F}] \textbf{Modeling the LA phonon spectral density}\\[-6mm]
\end{itemize}
\item[\textbf{S3}] \textbf{Impact of the system parameters}\\[-6mm]
\begin{itemize}
	\item[\textbf{S3.A}] \textbf{Spectrometer response}\\[-6mm]
	\item[\textbf{S3.B}] \textbf{Modeling the laser pulses}\\[-6mm]
	\item[\textbf{S3.C}] \textbf{Fitting PL spectra of the emitter}\\[-6mm]
	\item[\textbf{S3.D}] \textbf{Gaussian spectral jitter}\\[-6mm]
	\item[\textbf{S3.E}] \textbf{Summary of the parameters and relevant formulas}\\[-6mm]
\end{itemize}
\item[\textbf{S4}] \textbf{Simulations for different pulse shapes}\\[-6mm]
\begin{itemize}
	\item[\textbf{S4.A}] \textbf{Coherent control visibility dynamics}\\[-6mm]
	\item[\textbf{S4.B}] \textbf{Impact of LAs and laser asymmetry on the PLE}
\end{itemize}
\end{itemize}

\section{Experiment}
\subsection{Sample preparation}
\begin{figure}[h]
\centering
    \includegraphics[width=0.85\textwidth]{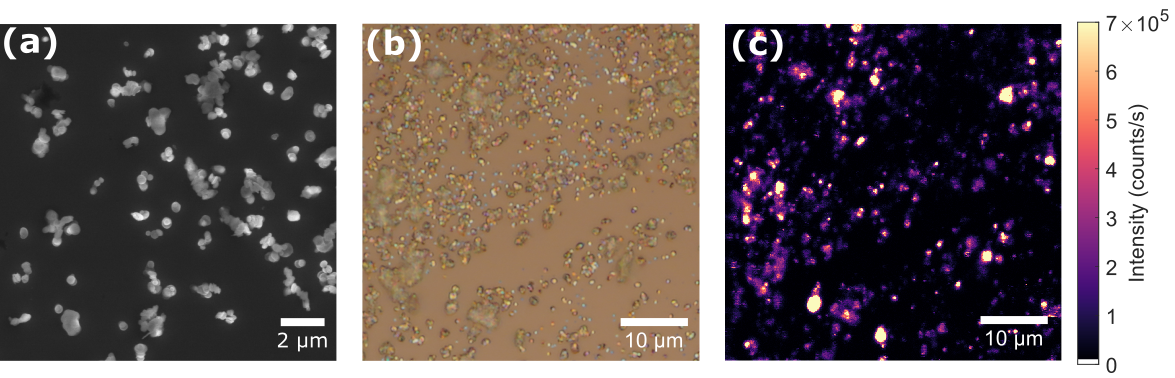}
	\caption{\textbf{Sample characterization.} (a) SEM, (b) white-light and (c) photoluminescence micrograph of a typical randomly scattered hBN nanopowder sample. Some of the bright spots in PL correspond to single-photon emitters in hBN.}
	\label{fig:randomprobe}
\end{figure}

We investigate single-photon emitters in commercially available hBN nanopowder (Sigma-Aldrich)~\cite{wigger2019phonon, preuss2021assembly}. The particles are disk-shaped with average diameters of $< 150$~nm. The powder is annealed at 850$^\circ$\,C for 30~min in an Ar environment to increase the number of single-photon emitters~\cite{tran2016quantum}. Subsequently, the powder is scattered onto a 500~\textmu m thick quartz substrate, on which the crystals stick via van der Waals forces. Figure \ref{fig:randomprobe} shows these randomly scattered hBN particles in a scanning electron microscope (SEM) (a), white-light (b), and photoluminescence (PL) micrograph (c) at different positions of a typical sample. The focused laser beam is used to individually address single particles or small clusters of particles.

\subsection{Experimental setup}\label{setup}
\begin{figure}[b]
	\centering
    \includegraphics[width=0.8\textwidth]{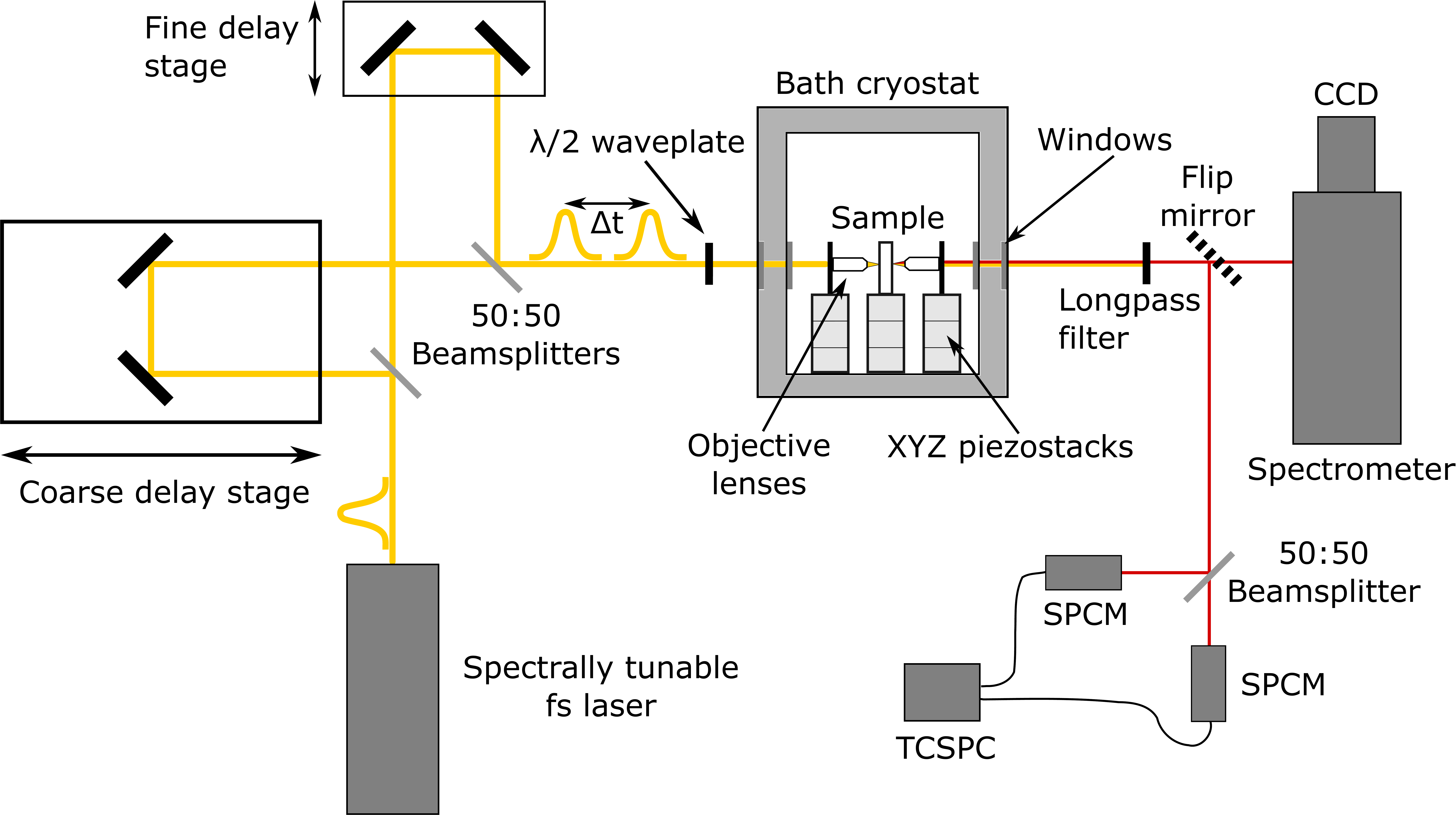}
	\caption{\textbf{Sketch of the experimental setup.}}
	\label{fig:setup}
\end{figure}

The sample is mounted in a variable temperature liquid helium bath optical cryostat. A low-temperature microscope inserted in the sample chamber of the bath cryostat with two high numerical aperture objective lenses ($\text{NA} = 0.82$) allows for high resolution optical measurements. The sample chamber is filled with helium (nitrogen) vapor, cooling the sample down to 8~K (80~K). A sketch of the experimental setup is shown in Fig.~\ref{fig:setup}.

Ultrafast laser pulses are provided by a femtosecond Er:fiber laser system. The pulses are used to generate a supercontinuum in a highly nonlinear fiber and are subsequently frequency doubled using a periodically poled lithium niobate (PPLN) crystal to obtain $\approx 230$~fs pulses that are tunable over the full visible spectrum with a spectral width of 5~meV~\cite{schmidt2016ultrafast}. Each pulse is split in a Mach-Zehnder interferometer at a 50:50 plate beamsplitter. The two pulses are delayed with respect to each other up to 1360~ps of coarse optical delay using a motorized translation stage (PI M-521.DD, 204~mm long) and up to 667~fs of fine optical delay with a piezo stage (PI P-621.1CD, 100~\textmu m scan range) before being rejoined by an identical 50:50 plate beam splitter. The polarization of the combined laser pulses can be adjusted using a $\lambda/2$ waveplate to match the dipole orientation of the single-photon emitter, optimizing the excitation efficiency and therefore maximizing the PL intensity of the emitter. The PL of the emitter is collected and focused into a spectrometer and cryogenically cooled CCD (Princeton Instruments Acton SpectraPro SP-2500 and  Pylon PYL100BRX). The resolution of the spectrometer is $<$0.1~nm when using 1200 lines/mm and 0.3~nm with 300 lines/mm diffraction gratings for high resolution or large spectral range measurements, respectively. For photon correlation measurements, the photoluminescence is routed into a single-mode fiber-coupled Hanbury Brown and Twiss setup. Using two Excelitas SPCM-AQRH-14 single-photon counting modules and a PicoQuant TimeHarp 260 time-correlated single-photon counting board, we detect the arrival times of single photons in a start-stop-type experiment and measure the pulsed second-order autocorrelation function $g^2(\tau)$, described in the following chapter.

\subsection{Single-photon emission characteristics}\label{g2}
\begin{figure}[h]
 	\centering
    \includegraphics[width=0.6\textwidth]{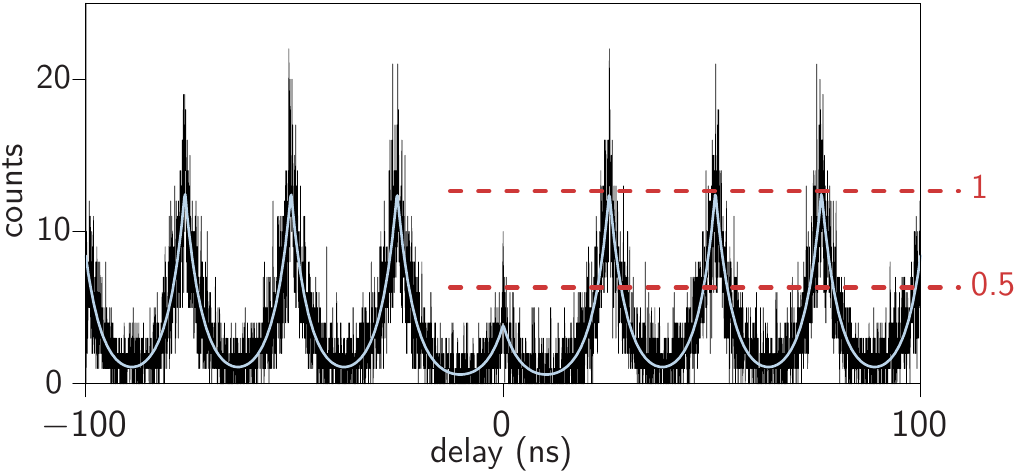}
	\caption{\textbf{Single-photon emission.} Second-order autocorrelation function $g^2(\tau)$ of the emitter shown in the main manuscript, measured at $T=8$~K. The emitter is excited resonantly. The peak height at $\tau=0$ time delay is reduced to 0.3 relative to the mean peak height. The data is not background corrected.}
	\label{fig:g2}
\end{figure}

To test for single-photon emission from the emitter shown in the manuscript, we perform a Hanbury Brown and Twiss experiment with pulsed laser excitation (see Fig. \ref{fig:setup}) to obtain the second-order autocorrelation function as described in chapter \ref{setup}. The measurement is carried out at $T=8$~K and the results are shown in Fig. \ref{fig:g2}. The pulsed laser excitation is resonant to the ZPL of the emitter. All photons in the spectral range of $-130$~meV to $-440$~meV detuned from the ZPL impinge on the single-photon detectors. This includes the first and second LO phonon sideband (PSB). No background correction is carried out. A set of exponential functions
\begin{align}
    A\exp\left(-\frac{|\tau|}{t_{\text{decay}}}\right) + \sum_n B\exp\left(-\frac{|\tau-n\cdot \tau_0|}{t_{\text{decay}}}\right)\,,\qquad n=\pm 1,\,\pm2,\,\pm3,\,\dots
\end{align} 
is fitted to the data, where $A$ and $B$ are the amplitudes, $t_{\text{decay}}$ is the decay time, $\tau$ is the time delay, $\tau_0$ the separation of the laser pulses. The relative amplitude at time delay $\tau=0$ is $A/B\approx 0.3$ which is below the single-photon threshold of 0.5. The decay time of the peaks is $t_{\text{decay}}\approx4$~ns, which is in agreement with the typical lifetime of a single-photon emitter in hBN~\cite{wigger2019phonon}.
 
\subsection{Data acquisition and analysis}
\begin{figure}[h]
\centering
    \includegraphics[width=0.75\textwidth]{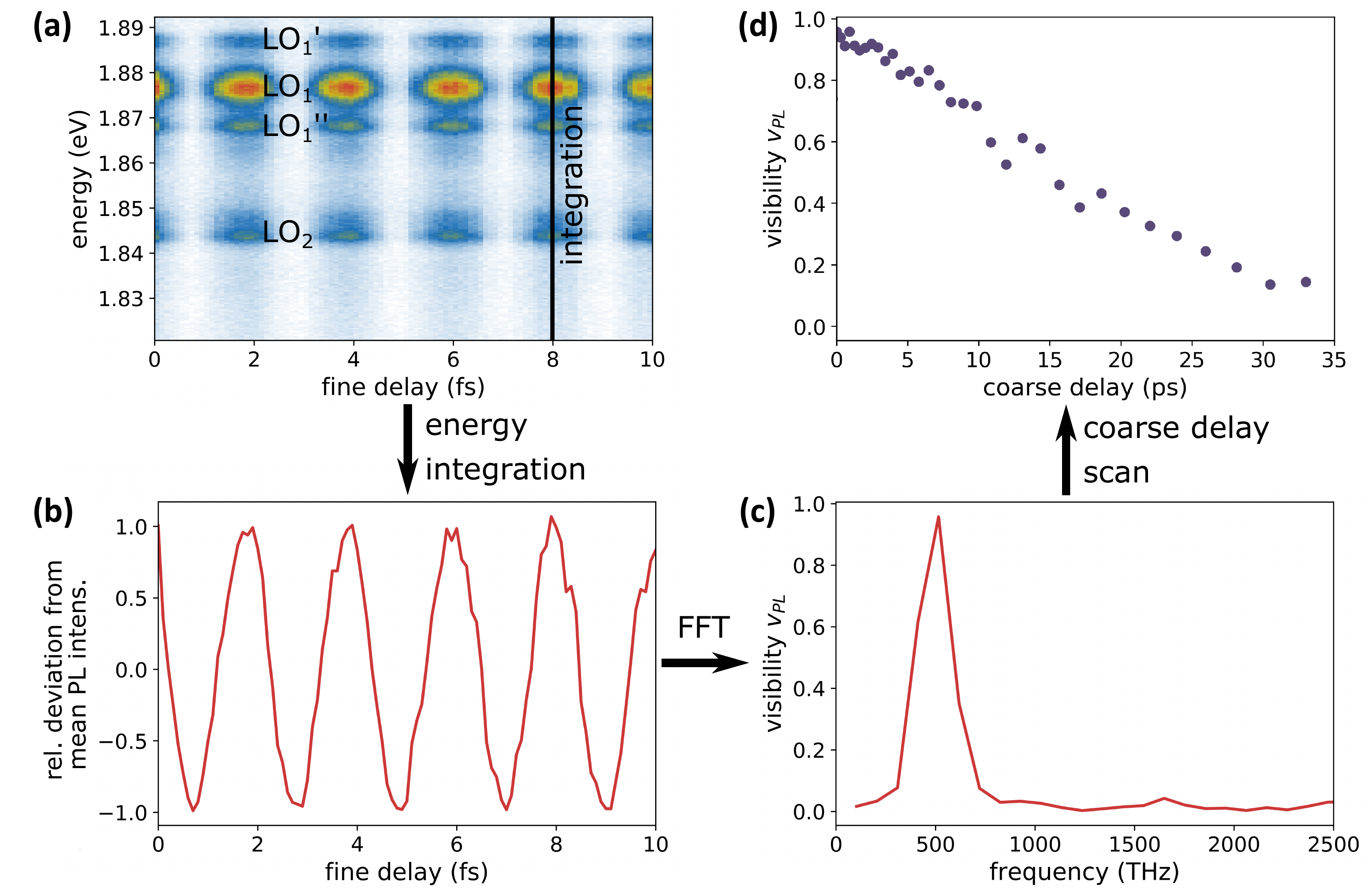}
	\caption{\textbf{Data aquisition and analysis.} (a) Raw PL spectrum of the LO phonon sidebands measured for 0 -- 10~fs of fine delay at a coarse delay of 1~ps. (b) Relative deviation of the signal in (a) from the mean PL intensity. (c) Fast Fourier transform (FFT) of the oscillations in (b). A peak at 492~THz is observable, corresponding to the transition energy of the emitter at 2.036~eV at $T=8$~K. The height of the peak is the visibility of the quantum interference at this coarse delay step. (d) Visibility data extracted from (c) for all coarse delays.}
	\label{fig:evaluation}
\end{figure}

To detect the coherent control signal of the hBN single-photon emitter, PL spectra are recorded for fine delays $\Delta t_f$ of the two pulses between 0 and 10~fs, with a step size of 0.1~fs. This is done at each coarse delay step with, depending on the experiment, varying step sizes from 0.1~ps to 2~ps. In the following discussion, all intensities also depend on the coarse delay, which is kept implicit. For resonant excitation, the PL of the PSB is collected, as shown in Fig. \ref{fig:evaluation}(a). For excitation via the PSB, the PL of the zero-phonon line (ZPL) is collected. The PL intensity data $i_\text{PL}(\Delta t_f, E)$ is then processed by integrating in the direction of the vertical black line shown in Fig. \ref{fig:evaluation}(a), i.e., over the energy $E$
$$
I_\text{PL}(\Delta t_f) = \int\limits_{E_\text{min}}^{E_\text{max}} i_\text{PL}(\Delta t_f,E)\, {\rm d}E
$$
and calculating the relative deviation from the mean PL intensity
$$
	\Delta I_\text{PL}(\Delta t_f) = \frac{I_\text{PL}(\Delta t_f) - \overline{I}_\text{PL} }{\overline{I}_\text{PL}} 
	\qquad \text{where}\qquad
	\overline{I}_\text{PL} = \frac{1}{\max(\Delta t_f)-\min(\Delta t_f)} \int\limits_{\min(\Delta t_f)}^{\max(\Delta t_f)} I_\text{PL}(\tau) \,{\rm d}\tau 
$$
to obtain the amplitude of the oscillation (Fig. \ref{fig:evaluation}(b)). With this definition, the coherent control visibility for a given coarse delay $\Delta t$ from Eq. (1) in the main text reads
$$
v_\text{PL} = \max[\Delta I_\text{PL}(\Delta t_f)]\,
$$
for a noise-free, perfectly sinusoidal signal.

To account for deviations from this idealized case, a fast Fourier transform (FFT) is carried out to extract the amplitude of the oscillation corresponding to the transition frequency of the emitter, which at $T=8$~K is 492~THz (2.036~eV), as shown in Fig. \ref{fig:evaluation}(c). This procedure allows for an accurate analysis even when the oscillation amplitude becomes close to the noise level of the experiment after an almost full decoherence of the two-level system. This amplitude corresponds to the PL visibility of the coherent control experiment as described in the main text. This analysis is repeated for each coarse delay step. The resulting visibility data set is shown in Fig. \ref{fig:evaluation}(d). It corresponds to the envelope of the PL oscillation and is related to the coherence of the system as described in the main text and the Theory section.

 \newpage
\subsection{Ruling out additional zero-phonon lines}
\begin{figure}[h!]
	\centering
    \includegraphics[width=0.65\textwidth]{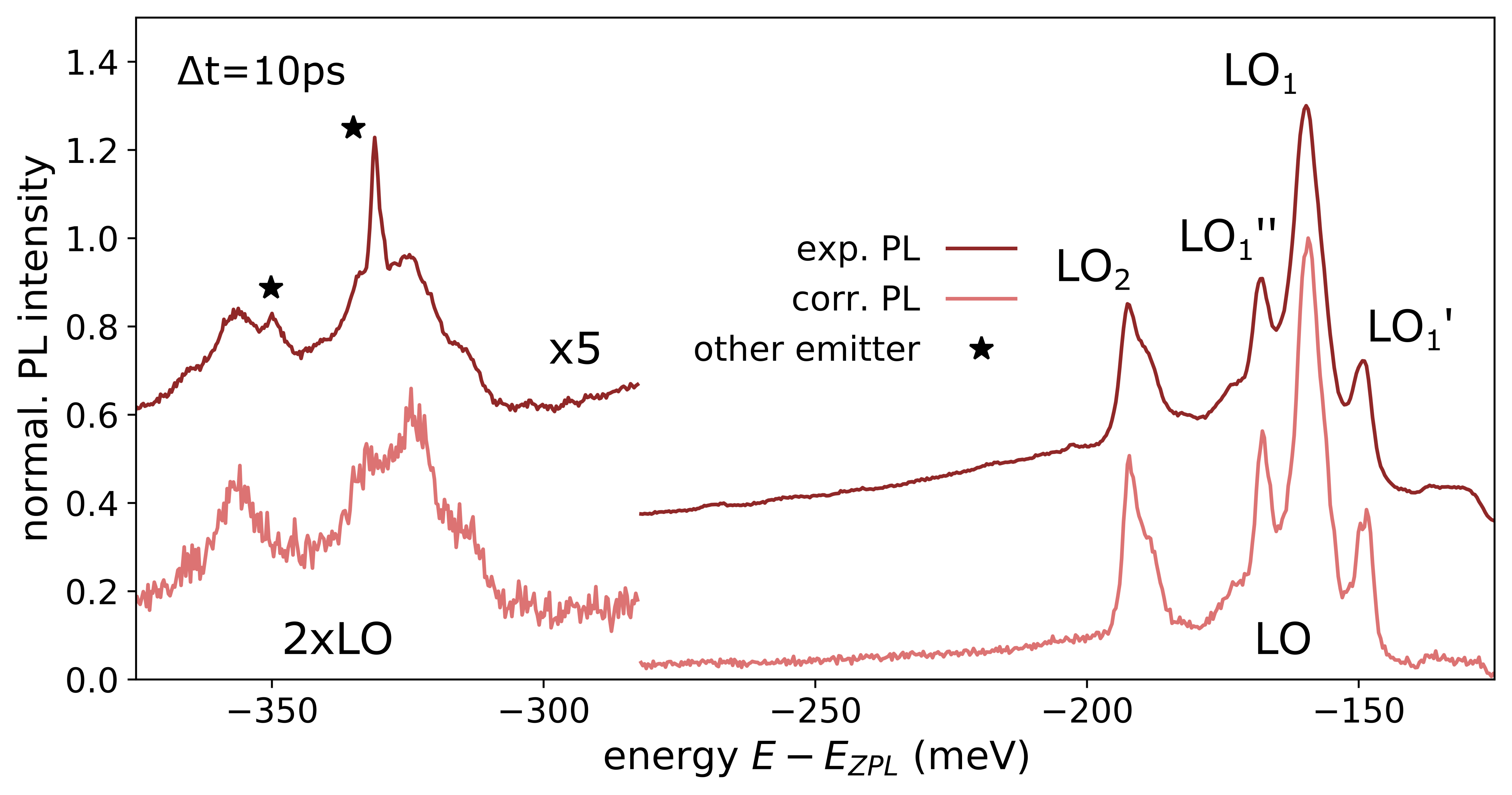}
	\caption{\textbf{Identification of other emitters in the PL spectrum.} Comparison of the single-pulse PL spectrum of the PSBs (upper curves, also shown in Fig.~1 in the main text) and the difference spectrum extracted from the coherent control experiment at $\Delta t=10$~ps (lower curves). The data is shifted vertically for clarity. The emitter is excited resonantly. The difference spectrum is obtained by subtracting the minimum of an oscillation from the maximum. Only parts of the spectrum related to the two-level system remain after this subtraction. We can rule out, that the narrow lines LO$_1$, LO$_1'$, LO$_1''$, and LO$_2$ stem from another emitter. The sharp lines marked with a star are not present in the difference spectrum and therefore belong to other emitters.}
	\label{fig:Truespec}
\end{figure}

The PL spectrum of the PSBs of the single-photon emitter exhibits a complex structure with narrow lines (see Fig. 1 of the main text and \ref{fig:Truespec}, upper spectrum). To identify possible contributions from other emitters in the spectrum, we utilize the PL spectra acquired during the coherent control experiments. We measure the PL spectra after resonant two-pulse excitation at time delays $\Delta t$ of the laser pulses without temporal overlap (e.g. 10~ps). By subtracting the PL spectrum at a minimum of the oscillation in Fig.~\ref{fig:evaluation}(b) from the spectrum at a maximum, we remove all incoherent background from the PL spectrum. Thereby, also contributions from other emitters are efficiently suppressed which are only non-resonantly excited and therefore do not show long coherence.

The resulting spectrum is depicted in the lower part of Fig. \ref{fig:Truespec}. Using this analysis, we find that LO$_1$, LO$_1'$, LO$_1''$, and LO$_2$ can be attributed to the two-level system (ZPL) under study, since they still exhibit a coherent control signal at 10~ps time delay. ZPLs from other emitters will not appear in the difference PL spectrum. In the single-pulse PL spectrum a sharp line in the second LO sideband at $E-E_\text{ZPL}\approx -335$~meV (marked by the star) is observed which is not reproduced by our simulation (see Fig. 1 of the main text). It is also not observed in the difference PL spectrum and therefore stems from a ZPL of another emitter. This is also the case for the smaller narrow line at -348~meV.

\subsection{Characterization of the Gaussian spectral jitter}\label{sec:jitter_exp}
\begin{figure}[h]
\centering
    \includegraphics[width=0.6\textwidth]{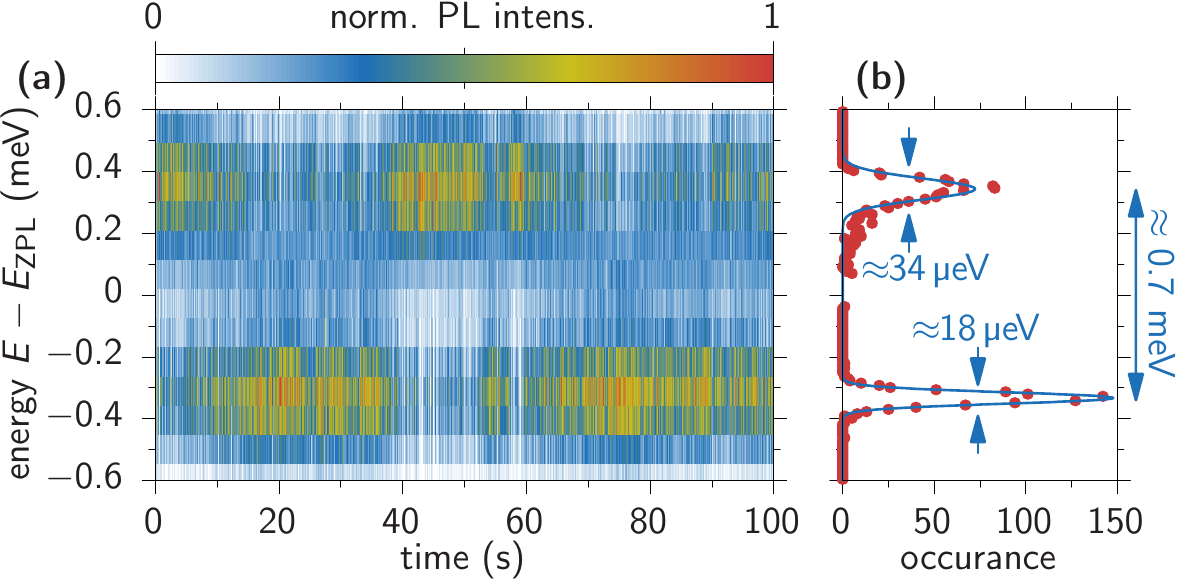}
	\caption{\textbf{Estimation of the Gaussian spectral jitter.} (a) Time series of PL spectra with 10~ms integration time. (b) Histogram of the peak energies as red dots and Gaussian fit as blue line.}
	\label{fig:jitter}
\end{figure}

In the main text we identify a spectral jitter on the timescale of the repetition rate of the coherent control experiment, which has an amplitude smaller than the linewidth of the ZPL. To determine the amplitude of this jitter, we measure a series of PL spectra depicted in Fig.~\ref{fig:jitter}(a) where each time step is integrated over 10~ms. The time required to read out the camera after each integration is $\approx 90$~ms, during which no data is acquired. As discussed in the main text and seen in the figure, the additional random telegraph noise results in a two-peak structure of the ZPL. We determine the energies of the peak maxima by fitting two Gaussians for each time step. From these 2000 energies for the peak positions we calculate the histogram depicted in Fig.~\ref{fig:jitter}(b) as red dots. We then determine the width of the resulting two maxima by fitting two Gaussians, which leads to the standard deviations 34~\textmu eV for the peak at $E-E_\text{ZPL}\approx + 0.35$~meV and 18~\textmu eV for the peak at $E-E_\text{ZPL}\approx - 0.35$~meV. $E_\text{ZPL}$ is in this case defined as the center between the two states. Note that each state in the time series is the ZPL at that time.

These two widths (18~\textmu eV and 34~\textmu eV) contain contributions from the spectral jitter and the uncertainty of the energy determination due to photon noise, which originates from the finite number of photons collected in each peak
\begin{align}\label{eq:noise}
    \sigma_\text{total}^2 = \sigma_\text{jitter}^2 + \sigma_\text{noise}^2\,.
\end{align}
$\sigma_\text{noise}$ can be estimated using the following formula, under the assumption that the measurement noise is dominated by shot noise~\cite{mortensen2010optimized}
\begin{align}
	\sigma_\text{noise} = \sqrt{\frac{16}{9}\,\frac{s^2 + a^2/12}{N}}\,.
\end{align}
Here, $N \approx 700$ is the number of photons detected per peak and frame, $s \approx 180$~\textmu eV is the standard deviation of the Gaussian peak fit of each time step and $a=97$~\textmu eV is the spectral width of a single pixel of the spectrometer-CCD combination. This results in a noise-induced uncertainty of the emitter energy of $\sigma_\text{noise} = 9$~\textmu eV.
Using \eqref{eq:noise}, we can estimate the jitter amplitude $\sigma_\text{jitter}$ to 16~\textmu eV and 33~\textmu eV for the two peaks, respectively.

Due to the finite temporal resolution of the series of spectra, the determined values are lower bounds for the spectral jitter. Nevertheless, they are in good agreement with the values obtained from the coherent control experiment shown in the main text.

\newpage
\section{Theory}
\subsection{Model of the color center}
As described in the main text, we model the hexagonal Boron Nitride (hBN) color center as a two-level system (TLS), coupled to phonon modes of the bulk crystal via the independent boson Hamiltonian~\cite{mahan2013many,krummheuer2002the,alicki2004pure,roszak2006path,nazir2016modelling}. The full Hamiltonian, that describes the color center driven by the electric field of a laser, is given by
\begin{equation}\label{eq:H_t}
H(t)=\underbrace{H_{\mathrm{IB}}+H_{\mathrm{anh}}+H_{\mathrm{pd}}}_{H_0}+H_I(t)\,,
\end{equation}
where $H_0$ describes the free dynamics in absence of laser pulses. $H_{\mathrm{IB}}$ describes the coupling of the phonon modes with the excited state of the TLS and is modeled via
\begin{equation}\label{eq:H_IB}
H_{\mathrm{IB}}=\hbar\omega_{\mathrm{TLS}} X^{\dagger}X+\sum_n \hbar\omega_n b_n^{\dagger}b_n^{}+X^{\dagger}X\sum_n \hbar\qty(g_n^{}b_n^{\dagger}+g_n^*b_n^{}).
\end{equation}
Here, $\hbar\omega_{\mathrm{TLS}}$ is the transition energy of the TLS and $X=\ket{G}\bra{X}$ is the annihilation operator of the excited state $\ket{X}$, where $\ket{G}$ denotes the ground state. The frequencies of the phonons are given by $\omega_n$ with associated bosonic annihilation operators $b_n$ and independent boson coupling parameters $g_n$. The index $n$ collectively denotes all quantum numbers of the phonon modes under consideration. We focus here on the longitudinal acoustic (LA) phonons and the longitudinal optical (LO) phonons coupling to the emitter. We assume that the LO modes are dispersion-less, such that we can collect all LO phonon modes with the same frequency in a single effective mode~\cite{stauber2000electron}.\\
We introduce a finite lifetime of the LO modes due to anharmonic coupling to a bath of low energy phonons via $H_{\mathrm{anh}}$~\cite{groll2021controlling}. When speaking of 'phonons' in the following, we do not mean this bath to which the LO phonons couple. Furthermore, we account for pure dephasing of the color center due to the coupling to some unspecified bath via $H_{\rm pd}$. We assume that both of these Hamiltonians lead to phenomenological Lindblad dissipators.\\
$H_I(t)$ describes the time-dependent interaction of the color center
with the external laser field. It is in the dipole and rotating wave approximation given by
\begin{equation}\label{eq:H_I}
H_I(t)=\frac{1}{2}\qty[\mathcal{E}(t)X^{\dagger}+\mathcal{E}(t)^*X]
\end{equation}
with $\mathcal{E}(t)=-2\bra{X}\bm{d}\cdot\bm{E}(t)\ket{G}$ denoting the (effective) laser field, where $\bm{d}$ is the dipole of the TLS and $\bm{E}(t)$ is the electric field of the laser at the color center's position.\\
Before the interaction of the color center with any laser pulse, the TLS is in its ground state and the phonons are in a thermal state $\rho_{\mathrm{phon}}$ at a given temperature $T$, such that the initial total density matrix  at $t\to-\infty$ is given by
\begin{equation}
\rho_{\mathrm{tot}}(-\infty)=\ket{G}\bra{G}\otimes\rho_{\mathrm{phon}}\otimes\rho_{\mathrm{pd}}\otimes\rho_{\mathrm{anh}}\,,
\end{equation}
where $\rho_{\mathrm{pd}}$ and $\rho_{\mathrm{anh}}$ are the density matrices of the baths inducing pure dephasing and anharmonic LO decay, respectively. Since we are interested in temperatures below room temperature and all LO phonons have energies $>150$~meV, they are assumed to be in their vacuum state, i.e., $b_n\rho_{\mathrm{phon}}=\rho_{\mathrm{phon}}b_n^{\dagger}=0$ for $n\in\qty{\mathrm{LO}}$.\\
Note, that in the present work we are interested in picosecond and sub-picosecond dynamics, such that we can neglect the decay of the excited state of the color center, which happens on a nanosecond-timescale~\cite{tran2016quantum}.
\subsection{Lindblad equation in the polaron frame}
In the following we will discuss the effective non-unitary dynamics of the coupled TLS-phonon system in the polaron frame. One moves to the polaron frame via the unitary polaron transformation, which diagonalizes $H_{\mathrm{IB}}$ and is given by~\cite{nazir2016modelling}
\begin{equation}
T_P=\ket{G}\bra{G}+\ket{X}\bra{X}B_+\,,
\end{equation}
where 
\begin{equation}\label{eq:B_pm_def}
B_{\pm}=\exp[\pm\sum_n \frac{1}{\omega_n}\qty(g_nb_n^{\dagger}-g_n^*b_n)]=\prod_n D_n\qty(\pm \frac{g_n}{\omega_n})
\end{equation}
are multi-mode displacement operators in phase space for the phonon modes from \eqref{eq:H_IB} and $D_n(\alpha)$ is the single-mode displacement operator of the $n$-th mode with coherent amplitude $\alpha$~\cite{glauber1963coherent}. With this unitary transformation, the independent boson Hamiltonian is transformed into
\begin{equation}
H_{\mathrm{IB},P}=T_P^{}H_{\mathrm{IB}}T_P^{\dagger}=\hbar\omega_{\mathrm{ZPL}}X^{\dagger}X+\sum_n\hbar\omega_nb_n^{\dagger}b_n\,,
\end{equation}
with
\begin{equation}
\omega_{\mathrm{ZPL}}=\omega_{\mathrm{TLS}}-\sum_n\frac{|g_n|^2}{\omega_n}=\omega_{\mathrm{TLS}}-\int\limits_0^{\infty} \dd\omega\,\frac{J(\omega)}{\omega}\label{eq:omega_ZPL}
\end{equation}
being the polaron shifted transition frequency of the color center. In the emission spectra discussed later, it is the frequency of the zero phonon line (ZPL) which is usually the dominant peak in the spectra. $J(\omega)$ is the so-called (phonon) spectral density, given by
\begin{equation}
J(\omega)=\sum_n |g_n|^2\delta(\omega-\omega_n)\,.
\end{equation}
The coupling of the LO phonons to the bath of low energy phonons with density matrix $\rho_{\mathrm{anh}}$ and the coupling of the TLS to the pure dephasing bath with density matrix $\rho_{\mathrm{pd}}$ is taken to be such, that the Born-Markov approximations hold and the total density matrix approximately factorizes at all times, i.e.,
\begin{equation}
\rho_{\mathrm{tot}}(t)\approx\rho(t)\otimes\rho_{\mathrm{pd}}\otimes\rho_{\mathrm{anh}}\,,
\end{equation}
where $\rho(t)$ is the reduced density matrix of the TLS-phonon system.	Under these conditions and due to the fact that the polaron transformation only acts on the TLS and the phonons, the total density matrix in the polaron frame is given by
\begin{equation}
T_P^{}\rho_{\mathrm{tot}}(t)T_P^{\dagger}\approx T_P^{}\rho(t)T_P^{\dagger}\otimes\rho_{\mathrm{pd}}\otimes\rho_{\mathrm{anh}}\equiv\rho_P(t)\otimes\rho_{\mathrm{pd}}\otimes\rho_{\mathrm{anh}}\,.
\end{equation}
Our goal is to derive an effective equation of motion for the reduced density matrix in the polaron frame $\rho_P$. In the absence of laser pulses, the propagation of the reduced density matrix in the polaron frame from an initial time $t_0$ to a time $t$ is given by
\begin{align}
\rho_P(t)&=\Tr_{\mathrm{pd}, \mathrm{anh}}\qty[T_P^{}\rho_{\mathrm{tot}}(t)T_P^{\dagger}]\overset{\mathcal{E}=0}{=}\Tr_{\mathrm{pd}, \mathrm{anh}}\qty[T_P^{}U_0^{}(t,t_0)\rho_{\mathrm{tot}}(t_0)U_0^{\dagger}(t,t_0)T_P^{\dagger}]\notag\\
&=\Tr_{\mathrm{pd}, \mathrm{anh}}\qty[T_P^{}U_0^{}(t,t_0)T_P^{\dagger}\rho_P(t_0)\otimes\rho_{\mathrm{pd}}\otimes\rho_{\mathrm{anh}} T_P^{}U_0^{\dagger}(t,t_0)T_P^{\dagger}]\notag\\
&=\Tr_{\mathrm{pd}, \mathrm{anh}}\qty[U_{0,P}^{}(t,t_0)\rho_P(t_0)\otimes\rho_{\mathrm{pd}}\otimes\rho_{\mathrm{anh}} U_{0,P}^{\dagger}(t,t_0)]\,.
\end{align}
Here, $\Tr_{\mathrm{pd}, \mathrm{anh}}[\dots]$, denotes the trace over the subspaces of the baths inducing pure dephasing and LO decay. $U_0(t,t_0)=\exp[-\frac{i}{\hbar}H_0(t-t_0)]$ is the free time evolution operator and $U_{0,P}(t,t_0)=T_P^{}U_0^{}(t,t_0)T_P^{\dagger}$ is its version in the polaron frame. Finally, we introduce the phenomenological Lindblad dissipators, that describe pure dephasing and LO decay, by stating that the relation between $\rho_P$ at different times in the absence of laser pulses is obtained by solving~\cite{groll2021controlling}
\begin{equation}\label{eq:eq_of_motion}
\dv{t}\rho_P(t)\overset{\mathcal{E}=0}{=}\mathcal{L}_{\mathrm{pol}}[{\rho}_P(t)]+\mathcal{L}_{\mathrm{phon}}[{\rho}_P(t)]\,,
\end{equation}
where
\begin{equation}\label{eq:Lindblad_polaron}
\mathcal{L}_{\mathrm{pol}}[{\rho}_P(t)]=-i\comm{\omega_{\mathrm{ZPL}}X^{\dagger}X}{{\rho}_P(t)}+\mathcal{D}_{\mathrm{pd}}[{\rho}_P(t)]
\end{equation}
is the Lindblad superoperator acting on the polaron system and
\begin{equation}\label{eq:Lindblad_phonon}
\mathcal{L}_{\mathrm{phon}}[{\rho}_P(t)]=-i\comm{\sum_n\omega_nb_n^{\dagger}b_n}{{\rho}_P(t)}+\mathcal{D}_{\mathrm{anh}}[{\rho}_P(t)]
\end{equation}
is the Lindblad superoperator acting on the phonon system. The phonon and polaron subsystems are thus not coupled by the time evolution induced by $H_{0,P}=T_PH_0T_P^{\dagger}$. The Lindblad dissipators for pure dephasing of the TLS and decay of the LO phonons are given by
\begin{equation}
\mathcal{D}_{\mathrm{pd}}(\rho_P)=\gamma_{\mathrm{pd}}\qty(X^{\dagger}X\rho_P X^{\dagger}X-\frac{1}{2}\acomm{X^{\dagger}X}{\rho_P})
\end{equation}
and
\begin{align}
\mathcal{D}_{\mathrm{anh}}(\rho_P)&=\sum_{n\in \qty{\mathrm{LO}}}\gamma_{\mathrm{LO}}^{(n)}\qty(b_n\rho_P b_n^{\dagger}-\frac{1}{2}\acomm{b_n^{\dagger}b_n}{\rho_P})\,,
\end{align}
respectively. Here, $\gamma_{\mathrm{pd}}$ is the pure dephasing rate and $\gamma_{\mathrm{LO}}^{(n)}$ is the decay rate of the respective LO mode $n$. The subscript $n\in\qty{\mathrm{LO}}$ under the sum indicates that the dissipator only acts on the LO modes.
\subsection{Spectra}
The central formula used in this section to calculate the spectra of the laser and the emitter is presented in Ref.~\cite{eberly1977time} as the time-dependent physical spectrum of light. It is given by
\begin{equation}\label{eq:spec_def_general}
I(t,\omega)=\int\limits_{-\infty}^{\infty}\dd t_1\,\int\limits_{-\infty}^{\infty}\dd t_2\,r^*(t-t_1,\omega)r(t-t_2,\omega)\expval{E^{(-)}(t_1)E^{(+)}(t_2)}\,,
\end{equation}
where $r(\tau,\omega)$ is the response function of the spectrometer, i.e., the outgoing field, when the incoming field is a $\delta$-pulse, and $\omega$ is the frequency, at which the spectrum is measured. $E^{(\pm)}$ is the positive/negative frequency part of the electric field operator at the position of the spectrometer, projected onto the detector's dipole.
\subsubsection{Photoluminescence of the emitter}
To compare between theory and experiment, we need to know the spectral densities of the LO and LA phonons. Information on these can be gained from photoluminescence (PL) spectra of the emitter. The time-dependent emission spectrum of the emitter is given by~\cite{mollow1969power,eberly1977time,groll2021controlling}
\begin{equation}\label{eq:spectrum}
I_{\mathrm{PL}}(t,\omega)=\int\limits_{-\infty}^{\infty}\dd t_1\,\int\limits_{-\infty}^{\infty}\dd t_2\,r^*(t-t_1,\omega)r(t-t_2,\omega)\expval{X^{\dagger}(t_1)X(t_2)}\,.
\end{equation}
We consider the case in which the emitter is excited by a laser pulse with a duration of a few hundred fs. The subsequent radiative decay happens on a nanosecond-timescale, whereas the phonons thermalize on a much faster timescale of a few picoseconds~\cite{cusco2018isotopic}. After the pulse interaction, only the part of the density matrix proportional to $\ket{X}\bra{X}$ contributes to the correlation function $\expval{X^{\dagger}(t_1)X(t_2)}$. In this subspace the phonons thermalize into a displaced thermal phonon state $B_-\rho_{\mathrm{phon}}B_+$~\cite{groll2021controlling}, such that the reduced density matrix $\ket{X}\bra{X}B_-\rho_{\mathrm{phon}}B_+$ is stationary under the effective equations of motion in \eqref{eq:eq_of_motion}. For the time-integrated PL spectrum, the initial non-equilibrium dynamics are irrelevant compared with the stationary dynamics due to the long radiative lifetime. Thus in calculating the PL spectrum we assume a stationary emitter with the total density matrix $\ket{X}\bra{X}B_-\rho_{\mathrm{phon}}B_+\otimes\rho_{\mathrm{anh}}\otimes\rho_{\mathrm{pd}}=B_-X^{\dagger}\rho_{\mathrm{tot}}(-\infty)XB_+$ and time evolution due to $H_0$. The correlation function $\expval{X^{\dagger}(t_1)X(t_2)}$ is calculated in Sec.~\ref{sec:correlation_functions} and is equal to $\tilde{C}(t_1,t_2)=\tilde{C}(t_1-t_2)$ from \eqref{eq:C_tilde_inital} as shown later.\\
For a stationary emitter, i.e., long after the interaction with any laser pulses, the time-integrated spectrum is given by
\begin{equation}
I_{\mathrm{PL}}(\omega)=\int\limits_{-\infty}^{\infty}\dd t_1\,\int\limits_{-\infty}^{\infty}\dd t_2\,r^*(t-t_1,\omega)r(t-t_2,\omega)\tilde{C}(t_1-t_2)\,, 
\end{equation}
where we omitted all irrelevant prefactors. This can be rewritten as
\begin{align}\label{eq:PL}
I_{\mathrm{PL}}(\omega)&=\int\limits_{-\infty}^{\infty}\dd t_1\,\int\limits_{-\infty}^{\infty}\dd t_2\,r^*(t_1,\omega)r(t_2,\omega)\tilde{C}(t_2-t_1)\notag\\
&\sim \int\limits_{-\infty}^{\infty}\dd \Omega\int\limits_{-\infty}^{\infty}\dd t\int\limits_{-\infty}^{\infty}\dd t_1\,\int\limits_{-\infty}^{\infty}\dd t_2\,r^*(t_1,\omega)r(t_2,\omega)\tilde{C}(t)e^{-i\Omega(t-t_2+t_1)}\notag\\
&=\int\limits_{-\infty}^{\infty}\dd \Omega \underbrace{\qty|\int\limits_{-\infty}^{\infty}\dd \tau\, r(\tau,\omega)e^{i\Omega\tau}|^2}_{R(\omega,\Omega)}\int\limits_{-\infty}^{\infty}\dd t\,\tilde{C}(t)e^{-i\Omega t}\,.
\end{align}
In the following section (\eqref{eq:spectrum_laser}) we will see that $R(\omega,\Omega)$, the spectral response of the spectrometer, is given by $\delta(\omega-\Omega)$ in the case of perfect resolution, such that
\begin{equation}\label{eq:PL_R=delta}
I_{\mathrm{PL}}^{R=\delta}(\omega)=\int\limits_{-\infty}^{\infty}\dd t\,\tilde{C}(t)e^{-i\omega t}
\end{equation}
is the ideal emission spectrum of the emitter.
\subsubsection{Ideal and detected laser spectrum}
To calculate the laser spectrum, we start again with \eqref{eq:spec_def_general} and use the fact, that for coherent laser light it is
$$\expval{E^{(-)}(t_1)E^{(+)}(t_2)}=\expval{E^{(-)}(t_1)}\expval{E^{(+)}(t_2)}=\expval{E^{(+)}(t_1)}^*\expval{E^{(+)}(t_2)}\,.$$
Employing Parseval's theorem, we can determine the detected, time-integrated laser spectrum
\begin{align}
I_{\mathrm{laser}}(\omega)&=\int\limits_{-\infty}^{\infty}\dd t\,I_{\mathrm{laser}}(t,\omega)\notag\\
&=\int\limits_{-\infty}^{\infty}\dd t\,\int\limits_{-\infty}^{\infty}\dd t_1\,\int\limits_{-\infty}^{\infty}\dd t_2\,r^*(t-t_1,\omega)r(t-t_2,\omega)\expval{E^{(+)}(t_1)}^*\expval{E^{(+)}(t_2)}\notag\\
&=\int\limits_{-\infty}^{\infty}\dd t\,\qty|\int\limits_{-\infty}^{\infty}\dd \tau\,r(\tau,\omega)\expval{E^{(+)}(t-\tau)}|^2\notag\\
&\sim \int\limits_{-\infty}^{\infty}\dd\Omega\, \qty|\int\limits_{-\infty}^{\infty}\dd t\,e^{i\Omega t}\int_{-\infty}^{\infty}\dd \tau\,r(\tau,\omega)\expval{E^{(+)}(t-\tau)}|^2\notag\\
&=\int\limits_{-\infty}^{\infty}\dd\Omega\, \qty|\int\limits_{-\infty}^{\infty}\dd t\,\expval{E^{(+)}(t)}e^{i\Omega t}\int\limits_{-\infty}^{\infty}\dd \tau\,r(\tau,\omega)e^{i\Omega \tau}|^2\notag\\
&=\int\limits_{-\infty}^{\infty}\dd \Omega\, R(\omega,\Omega)\qty|\int\limits_{-\infty}^{\infty}\dd t\,\expval{E^{(+)}(t)}e^{i\Omega t}|^2\,.\label{eq:spectrum_laser}
\end{align}
We see, that the detected spectrum is given by the absolute square of the Fourier transform of the electric field, i.e., the ideal laser spectrum, if $R(\omega,\Omega)=\delta(\omega-\Omega)$, which suggests to call this the case of perfect resolution. The spectral response $R(\omega,\Omega)$ can be determined by detecting the spectrum of a spectrally narrow light source approaching the limit of having a single frequency component $\sim\exp(-i\omega_l t)$, such that the detected spectrum is proportional to $R(\omega,\omega_l)$. If we assume a homogeneous response over the frequency interval of interest, i.e., $R(\omega,\Omega)=R(\omega-\Omega)$, we can thus determine the spectral response entirely.\\
The ideal laser spectrum
\begin{equation}\label{eq:laser_spectrum_R=delta}
I_{\mathrm{laser}}^{R=\delta}(\omega)=\qty|\int\limits_{-\infty}^{\infty}\dd t\,\expval{E^{(+)}(t)}e^{i\omega t}|^2
\end{equation}
has a direct connection to the autocorrelation of the electric field of the laser. This can be seen by calculating the Fourier transform of the ideal spectrum
\begin{align}\label{eq:autocorrelation_from_spectrum}
\int\limits_{-\infty}^{\infty}\dd \omega\,I_{\mathrm{laser}}^{R=\delta}(\omega)e^{-i\omega t}&=\int\limits_{-\infty}^{\infty}\dd \omega\,\int\limits_{-\infty}^{\infty}\dd{\tau_1}\,\int\limits_{-\infty}^{\infty}\dd{\tau_2}\,e^{i\omega(\tau_1-\tau_2-t)}\expval{E^{(+)}(\tau_1)}\expval{E^{(+)}(\tau_2)}^*\notag\\
&=2\pi\int\limits_{-\infty}^{\infty}\dd{\tau_1}\expval{E^{(+)}(\tau_1)}\expval{E^{(+)}(\tau_1-t)}^*=2\pi\qty(\expval{E^{(+)}}\star \expval{E^{(+)}})(t)\,,
\end{align}
where the $\star$-operator denotes the autocorrelation of the electric field $\expval{E^{(+)}(\tau_1)}$.
\subsection{Modeling the coherent control experiment}
In the coherent control experiment, two laser pulses emitted from the same source and with a relative delay $\Delta t$ interact with the color center. The first pulse creates a coherence, which is then probed by the second pulse. The PL after the second pulse is the observable in the experiment. An optical excitation of the coupled phonon-TLS systems induces non-equilibrium dynamics that last for several picoseconds. In the following, the entire system is in a quasi-equilibrium, where excited and ground state subspaces are in their respective thermal equilibrium and the excited state decays on a nanosecond time scale. During the quasi-equilibrium stage, the time-dependent PL spectrum, integrated over an arbitrary frequency interval, yields the occupation of the excited state~\cite{groll2021controlling}. Since the decay dynamics takes much longer than the initial non-equilibrium dynamics, the time-integrated PL is basically proportional to the occupation of the emitter after the laser pulse interaction.\\
For the simulation of the coherent control signal (Ramsey fringes), this implies that we need to calculate the occupation of the excited state $\ket{X}$ after the interaction with the two laser pulses. In our case, since we neglect the decay of the emitter during the short non-equilibrium stage of the dynamics, this means that we can just calculate the occupation of the excited state at time $t=\infty$. Consequently, the observable is the TLS's excited state occupation
\begin{equation}\label{eq:f_def}
f(\Delta t)\equiv f(\infty;\Delta t)=\expval{X^{\dagger}X}(\infty)=\Tr\qty[X^{\dagger}XU(\infty,-\infty)\rho_{\mathrm{tot}}(-\infty)U^{\dagger}(\infty,-\infty)]\,,
\end{equation}
where $U(t,t_0)$ is the time evolution operator associated with the full Hamiltonian $H(t)$ from \eqref{eq:H_t}, i.e.,
\begin{equation}\label{eq:U}
U(t,t_0)=\hat{T}\exp[-\frac{i}{\hbar}\int\limits_{t_0}^t\dd\tau\,H(\tau)]\,,
\end{equation}
and $\hat{T}$ is the time-ordering operator. The effective laser field of the double pulse reads
\begin{equation}\label{eq:E}
\mathcal{E}(t)=e^{-i\omega_l t}\qty(\tilde{\mathcal{E}}(t)+e^{i\omega_l\Delta t}\tilde{\mathcal{E}}(t-\Delta t))
\end{equation}
and introduces a dependence of the final state occupation $f$ on the delay $\Delta t$ between the pulses via the interaction $H_I(t)$ from \eqref{eq:H_I}. The carrier frequency of the laser is given by $\omega_l$ and $\tilde{\mathcal{E}}(t)$ is the potentially still complex envelope of a single pulse.
\subsubsection{Perturbation expansion}
In the following we perform a perturbation expansion of our observable in \eqref{eq:f_def} in terms of the exciting laser field. We will keep the calculation to the lowest non-vanishing order, which is the main contribution at small pulse areas, i.e., low intensities of the exciting laser pulses. As it turns out, this contribution is sufficient to reproduce the experimental data.\\
We start by writing down a perturbation expansion of the full time evolution operator in terms of the interaction Hamiltonian $H_I$
\begin{equation}
U(t,t_0)=U_0(t,t_0)-\frac{i}{\hbar}\int\limits_{t_0}^t\dd{\tau}\,U_0(t,\tau)H_I(\tau)U_0(\tau,t_0)+\dots \,.\label{eq:u_approx}
\end{equation}
Since $U_0$ does not change the occupation of the emitter, i.e., $\bra{G}U_0\ket{X}=0$, and $\rho_{\mathrm{tot}}(-\infty)\sim\ket{G}\bra{G}$ describes the emitter in its ground state, the lowest order in \eqref{eq:u_approx} is irrelevant and the first contribution to the observable is given by
\begin{align}
f(\Delta t)&\approx\frac{1}{\hbar^2} \int\limits_{-\infty}^{\infty}\dd{\tau_1}\int\limits_{-\infty}^{\infty}\dd{\tau_2}\mathrm{Tr} \big[X^{\dagger}XU_0(\infty,\tau_1)H_I(\tau_1)U_0(\tau_1,-\infty)\notag\\
&\hspace{4cm} \times \rho_{\mathrm{tot}}(-\infty)U_0(-\infty,\tau_2)H_I(\tau_2)U_0(\tau_2,\infty) \big]\,.
\end{align}
Again, since $U_0$ does not alter the occupation of the emitter, this reduces to
\begin{equation}\label{eq:f2_approx}
f(\Delta t)\approx \frac{1}{4\hbar^2}\int\limits_{-\infty}^{\infty}\dd{\tau_1}\int\limits_{-\infty}^{\infty}\dd{\tau_2}\mathcal{E}(\tau_1)\mathcal{E}^*(\tau_2)C(\tau_2,\tau_1)\equiv f^{(2)}(\Delta t)\,,
\end{equation}
where the correlation function
\begin{align}
C(\tau_2,\tau_1)&=C(\tau_2-\tau_1)\notag\\
	&=\Tr\qty[X^{\dagger}XU_0(\infty,\tau_1)X^{\dagger}U_0(\tau_1,-\infty)\rho_{\mathrm{tot}}(-\infty)U_0(-\infty,\tau_2)XU_0(\tau_2,\infty)]
\end{align}
is calculated in Sec.~\ref{sec:correlation_functions} (see \eqref{eq:C_inital}). $f^{(2)}$ denotes the second order perturbation theory contribution to the full observable $f$.
\subsubsection{Representation of the observable in terms of the spectra of the emitter and the laser pulse}
In this section we show that the second-order contribution $f^{(2)}$ is essentially determined by the spectrum of the applied laser pulse and by the absorption spectrum of the emitter. From \eqref{eq:f2_approx} we see that the second-order contribution to the full coherent control signal can be written as
\begin{align}
f^{(2)}(\Delta t)&=\frac{1}{4\hbar^2}\int\limits_{-\infty}^{\infty}\dd\tau_1\,\int\limits_{-\infty}^{\infty}\dd\tau_2\,\mathcal{E}^*(\tau_2)\mathcal{E}(\tau_1)C(\tau_2-\tau_1)\notag\\
&=\frac{1}{4\hbar^2}\qty[\int\limits_{-\infty}^{\infty}\dd\tau_2\,\int\limits_{-\infty}^{\tau_2}\dd\tau_1+\int\limits_{-\infty}^{\infty}\dd\tau_2\,\int\limits_{\tau_2}^{\infty}\dd\tau_1]\,\mathcal{E}^*(\tau_2)\mathcal{E}(\tau_1)C(\tau_2-\tau_1)\notag\\
&=\frac{1}{4\hbar^2}\qty[\int\limits_{-\infty}^{\infty}\dd\tau_2\,\int\limits_{-\infty}^{\tau_2}\dd\tau_1+\int\limits_{-\infty}^{\infty}\dd\tau_1\,\int\limits_{-\infty}^{\tau_1}\dd\tau_2]\,\mathcal{E}^*(\tau_2)\mathcal{E}(\tau_1)C(\tau_2-\tau_1)\notag\\
&=\frac{1}{4\hbar^2}\int\limits_{-\infty}^{\infty}\dd\tau_2\,\int\limits_{-\infty}^{\tau_2}\dd\tau_1\,\qty[\mathcal{E}^*(\tau_2)\mathcal{E}(\tau_1)C(\tau_2-\tau_1)+\mathcal{E}^*(\tau_1)\mathcal{E}(\tau_2)C(\tau_1-\tau_2)]\,.
\end{align}
Introducing the variables $\tau=\tau_2-\tau_1$ and $T=\tau_2/2+\tau_1/2$, such that $\dd{\tau_1}\dd{\tau_2}=\dd T\dd{\tau}$, we obtain
\begin{align}
&f^{(2)}(\Delta t)\notag\\
&=\frac{1}{4\hbar^2}\int\limits_{-\infty}^{\infty}\dd T\,\int\limits_{0}^{\infty}\dd\tau\,\qty[\mathcal{E}^*(T+\tau/2)\mathcal{E}(T-\tau/2)C(\tau)+\mathcal{E}^*(T-\tau/2)\mathcal{E}(T+\tau/2)C(-\tau)]\notag\\
&=\frac{1}{4\hbar^2}\int_{\mathbb{R}^2}\dd T \dd{\tau}\,\mathcal{E}^*(T+\tau/2)\mathcal{E}(T-\tau/2)C(\tau)\notag\\
&=\frac{1}{4\hbar^2}\int_{\mathbb{R}^2}\dd T \dd{\tau}\,\mathcal{E}(T)\mathcal{E}^*(T+\tau)C(\tau)\,.
\end{align}
Now we introduce the representation of a $\delta$-function
\begin{align}
f^{(2)}(\Delta t)&=\frac{1}{8\pi\hbar^2}\int_{\mathbb{R}^4}\dd\Omega\dd t\dd T\dd\tau\,e^{-i\Omega (t-T-\tau)}\mathcal{E}(T)\mathcal{E}^*(t)C(\tau)\notag\\
&=\frac{1}{8\pi\hbar^2}\int_{\mathbb{R}^4}\dd\Omega\dd t\dd T\dd\tau\,e^{i\Omega T}\mathcal{E}(T)e^{-i\Omega t}\mathcal{E}^*(t)e^{i\Omega\tau}C(\tau)\notag\\
&=\frac{1}{8\pi\hbar^2}\int_{\mathbb{R}}\dd\Omega\,\qty|\int_{\mathbb{R}}\dd T\,e^{i\Omega T}\mathcal{E}(T)|^2\int_{\mathbb{R}}\dd\tau\,e^{i\Omega\tau}C(\tau)\,.
\end{align}
By looking at \eqref{eq:laser_spectrum_R=delta}, we see that this expression contains the ideal spectrum of the double pulse used in the coherent control experiment. Note that for linearly polarized light, as is used in the experiment, we can assume $\mathcal{E}(t)\sim \expval{E^{(+)}(t)}$. To make the dependence on the delay $\Delta t$ obvious, we will now express the double pulse spectrum via the single pulse spectrum. Using the representation of the laser field in terms of the two laser pulses from \eqref{eq:E}, we find
\begin{align}
&\qty|\int_{\mathbb{R}}\dd T\,e^{i\Omega T}\mathcal{E}(T)|^2\notag\\
&=\qty|\int_{\mathbb{R}}\dd T\,e^{i(\Omega-\omega_l) T}\qty[\tilde{\mathcal{E}}(T)+e^{i\omega_l\Delta t}\tilde{\mathcal{E}}(T-\Delta t)]|^2\notag\\
&=\int_{\mathbb{R}^2}\dd T\dd T'\,e^{i(\Omega-\omega_l) T}\qty[\tilde{\mathcal{E}}(T)+e^{i\omega_l\Delta t}\tilde{\mathcal{E}}(T-\Delta t)]e^{-i(\Omega-\omega_l) T'}\qty[\tilde{\mathcal{E}}^*(T')+e^{-i\omega_l\Delta t}\tilde{\mathcal{E}}^*(T'-\Delta t)]\notag\\
&=\qty|\int_{\mathbb{R}}\dd T\,e^{i(\Omega-\omega_l) T}\tilde{\mathcal{E}}(T)|^2+\qty|\int_{\mathbb{R}}\dd T\,e^{i(\Omega-\omega_l) T}\tilde{\mathcal{E}}(T-\Delta t)|^2\notag\\
&+\int_{\mathbb{R}^2}\dd T\dd T'\,e^{i(\Omega-\omega_l) T}e^{-i(\Omega-\omega_l) T'}\qty[\tilde{\mathcal{E}}(T)e^{-i\omega_l\Delta t}\tilde{\mathcal{E}}^*(T'-\Delta t)+e^{i\omega_l\Delta t}\tilde{\mathcal{E}}(T-\Delta t)\tilde{\mathcal{E}}^*(T')]\notag\\
&=2\qty|\int_{\mathbb{R}}\dd T\,e^{i(\Omega-\omega_l) T}\tilde{\mathcal{E}}(T)|^2\notag\\
&+\int_{\mathbb{R}^2}\dd T\dd T'\,e^{i(\Omega-\omega_l) T}e^{-i(\Omega-\omega_l) T'}\qty[\tilde{\mathcal{E}}(T)\tilde{\mathcal{E}}^*(T')e^{-i\Omega\Delta t}+\tilde{\mathcal{E}}(T)\tilde{\mathcal{E}}^*(T')e^{i\Omega\Delta t}]\notag\\
&=\qty[2+2\cos(\Omega \Delta t)]\underbrace{\qty|\int_{\mathbb{R}}\dd T\,e^{i(\Omega-\omega_l) T}\tilde{\mathcal{E}}(T)|^2}_{I_{\text{laser}}^{\mathrm{ideal}}(\Omega)}\label{eq:spectrum_double_pulse}\,.
\end{align}
The expression inside the absolute value is proportional to the ideal ($R=\delta$) spectrum of a single laser pulse, as can be seen by examining \eqref{eq:laser_spectrum_R=delta} together with \eqref{eq:E}, and is therefore named $I_{\text{laser}}^{\mathrm{ideal}}(\Omega)$. With this intermediate result, we can write
\begin{equation}
f^{(2)}(\Delta t)=\frac{1}{2\pi\hbar^2}\int_{\mathbb{R}}\dd{\Omega}\,\cos^2\qty(\frac{\Omega\Delta t}{2})I_{\mathrm{laser}}^{\mathrm{ideal}}(\Omega)\int_{\mathbb{R}}\dd\tau\,e^{i\Omega\tau}C(\tau)\,.
\end{equation}
Now we turn to the integral with the correlation function $C(\tau)$. From Eqs. (\ref{eq:C_final}), (\ref{eq:C_tilde_final}), and (\ref{eq:PL_R=delta}) we know, that 
\begin{equation}
I_{\mathrm{PL}}^{R=\delta}(\Omega)=\int_{\mathbb{R}}\dd{\tau}e^{-i\Omega\tau}\tilde{C}(\tau)=\int_{\mathbb{R}}\dd{\tau}e^{-i\Omega\tau}e^{2i\omega_{\mathrm{ZPL}}\tau}C(\tau)
\end{equation}
is the ideal emission spectrum of the emitter. Therefore,
\begin{equation}\label{eq:spec_abs}
\int_{\mathbb{R}}\dd\tau\,e^{i\Omega\tau}C(\tau)=I_{\mathrm{PL}}^{R=\delta}(2\omega_{\mathrm{ZPL}}-\Omega)\equiv I_{\mathrm{abs}}(\Omega)
\end{equation}
is the ideal emission spectrum, mirrored at the ZPL, or in other terms: the absorption spectrum~\cite{mahan2013many,wigger2019phonon}. Consequently, the second order contribution to the observable for the coherent control experiment can be written as
\begin{equation}\label{eq:f2_from_spectra}
f^{(2)}(\Delta t)=\frac{1}{2\pi\hbar^2}\int_{\mathbb{R}}\dd{\Omega}\,\cos^2\qty(\frac{\Omega\Delta t}{2})I_{\mathrm{laser}}^{\mathrm{ideal}}(\Omega)I_{\mathrm{abs}}(\Omega)
\end{equation}
and depends on the ideal spectrum of a single laser pulse and on the absorption spectrum of the color center. It is not sensitive to chirp, which would be a frequency-dependent phase for the Fourier transform of $\mathcal{E}(t)$ and thus cancels in \eqref{eq:laser_spectrum_R=delta}. At higher orders of the expansion we expect chirp to have an impact. Since the laser spectrum is centered around the carrier frequency $\omega_l$, it is obvious from \eqref{eq:f2_from_spectra} that the observable will be oscillatory with respect to the delay $\Delta t$ with a frequency determined by the weight of the overlap of the ideal laser spectrum and the absorption spectrum. Due to the similarity with Ramsey interferometry, these oscillations are often called Ramsey fringes~\cite{press2008complete}.
\subsubsection{Relation between visibility of the Ramsey fringes and the phonon correlation function}
One interesting quantity to obtain from the coherent control experiment is the phonon correlation function $G_{-+}(\tau)$ (see \eqref{eq:G_mp} in Sec.~\ref{sec:phonon_corr}), which contains information on the emitter-phonon coupling. We want to gain information on this quantity from our observable $f(\Delta t)$. To this aim, let us look at the contribution of lowest order in the electric field of the laser pulse from \eqref{eq:f2_from_spectra}. The spectrum of the laser $I_{\mathrm{laser}}^{\mathrm{ideal}}$ is peaked around the carrier frequency $\omega_l$. The absorption spectrum $I_{\mathrm{abs}}$ is peaked at the ZPL frequency $\omega_{\mathrm{ZPL}}$, which roughly corresponds to $2$eV but also contains dominant peaks at the LO phonon sidebands (PSBs). The time-dependence of $f^{(2)}(\Delta t)$ is dominated by a high-frequency oscillation, as can be seen by shifting the frequency integration by the mean frequency of the laser and absorption spectrum overlap
\begin{equation}
\overline{\Omega}=\frac{\int_{\mathbb{R}}\dd\Omega\,\Omega I_{\mathrm{laser}}^{\mathrm{ideal}}(\Omega)I_{\mathrm{abs}}(\Omega)}{\int_{\mathbb{R}}\dd\Omega\,I_{\mathrm{laser}}^{\mathrm{ideal}}(\Omega)I_{\mathrm{abs}}(\Omega)}\,,
\end{equation}
such that
\begin{equation}
f^{(2)}(\Delta t)=\frac{1}{2\pi\hbar^2}\int_{\mathbb{R}}\dd{\Omega}\,\cos^2\qty(\frac{(\Omega+\overline{\Omega})\Delta t}{2})\underbrace{I_{\mathrm{laser}}^{\mathrm{ideal}}(\Omega+\overline{\Omega})I_{\mathrm{abs}}(\Omega+\overline{\Omega})}_{I_0(\Omega)}\,.
\end{equation}
The quantity $I_0(\Omega)$ is a distribution with mean frequency $0$ and the width $\hbar\Delta \Omega$ of this distribution is given by the width of the overlap between laser spectrum and absorption spectrum, which is in our case on the order of a few meV. The mean frequency $\bar{\Omega}$ however, is on the order of 2~eV$/\hbar$, as it is mainly determined by the frequencies $\omega_{\mathrm{ZPL}}$ and $\omega_l$. If the laser excites resonantly on the ZPL, we have $\overline{\Omega}\approx\omega_{\mathrm{ZPL}}$. If the laser excites on a LO-PSB for a LO phonon with frequency $\omega_{\mathrm{LO}}$, we have $\overline{\Omega}\approx\omega_{\mathrm{ZPL}}+\omega_{\mathrm{LO}}$. Therefore $f^{(2)}(\Delta t)$ shows an oscillatory behavior with the mean frequency $\overline{\Omega}$. The envelope of this oscillation varies on a timescale which is long compared to the oscillation period $2\pi/\overline{\Omega}$, as it is determined by the comparably small width of the distribution $I_0$ in the meV range.\\
As we will show in the following the visibility of the oscillations has a direct relation with the correlation function $G_{-+}(\tau)$. To demonstrate this we rewrite $f^{(2)}$, using $\cos^2(x)=\frac{1}{2}\qty[1+\cos(2x)]$, as
\begin{equation}\label{eq:f2_separation}
f^{(2)}(\Delta t)=\frac{1}{4\pi\hbar^2}\int_{\mathbb{R}}\dd{\Omega}\,I_0(\Omega)+\frac{1}{4\pi\hbar^2}\Re \Bigg[e^{i\bar{\Omega}\Delta t}\underbrace{\int_{\mathbb{R}}\dd{\Omega}\,e^{i\Omega\Delta t}I_0(\Omega)}_{\tilde{I}_0(\Delta t)}\Bigg]\,.
\end{equation}
The first term does not depend on the delay $\Delta t$, being the average of the oscillations, whereas the second term consists of a fast oscillatory contribution $\exp(i\bar{\Omega}\Delta t)$ and a slowly varying one $\tilde{I}_0(\Delta t)$. According to the harmonic addition theorem~\cite{nahin1995science}, we can rewrite the second term as
\begin{align}
\Re \qty[e^{i\overline{\Omega}\Delta t}\tilde{I}_0(\Delta t)]&=\Re\qty[\tilde{I}_0(\Delta t)]\cos(\overline{\Omega}\Delta t)-\Im\qty[\tilde{I}_0(\Delta t)]\sin(\overline{\Omega}\Delta t)\notag\\
&=\text{sgn}\qty{\Re\qty[\tilde{I}_0(\Delta t)]}\left|\tilde{I}_0(\Delta t)\right|\cos\qty(\overline{\Omega}\Delta t+\tan^{-1}\qty{\frac{\Im\qty[\tilde{I}_0(\Delta t)]}{\Re\qty[\tilde{I}_0(\Delta t)]}})\,.
\end{align}
Thus, since $\tilde{I}_0(\Delta t)$ is slowly varying, the visibility $v$ of the oscillation of $f^{(2)}$ is given by
\begin{equation}\label{eq:visibility}
v(\Delta t)=\frac{f^{(2)}_{\mathrm{max}}(\Delta t)-f^{(2)}_{\mathrm{min}}(\Delta t)}{f^{(2)}_{\mathrm{max}}(\Delta t)+f^{(2)}_{\mathrm{min}}(\Delta t)}=\frac{\left|\tilde{I}_0(\Delta t)\right|}{\int_{\mathbb{R}}\dd\Omega\,I_0(\Omega)}=\frac{\left|\int_{\mathbb{R}}\dd\Omega\,e^{i\Omega\Delta t}I_0(\Omega)\right|}{\int_{\mathbb{R}}\dd\Omega\,I_0(\Omega)}\,,
\end{equation}
where $f^{(2)}_{\mathrm{max}}(\Delta t)=\max\limits_{\tau\in[0,2\pi/\overline{\Omega}]}f^{(2)}(\Delta t+\tau)$, and analog for $f^{(2)}_{\mathrm{min}}(\Delta t)$. The visibility has the property $v(0)=1$ and $v(\Delta t)\leq v(0)$, as it is indeed the absolute value of the characteristic function $\expval{\exp(i \Omega \Delta t)}$ of the distribution $I_0(\Omega)$ which can be seen in \eqref{eq:visibility}. Furthermore it is a direct measure of 
\begin{equation}
\left|\tilde{I}_0(\Delta t)\right|=\qty|e^{i\overline{\Omega}\Delta t}\int_{\mathbb{R}}\dd{\Omega}\,e^{i\Omega\Delta t}I_0(\Omega)|=\qty|\int_{\mathbb{R}}\dd{\Omega}\,e^{i\Omega\Delta t}I_{\mathrm{laser}}^{\mathrm{ideal}}(\Omega)I_{\mathrm{abs}}(\Omega)|\,,
\end{equation}
which is the absolute value of the Fourier transform of the overlap of the ideal laser spectrum and the absorption spectrum.\\
In the case of $\delta$-shaped laser pulses, i.e., a flat laser spectrum, this quantity is proportional to the absolute value of the Fourier transform of the absorption spectrum
\begin{equation}
\left|\tilde{I}_0(\Delta t)\right|\sim\qty|\int_{\mathbb{R}}\dd{\Omega}\,e^{i\Omega\Delta t}I_{\mathrm{abs}}(\Omega)|\,.
\end{equation}
This in turn is given by the ZPL-mirrored PL spectrum (see \eqref{eq:spec_abs}) in the case of a perfect spectrometer, i.e., 
\begin{align}\label{eq:I_0_abs_convolution}
\left|\tilde{I}_0(\Delta t)\right|&\sim\qty|\int_{\mathbb{R}}\dd{\Omega}\,e^{i\Omega\Delta t}I_{\mathrm{PL}}^{R=\delta}(2\omega_{\mathrm{ZPL}}-\Omega)|\notag\\
&\sim\qty|\int_{\mathbb{R}}\dd{\Omega}\,e^{i\Omega\Delta t}\int\limits_{\mathbb{R}}\dd\tau\, G_{-+}(\tau)e^{i(\Omega-\omega_{\mathrm{ZPL}})\tau}e^{-\frac{\gamma_{\mathrm{pd}}}{2}|\tau|}|\notag\\
&\sim \qty|G_{-+}(\Delta t)e^{-\frac{\gamma_{\mathrm{pd}}}{2}|\Delta t|}|\,,
\end{align}
where we used \eqref{eq:PL_R=delta} and \eqref{eq:C_final} to express the PL spectrum in terms of the phonon correlation function $G_{-+}$. Thus, we see that the visibility is a measure of the pure dephasing and of the phonon correlation function, which contains the information about the color-center--phonon coupling and the LO decay.\\
The case of finite temporal pulse widths is slightly more complicated, leading to a finite time resolution. In this case, the visibility is no direct measure of the phonon correlation function since the finite pulse width leads to smeared-out visibility dynamics. Using the convolution theorem for Fourier transformations and the fact that the inverse Fourier transform of the laser spectrum is proportional to the autocorrelation of a single laser pulse (see \eqref{eq:autocorrelation_from_spectrum})
\begin{align}
\int_{\mathbb{R}}\dd\Omega\,e^{i\Omega\Delta t}I_{\mathrm{laser}}^{\mathrm{ideal}}(\Omega)&\sim\qty[\mathcal{E}_s\star\mathcal{E}_s](-\Delta t)\,,\\
\mathcal{E}_s(t)=e^{-i\omega_l t}\tilde{\mathcal{E}}(t)
\end{align}
we arrive at
\begin{equation}\label{eq:I_0_abs}
\left|\tilde{I}_0(\Delta t)\right|\sim\qty|\qty{\qty[\mathcal{E}_s\star\mathcal{E}_s](\tau)*\qty[G_{-+}(\tau)e^{-i\omega_{\mathrm{ZPL}}\tau}e^{-\frac{\gamma_{\mathrm{pd}}}{2}|\tau|}]}(\Delta t)|\,,
\end{equation}
which is the convolution between the laser pulse autocorrelation and the correlation function of the phonons $G_{-+}$ times the correlation function $e^{-i\omega_{\mathrm{ZPL}}\tau}e^{-\frac{\gamma_{\mathrm{pd}}}{2}|\tau|}$ of the polaron. Thus, a finite laser pulse width leads to a finite time resolution due to the convolution of the quantities of interest, i.e., the correlation functions, with the autocorrelation of the laser pulse. We note here, that there is at this level of approximation, i.e., lowest order in the electric field contribution, no special emphasis on the delays for which the pulses overlap. The convolution with the autocorrelation of the pulse leads to the same smearing out for every value of the delay $\Delta t$.
\subsubsection{Approximation for the visibility on short timescales for small detunings}\label{sec:vis_LA}
In the context of slightly detuned excitation around the ZPL aiming for the influence of LA phonons on the visibility dynamics on short timescales, it is useful to have an approximative formula at hand to understand the data. We derive this here by first separating the absorption spectrum into a absorption via the ZPL and an phonon-assisted absorption via the PSBs
\begin{align}
I_{\mathrm{abs}}(\omega)&=\int\dd{\tau}\, e^{i\omega\tau}C(\tau)=\int\dd{\tau}\, e^{i(\omega-\omega_{\mathrm{ZPL}})\tau}e^{-\frac{\gamma_{\mathrm{pd}}}{2}|\tau|}G_{-+}(\tau)\notag\\
&=\underbrace{\int\dd{\tau}\, e^{i(\omega-\omega_{\mathrm{ZPL}})\tau}e^{-\frac{\gamma_{\mathrm{pd}}}{2}|\tau|}\qty[G_{-+}(\tau)-B^2]}_{I_{\mathrm{abs,PSB}}}+\underbrace{B^2\int\dd{\tau}\, e^{i(\omega-\omega_{\mathrm{ZPL}})\tau}e^{-\frac{\gamma_{\mathrm{pd}}}{2}|\tau|}}_{I_{\mathrm{abs,ZPL}}}\,.
\end{align}
Note, that for super-ohmic LA spectral densities, in particular $J_{\mathrm{LA}}(\omega)\sim\omega^3$ at $\omega\rightarrow 0$, as is the case in the present work (see Sec.~\ref{sec:la_specdens}), the correlation function $G_{-+}(\tau)$ is stationary for $\tau\rightarrow\pm\infty$ with $G_{-+}(\pm\infty)=B^2$, where $B=\Tr\qty[B_{\pm}\rho_{\mathrm{phon}}]$ is the thermal expectation value of the multi-mode displacement operators. This stationary value leads to a sharp peak, i.e., the ZPL, in the spectrum. The remaining part of the phonon correlation function, i.e., $(G_{-+}-B^2)$ leads to the PSBs. Therefore, we chose this separation of the absorption spectrum into the absorption via ZPL $I_{\mathrm{abs,ZPL}}$ and the absorption via PSBs $I_{\mathrm{abs,PSB}}$.\\
Since we are interested in short timescales, we will assume $\gamma_{\mathrm{pd}}=0$ in the derivation of the approximative formula. The visibility from \eqref{eq:visibility} can then be written as
\begin{align}
v(\Delta t)&=\frac{\left|\tilde{I}_0(\Delta t)\right|}{\int_{\mathbb{R}}\dd\Omega\,I_0(\Omega)}=\frac{\qty|\int_{\mathbb{R}}\dd\Omega\,e^{i\Omega\Delta t}I_{\mathrm{laser}}^{\mathrm{ideal}}(\Omega)I_{\mathrm{abs}}(\Omega)|}{\int_{\mathbb{R}}\dd\Omega\,I_{\mathrm{laser}}^{\mathrm{ideal}}(\Omega)I_{\mathrm{abs}}(\Omega)}\notag\\
&=\frac{\qty|\int_{\mathbb{R}}\dd\Omega\,e^{i\Omega\Delta t}I_{\mathrm{laser}}^{\mathrm{ideal}}(\Omega)I_{\mathrm{abs,PSB}}(\Omega)+\int_{\mathbb{R}}\dd\Omega\,e^{i\Omega\Delta t}I_{\mathrm{laser}}^{\mathrm{ideal}}(\Omega)I_{\mathrm{abs,ZPL}}(\Omega)|}{\int_{\mathbb{R}}\dd\Omega\,I_{\mathrm{laser}}^{\mathrm{ideal}}(\Omega)I_{\mathrm{abs,PSB}}(\Omega)+\int_{\mathbb{R}}\dd\Omega\,I_{\mathrm{laser}}^{\mathrm{ideal}}(\Omega)I_{\mathrm{abs,ZPL}}(\Omega)}\,.
\end{align}
We are interested in excitation of the emitter close to the ZPL, i.e., $\omega_l\approx \omega_{\mathrm{ZPL}}$. The spectral width of the laser pulse shall be of a few meV, such that for our purposes the influence of LO phonons in $I_{\mathrm{abs,PSB}}$ can be neglected, since they have energies of $>150$~meV. The LA PSB in hBN is rather flat on a scale of a few meV, such that we make the crude approximation $I_{\mathrm{laser}}^{\mathrm{ideal}}(\Omega)I_{\mathrm{abs,PSB}}(\Omega)\approx I_{\mathrm{laser}}^{\mathrm{ideal}}(\Omega)I_{\mathrm{abs,PSB}}(\omega_l)$. The visibility is then of the form
\begin{align}
v(\Delta t)&\approx\frac{\qty|\int_{\mathbb{R}}\dd\Omega\,e^{i\Omega\Delta t}I_{\mathrm{laser}}^{\mathrm{ideal}}(\Omega)I_{\mathrm{abs,PSB}}(\omega_l)+2\pi B^2e^{i\omega_{\mathrm{ZPL}}\Delta t}I_{\mathrm{laser}}^{\mathrm{ideal}}(\omega_{\mathrm{ZPL}})|}{\int_{\mathbb{R}}\dd\Omega\,I_{\mathrm{laser}}^{\mathrm{ideal}}(\Omega)I_{\mathrm{abs,PSB}}(\omega_l)+2\pi B^2I_{\mathrm{laser}}^{\mathrm{ideal}}(\omega_{\mathrm{ZPL}})}\,,
\end{align}
where we inserted $I_{\mathrm{abs,ZPL}}(\Omega)=2\pi B^2 \delta(\Omega-\omega_{\mathrm{ZPL}})$ since $\gamma_{\mathrm{pd}}=0$ is assumed. Using now the fact that the Fourier transform of the ideal laser spectrum is the autocorrelation of the pulse (see \eqref{eq:autocorrelation_from_spectrum}), we arrive at
\begin{equation}\label{eq:visibility_approx}
v(\Delta t)\approx\frac{\qty|\qty[\mathcal{E}_s\star\mathcal{E}_s](\Delta t)I_{\mathrm{abs,PSB}}(\omega_l)+B^2e^{-i\omega_{\mathrm{ZPL}}\Delta t}I_{\mathrm{laser}}^{\mathrm{ideal}}(\omega_{\mathrm{ZPL}})|}{\qty[\mathcal{E}_s\star\mathcal{E}_s](0)I_{\mathrm{abs,PSB}}(\omega_l)+ B^2I_{\mathrm{laser}}^{\mathrm{ideal}}(\omega_{\mathrm{ZPL}})}\,,
\end{equation}
where we took the complex conjugate of the expression in the absolute value. To understand this formula a little better, we take a look at two aspects. (A) The dynamics on timescales, where the absolute value of the autocorrelation $\qty|\qty[\mathcal{E}_s\star\mathcal{E}_s](\Delta t)|$ is more or less constant and (B) the dynamics for large delays when the autocorrelation has decayed. In the experiment, the autocorrelation decays within 0.5~ps to 1~ps. Up to this time, it is approximately given by $\qty[\mathcal{E}\star\mathcal{E}](\Delta t)\approx\qty|\qty[\mathcal{E}\star\mathcal{E}](\Delta t)|e^{-i\omega_l\Delta t}$. This is strictly true for symmetric laser spectra and only approximately true for asymmetric spectra. The carrier frequency of the single pulse is also approximately the carrier frequency of the autocorrelation. Focusing now on aspect (A), we approximate the autocorrelation further as $\qty[\mathcal{E}_s\star\mathcal{E}_s](\Delta t)\approx\qty|\qty[\mathcal{E}_s\star\mathcal{E}_s](0)|e^{-i\omega_l\Delta t}$. Then the visibility becomes
\begin{align}
v(\Delta t)\overset{\Delta t (\mathrm{small})}{\approx}\frac{\qty|\qty[\mathcal{E}_s\star\mathcal{E}_s](0)e^{-i\omega_l\Delta t}I_{\mathrm{abs,PSB}}(\omega_l)+B^2e^{-i\omega_{\mathrm{ZPL}}\Delta t}I_{\mathrm{laser}}^{\mathrm{ideal}}(\omega_{\mathrm{ZPL}})|}{\qty[\mathcal{E}_s\star\mathcal{E}_s](0)I_{\mathrm{abs,PSB}}(\omega_l)+ B^2I_{\mathrm{laser}}^{\mathrm{ideal}}(\omega_{\mathrm{ZPL}})}\,.
\end{align}
Thus we see that it is of the form $\frac{\qty|ae^{-i\delta\Delta t}+b|}{a+b}$, where $\delta=\omega_l-\omega_{\mathrm{ZPL}}$ is the detuning between the carrier frequency of the laser and the frequency of the ZPL. The visibility therefore exhibits a beat with the beat frequency determined by the detuning. The origin of this beat is the separate excitation of the polaron at frequency $\omega_{\mathrm{ZPL}}$ and of polaron-phonon states at frequency $\omega_l=\omega_{\mathrm{ZPL}}+\delta$, where the detuning $\delta$ together with the spectral width of the pulse determines the involved phonon frequencies.\\
Now we turn to aspect (B), which concerns the long time behavior. By this we mean the behavior after the autocorrelation of the pulse has vanished but before the signal is diminished by pure dephasing, which we have neglected here. We will denote this limit by $\Delta t(\mathrm{large})$ and the visibility then approaches the stationary value
\begin{equation}
v(\Delta t)\overset{\Delta t(\mathrm{large})}{\approx}\frac{B^2I_{\mathrm{laser}}^{\mathrm{ideal}}(\omega_{\mathrm{ZPL}})}{\qty[\mathcal{E}_s\star\mathcal{E}_s](0)I_{\mathrm{abs,PSB}}(\omega_l)+ B^2I_{\mathrm{laser}}^{\mathrm{ideal}}(\omega_{\mathrm{ZPL}})}\,.
\end{equation}
The stationary value of the visibility is thus of the form $\frac{b}{a+b}$. We can take a look at two limiting cases. The first is $a\gg b$, such that the visibility approaches zero as stationary value. This limiting case can be reached by two different routes. The first is $I_{\mathrm{laser}}^{\mathrm{ideal}}(\omega_{\mathrm{ZPL}})\rightarrow 0$, i.e., an excitation that is strongly detuned such that the overlap of the laser with the ZPL is negligible. In this case the polaron is not excited directly and no coherence remains after the autocorrelation of the laser has vanished. The other route is $B^2\rightarrow 0$, i.e., strong phonon coupling leading to dominant LA sidebands. In this case also the absorption is dominantly via the PSB and not the ZPL, such that the polaron is only excited via phonon-assisted absorption resulting in a strong phonon induced dephasing. So the case $v(\Delta t)\overset{\Delta t(\mathrm{large})}{\approx}0$ is a signature of negligible direct excitation of the polaron. Keep in mind that we neglected the pure dephasing here, which would lead to a vanishing of the visibility in any case.\\
The other limiting case of the stationary value of the visibility is $b\gg a$, such that the visibility approaches unity. For this limiting case we need a strong overlap of the laser spectrum with the ZPL and weak LA PSBs. In this way we excite the polaron alone, whose coherence later decays by pure dephasing.\\
The above discussion shows that the short time visibility dynamics for excitation close to the ZPL are very sensitive to the detuning, which determines the overlap of the laser pulse spectrum with the ZPL. It is also very sensitive to the form of the laser spectrum, as this form is scanned by the ZPL when varying the detuning. Furthermore the stationary value of the visibility is determined by the relative strength between PSB excitation and ZPL excitation and thus the form of the LA PSB is important for the correct detuning dependence of the visibility.
\subsubsection{Slow spectral jitter}\label{sec:jitter_theory}
The color centers in hBN are sometimes subject to telegraph-noise type jitter and are usually subject to slow gaussian jitter~\cite{spokoyny2020effect}. Here we discuss shortly how to incorporate spectral jitter into the calculation of the coherent control visibility. We consider only the case of slow jitter, i.e., the system jumps between different values for the ZPL frequency $\omega_{\mathrm{ZPL}}$ on a long timescale, such that every repetition of the coherent control experiment is performed on a fixed ZPL frequency and the emitter jumps between frequencies only in between the experiments. Thus, in the experiment the contributions for different ZPL frequencies are just integrated, such that the observable becomes (see \eqref{eq:f2_from_spectra})
\begin{equation}\label{eq:jitter}
f^{(2)}(\Delta t)=\int\dd{\omega_{\mathrm{ZPL}}}\,\frac{1}{2\pi\hbar^2}\int_{\mathbb{R}}\dd{\Omega}\,\cos^2\qty(\frac{\Omega\Delta t}{2})I_{\mathrm{laser}}^{\mathrm{ideal}}(\Omega)I_{\mathrm{abs}}(\Omega,\omega_{\mathrm{ZPL}})p(\omega_{\mathrm{ZPL}})\,,
\end{equation}
where the dependence of the absorption spectrum on the ZPL frequency is explicitly given in the argument. $p(\omega_{\mathrm{ZPL}})$ is the probability distribution of finding the color center with the frequency $\omega_{\mathrm{ZPL}}$ in a repetition of the experiment. Defining the average absorption spectrum as
\begin{equation}
\expval{I_{\mathrm{abs}}(\Omega)}=\int\dd{\omega_{\mathrm{ZPL}}}\,I_{\mathrm{abs}}(\Omega,\omega_{\mathrm{ZPL}})p(\omega_{\mathrm{ZPL}})\,,
\end{equation}
we see that \eqref{eq:f2_from_spectra} and \eqref{eq:jitter} are equivalent, but $I_{\mathrm{abs}}(\Omega,\omega_{\mathrm{ZPL}})$ is replaced with its $\omega_{\mathrm{ZPL}}$-averaged value. Thus we can calculate the visibility in the same way as in \eqref{eq:visibility} but with the averaged absorption spectrum. In this case, we need to calculate
\begin{align}\label{eq:I_0_jitter}
\left|\tilde{I}_0(\Delta t)\right|&=\qty|\int \dd{\omega_{\mathrm{ZPL}}}\,\int_{\mathbb{R}}\dd{\Omega}\,e^{i\Omega\Delta t}I_{\mathrm{laser}}^{\mathrm{ideal}}(\Omega)I_{\mathrm{abs}}(\Omega,\omega_{\mathrm{ZPL}})p(\omega_{\mathrm{ZPL}})|\notag\\
&\sim \qty|\qty{\qty[\mathcal{E}_s\star\mathcal{E}_s](\tau)*\qty[G_{-+}(\tau)e^{-\frac{\gamma_{\mathrm{pd}}}{2}|\tau|}\int\dd{\omega_{\mathrm{ZPL}}}\,e^{-i\omega_{\mathrm{ZPL}}\tau}p(\omega_{\mathrm{ZPL}})]}(\Delta t)|
\end{align}
and normalize it. In the case of random telegraph noise, $p(\omega_{\mathrm{ZPL}})$ is a sum of two delta functions. In the case of Gaussian jitter, $p(\omega_{\mathrm{ZPL}})$ is a Gaussian.\\
In the following we explicitly consider a Gaussian jitter around $\omega_{\mathrm{ZPL}}^{(0)}$, whose frequency spread is much smaller than the typical width of the laser spectrum. The probability distribution is given by
\begin{equation}
p(\omega_{\mathrm{ZPL}})\sim \exp(-\frac{(\omega_{\mathrm{ZPL}}-\omega_{\mathrm{ZPL}}^{(0)})^2}{2\sigma_j^2})\,,
\end{equation}
where $\sigma_j$ is the spectral standard deviation. For the case of spectrally narrow Gaussian jitter we see from \eqref{eq:I_0_jitter} that the visibility is given by normalizing
\begin{align}
\left|\tilde{I}_0(\Delta t)\right|&\sim\qty|\qty{\qty[\mathcal{E}_s\star\mathcal{E}_s](\tau)*\qty[G_{-+}(\tau)e^{-\frac{\gamma_{\mathrm{pd}}}{2}|\tau|}\int\dd{\omega_{\mathrm{ZPL}}}\,e^{-i\omega_{\mathrm{ZPL}}\tau}p(\omega_{\mathrm{ZPL}})]}(\Delta t)|\notag\\
&\sim\qty|\qty{\qty[\mathcal{E}_s\star\mathcal{E}_s](\tau)*\qty[G_{-+}(\tau)e^{-\frac{\gamma_{\mathrm{pd}}}{2}|\tau|}e^{-i\omega_{\mathrm{ZPL}}^{(0)}\tau}\exp(-\frac{\tau^2\sigma_j^2}{2})]}(\Delta t)|\notag\\
&\sim \qty|\int \dd{\tau}\qty[\mathcal{E}_s\star\mathcal{E}_s](\tau)\exp(-\frac{(\Delta t-\tau)^2\sigma_j^2}{2})G_{-+}(\Delta t-\tau)e^{-\frac{\gamma_{\mathrm{pd}}}{2}|\Delta t-\tau|}e^{-i\omega_{\mathrm{ZPL}}^{(0)}(\Delta t-\tau)}|\notag\\
&\approx \exp(-\frac{\Delta t^2\sigma_j^2}{2})\qty|\qty{\qty[\mathcal{E}_s\star\mathcal{E}_s](\tau)*\qty[G_{-+}(\tau)e^{-\frac{\gamma_{\mathrm{pd}}}{2}|\tau|}e^{-i\omega_{\mathrm{ZPL}}^{(0)}\tau}]}(\Delta t)|\,.\label{eq:Gauss_effect}
\end{align}
In the last step we used that the timescale $1/\sigma_j$ of the Gaussian is much longer than the typical timescale of the laser pulse autocorrelation, i.e.,
\begin{equation}
\qty[\mathcal{E}_s\star\mathcal{E}_s](\tau)\exp(-\frac{(\Delta t-\tau)^2\sigma_j^2}{2})\approx \qty[\mathcal{E}_s\star\mathcal{E}_s](\tau)\exp(-\frac{\Delta t^2\sigma_j^2}{2})\,.
\end{equation}
Therefore, a spectrally narrow Gaussian jitter leads to a Gaussian envelope of the visibility on a long timescale and can be incorporated just by an additional Gaussian factor. The result still holds in the case of additional telegraph noise, which can here be incorporated by averaging $I_{\mathrm{abs}}\qty(\Omega,\omega_{\mathrm{ZPL}}^{(0)})$ over another probability distribution for $\omega_{\mathrm{ZPL}}^{(0)}$. As discussed in the main text the telegraph noise results in a beat of the visibility with a frequency corresponding to the frequency jump.
\subsubsection{Photoluminescence excitation}
Closely related, at least on a conceptional level, to the coherent control experiment is the photoluminescence excitation (PLE) experiment, in which the emitter is excited by a laser pulse and the integrated photoluminescence is detected. By tuning the carrier frequency of the laser pulse one can gain information on the absorption properties of the color center. The observable is again the emitter occupation after the pulse and we can interpret the single pulse as a double pulse with zero delay, such that for the PLE we need to calculate $f(\Delta t=0)$ for different settings of the carrier frequency of the laser. In lowest perturbation order, the PLE signal is therefore (see \eqref{eq:f2_from_spectra})
\begin{equation}\label{eq:PLE}
f^{(2)}(\Delta t=0; \omega_l)=\frac{1}{2\pi\hbar^2}\int_{\mathbb{R}}\dd{\Omega}\,I_{\mathrm{laser}}^{\mathrm{ideal}}(\Omega,\omega_l)I_{\mathrm{abs}}(\Omega)\equiv f^{(2)}(\omega_l)\,,
\end{equation}
where we explicitly write the dependence on the carrier frequency $\omega_l$. Since the ideal laser spectrum is a function of the difference $\omega_l-\Omega$, we see that the PLE is determined by the convolution of the absorption spectrum with the laser spectrum. From Eqs. (\ref{eq:f2_separation}) and (\ref{eq:I_0_abs_convolution}) we see, that
\begin{equation}
f^{(2)}(\omega_l)\sim \tilde{I}_0(\Delta t=0)\sim \qty|\qty{\qty[\mathcal{E}_s\star\mathcal{E}_s](\tau)*\qty[G_{-+}(\tau)e^{-i\omega_{\mathrm{ZPL}}\tau}e^{-\frac{\gamma_{\mathrm{pd}}}{2}|\tau|}]}(\Delta t=0)|\,.
\end{equation}
We thus need to perform essentially the same numerical calculations for the visibility dynamics and the PLE signal, namely: determine $I_0(\Delta t)$. The value at $\Delta t=0$ yields the PLE, whereas taking the absolute value and normalizing yields the visibility, as we see from \eqref{eq:visibility}.
\subsubsection{Laser interference}
To obtain direct experimental information on the autocorrelation of the laser pulse, one can send a double pulse directly onto the spectrometer. The time and frequency integrated spectrum of the double pulse is given by (see Eqs.~(\ref{eq:spectrum_laser}) and (\ref{eq:spectrum_double_pulse}))
\begin{align}
\int\limits_{-\infty}^{\infty}\dd{\omega}\,I_{\mathrm{laser}}(\omega)&=\int\limits_{-\infty}^{\infty}\dd{\omega}\,\int\limits_{-\infty}^{\infty}\dd \Omega\, R(\omega-\Omega)\qty|\int\limits_{-\infty}^{\infty}\dd t\,\expval{E^{(+)}(t)}e^{i\Omega t}|^2\notag\\
&\sim \int\limits_{-\infty}^{\infty}\dd \Omega\,\qty|\int\limits_{-\infty}^{\infty}\dd t\,\expval{E^{(+)}(t)}e^{i\Omega t}|^2\notag\\
&\sim \int\limits_{-\infty}^{\infty}\dd \Omega\,\qty[2+2\cos(\Omega \Delta t)]\underbrace{\qty|\int_{\mathbb{R}}\dd T\,e^{i(\Omega-\omega_l) T}\tilde{\mathcal{E}}(T)|^2}_{I_{\text{laser}}^{\mathrm{ideal}}(\Omega)}\,.
\end{align}
where we assumed a homogeneous spectral response $R(\omega,\Omega)=R(\omega-\Omega)$ in the relevant frequency interval and $\expval{E^{(+)}(t)}\sim\mathcal{E}(t)$. We see that this signal has the same form as $f^{(2)}(\Delta t)$ from \eqref{eq:f2_from_spectra}, but with $I_{\mathrm{abs}}(\Omega)\sim 1$. Therefore the time and frequency integrated double pulse spectrum also oscillates with the delay $\Delta t$, like $f^{(2)}(\Delta t)$. The oscillation frequency is however here directly given by $\omega_l$ and not determined by the overlap with the absorption spectrum of the emitter. We can retrace the steps leading to the visibility of the oscillations and find that it is here given by normalizing the expression
\begin{equation}
\qty|\int_{\mathbb{R}}\dd{\Omega}\,e^{i\Omega\Delta t}I_{\mathrm{laser}}^{\mathrm{ideal}}(\Omega)|\sim \qty|\qty[\mathcal{E}_s\star\mathcal{E}_s](\Delta t)|\,.\label{eq:laser_interference}
\end{equation}
Thus, in this way we can gain information on the absolute value of the autocorrelation of the single laser pulse used in the coherent control experiment. Apart from its spectrum, this gives additional valuable information on the laser pulse and aids in finding a sufficiently good theoretical fit.

\subsection{Correlation functions}\label{sec:correlation_functions}
In the calculation of the optical spectra of the color center and of the observable of the coherent control experiment we encountered two related correlation functions (see Eqs.~(\ref{eq:spectrum}) and (\ref{eq:f2_approx})), that are calculated here. We will see that the phonon spectral density is the central quantity to determine these correlation functions and thus also of central importance for the calculation of the considered optical signals.\\
First we note that the color center is assumed to be initially in its ground state $\ket{G}$ and the phonons to be in a thermal state $\rho_{\mathrm{phon}}$ at a given temperature $T$. Thus, the initial total density matrix is of the form
\begin{equation}
\rho_{\mathrm{tot}}(-\infty)=\ket{G}\bra{G}\otimes\rho_{\mathrm{phon}}\otimes\rho_{\mathrm{pd}}\otimes\rho_{\mathrm{anh}}=T_P^{}\rho_{\mathrm{tot}}(-\infty)T_P^{\dagger}\,,
\end{equation}
which is not affected by going into the polaron frame. The LO phonons are assumed to be in their vacuum state, i.e., $b_n\rho_{\mathrm{phon}}=\rho_{\mathrm{phon}}b_n^{\dagger}=0$ for $n\in\qty{\mathrm{LO}}$, since they have energies $>~150$~meV and we are interested in temperatures below room temperature. The two correlation functions that we encountered are
\begin{align}\label{eq:C_inital}
C(\tau_2,\tau_1)&=\Tr\qty[X^{\dagger}XU_0(\infty,\tau_1)X^{\dagger}U_0(\tau_1,-\infty)\rho_{\mathrm{tot}}(-\infty)U_0(-\infty,\tau_2)XU_0(\tau_2,\infty)]\,,\\
\tilde{C}(\tau_2,\tau_1)&=\Tr\qty[U_0(-\infty,\tau_2)X^{\dagger}U_0(\tau_2,\tau_1)XU_0(\tau_1,-\infty)X^{\dagger}B_-\rho_{\mathrm{tot}}(-\infty)XB_+]\,.\label{eq:C_tilde_inital}
\end{align}
Since it holds that $C(\tau_2,\tau_1)^*=C(\tau_1,\tau_2)$ (same for $\tilde{C}$) we only need to calculate them for $\tau_2\geq\tau_1$, while the other case can be inferred by complex conjugation. Both correlation functions can be evaluated by going first into the polaron frame and then employing the quantum regression theorem together with \eqref{eq:eq_of_motion}~\cite{breuer2002theory,groll2021controlling}, which we will do here explicitly.\\
First we bring $C$ in a form more similar to $\tilde{C}$ by noting that $H_0$ and thus $U_0$ does not alter the occupation of the emitter and commutes with $X^{\dagger}X$. This can be seen by observing that $H_{IB}$ commutes explicitly with $X^{\dagger}X$, $H_{\mathrm{pd}}$ leads to pure dephasing, which does not influence the occupation and $H_{\mathrm{anh}}$ acts only on LO phonon degrees of freedom. Thus with $X^{\dagger}XU_0X^{\dagger}=U_0X^{\dagger}$ and employing the cyclic property of the trace we can write
\begin{equation}
C(\tau_2,\tau_1)=\Tr\qty[U_0(-\infty,\tau_2)XU_0(\tau_2,\tau_1)X^{\dagger}U_0(\tau_1,-\infty)\rho_{\mathrm{tot}}(-\infty)]\,.
\end{equation}
Now we go to the polaron frame, which yields
\begin{align}
&C(\tau_2,\tau_1)\notag\\
&=\Tr\qty[U_0(-\infty,\tau_2)XU_0(\tau_2,\tau_1)X^{\dagger}U_0(\tau_1,-\infty)\rho_{\mathrm{tot}}(-\infty)]\notag\\
&=\Tr\qty{U_{0,P}(-\infty,\tau_2)XB_-U_{0,P}(\tau_2,\tau_1)X^{\dagger}B_+U_{0,P}(\tau_1,-\infty)\qty[\rho_P(-\infty)\otimes\rho_{\mathrm{pd}}\otimes\rho_{\mathrm{anh}}]}\,,\\
&\tilde{C}(\tau_2,\tau_1)\notag\\
&=\Tr\qty[U_0(-\infty,\tau_2)X^{\dagger}U_0(\tau_2,\tau_1)XU_0(\tau_1,-\infty)X^{\dagger}B_-\rho_{\mathrm{tot}}(-\infty)XB_+]\notag\\
&=\Tr\qty{U_{0,P}(-\infty,\tau_2)X^{\dagger}B_+U_{0,P}(\tau_2,\tau_1)XB_-U_{0,P}(\tau_1,-\infty)\qty[X^{\dagger}\rho_P(-\infty)X\otimes\rho_{\mathrm{pd}}\otimes\rho_{\mathrm{anh}}]}\,,
\end{align}
where $\rho_P(-\infty)=\ket{G}\bra{G}\otimes\rho_{\mathrm{phon}}$ is the initial reduced density matrix of the TLS-phonon system in the polaron frame. Now we will focus on the case $\tau_2\geq\tau_1$ and perform similar steps for both correlation functions. First we propagate $\rho_P$ and $X^{\dagger}\rho_P X$ from $-\infty$ to $\tau_1$ with \eqref{eq:eq_of_motion}. This propagation leaves the phonon density matrix unaffected since the LA phonons are in a thermal state and the LO phonons in their vacuum state. Thus, $\mathcal{L}_{\mathrm{phon}}[\rho_{\mathrm{phon}}]=0$. Likewise, $\mathcal{L}_{\mathrm{pol}}[\ket{G}\bra{G}]=\mathcal{L}_{\mathrm{pol}}[\ket{X}\bra{X}]=0$, such that the reduced density matrices $\rho_P(-\infty)$ and $X^{\dagger}\rho_P(-\infty) X$ remain unchanged.\\
Then we need to propagate $X^{\dagger}B_+\rho_P(-\infty)=X^{\dagger}B_+\rho_{\mathrm{phon}}$ from $\tau_1$ to $\tau_2$ in the case of $C$ and $XB_-\rho_{\mathrm{phon}}$ in the case of $\tilde{C}$. Observing that
\begin{equation}
\mathcal{L}_{\mathrm{pol}}[X]=\qty(+ i\omega_{\mathrm{ZPL}}-\frac{\gamma_{\mathrm{pd}}}{2})X\,,\quad \mathcal{L}_{\mathrm{pol}}[X^{\dagger}]=\qty(- i\omega_{\mathrm{ZPL}}-\frac{\gamma_{\mathrm{pd}}}{2})X^{\dagger}\,,
\end{equation}
the polaron part of the propagation just yields a factor $\exp\qty[\qty(\pm i\omega_{\mathrm{ZPL}}-\frac{\gamma_{\mathrm{pd}}}{2})(\tau_2-\tau_1)]$. The phonon part is more involved, but can be rewritten by remembering that the LO phonons are in their vacuum state, i.e., $\rho_{\mathrm{phon}}b_n^{\dagger}=0$ for $n\in\qty{\mathrm{LO}}$. Thus, 
\begin{equation}
\dv{t}B_{\pm}\rho_{\mathrm{phon}}=\mathcal{L}_{\mathrm{phon}}\qty[B_{\pm}\rho_{\mathrm{phon}}]=-i\comm{\sum_n\omega_nb_n^{\dagger}b_n-\frac{i}{2}\sum_{n\in \qty{\mathrm{LO}}}\gamma_{\mathrm{LO}}^{(n)}b_n^{\dagger}b_n\ }{\ B_{\pm}\rho_{\mathrm{phon}}}\,.
\end{equation}
This can be solved by introducing an effective non-unitary time evolution operator
\begin{equation}\label{eq:U_eff}
\eval{U_{\mathrm{eff}}(\tau_2,\tau_1)}_{\tau_2\geq\tau_1}=\exp\qty[-i\qty(\sum_n\omega_nb_n^{\dagger}b_n-i\sum_{n\in\qty{\mathrm{LO}}}\frac{\gamma_{\mathrm{LO}}^{(n)}}{2}b_n^{\dagger}b_n)(\tau_2-\tau_1)]\,.
\end{equation}
Thus the solution of \eqref{eq:eq_of_motion} at time $\tau_2$, when the initial reduced density matrix at $\tau_1$ is given by $XB_-\rho_{\mathrm{phon}}$, is
\begin{equation}
e^{\qty(\mathcal{L}_{\mathrm{pol}}+\mathcal{L}_{\mathrm{phon}})(\tau_2-\tau_1)}\qty[XB_-\rho_{\mathrm{phon}}]=e^{i\qty(\omega_{\mathrm{ZPL}}+i\frac{\gamma_{\mathrm{pd}}}{2})(\tau_2-\tau_1)}XU_{\mathrm{eff}}(\tau_2,\tau_1)B_-\rho_{\mathrm{phon}}U_{\mathrm{eff}}^{(-1)}(\tau_2,\tau_1)\,,
\end{equation}
while for the other case of $X^{\dagger}B_+\rho_{\mathrm{phon}}$ at $\tau_1$ it is given by
\begin{equation}
e^{\qty(\mathcal{L}_{\mathrm{pol}}+\mathcal{L}_{\mathrm{phon}})(\tau_2-\tau_1)}\qty[X^{\dagger}B_+\rho_{\mathrm{phon}}]=e^{-i\qty(\omega_{\mathrm{ZPL}}-i\frac{\gamma_{\mathrm{pd}}}{2})(\tau_2-\tau_1)}X^{\dagger}U_{\mathrm{eff}}(\tau_2,\tau_1)B_+\rho_{\mathrm{phon}}U_{\mathrm{eff}}^{(-1)}(\tau_2,\tau_1)\,.
\end{equation}
With this, the correlation functions take the form
\begin{align}\label{eq:C_sol1}
\eval{C(\tau_2,\tau_1)}_{\tau_2\geq\tau_1}&=	e^{-i\qty(\omega_{\mathrm{ZPL}}-i\frac{\gamma_{\mathrm{pd}}}{2})(\tau_2-\tau_1)}G_{-+}(\tau_2-\tau_1)\,,\\
\eval{\tilde{C}(\tau_2,\tau_1)}_{\tau_2\geq\tau_1}&=e^{i\qty(\omega_{\mathrm{ZPL}}+i\frac{\gamma_{\mathrm{pd}}}{2})(\tau_2-\tau_1)}G_{+-}(\tau_2-\tau_1)\,.\label{eq:C_tilde_sol1}
\end{align}
Here, $G_{\mp,\pm}(\tau)$ are phonon correlation functions, given by
\begin{equation}
\eval{G_{\mp,\pm}(\tau)}_{\tau\geq 0}=\Tr\qty[B_{\mp}U_{\mathrm{eff}}(\tau,0)B_{\pm}\rho_{\mathrm{phon}}U_{\mathrm{eff}}^{(-1)}(\tau,0)]\,.
\end{equation}
The properties $C(\tau_2,\tau_1)^*=C(\tau_1,\tau_2)$ and $\tilde{C}(\tau_2,\tau_1)^*=\tilde{C}(\tau_1,\tau_2)$ lead to $G_{\mp,\pm}(\tau)^*=G_{\mp,\pm}(-\tau)$, which can be seen in \eqref{eq:C_sol1} for the special case of $\gamma_{\mathrm{pd}}=0$ and will be taken as the definition of $G_{\mp,\pm}(\tau)$ in the case of $\tau<0$.
\subsubsection{Phonon correlation function $G_{-+}(\tau)$}
Here we will take a look at the phonon correlation function 
\begin{equation}\label{eq:G_mp_def}
\eval{G_{-+}(\tau_2-\tau_1)}_{\tau_2\geq\tau_1}=\Tr\qty[B_-U_{\mathrm{eff}}(\tau_2,\tau_1)B_+\rho_{\mathrm{phon}}U_{\mathrm{eff}}^{(-1)}(\tau_2,\tau_1)]
\end{equation}
with $U_{\mathrm{eff}}$ from \eqref{eq:U_eff}. We can split the multimode displacement operators $B_{\pm}$ into a product of LO and LA  multimode displacement operators
\begin{equation}
B_{\pm}=B_{\pm,\mathrm{LO}}B_{\pm,\mathrm{LA}}\,,\quad B_{\pm,\mathrm{LO}}=\prod_{n\in\qty{\mathrm{LO}}}D_n\qty(\pm\frac{g_n}{\omega_n})\,,\quad B_{\pm,\mathrm{LA}}=\prod_{n\notin\qty{\mathrm{LO}}}D_n\qty(\pm\frac{g_n}{\omega_n})\,.
\end{equation}
Since $\rho_{\mathrm{phon}}=\rho_{\mathrm{LO}}\otimes \rho_{\mathrm{LA}}$ is a thermal state, which is a product state of LO and LA states, the correlation function factorizes in an LO and an LA part. The LA part is given by the usual phonon correlation function~\cite{mahan2013many,wigger2019phonon}, whereas the LO part is modified by their decay. In total we have
\begin{align}
\eval{G_{-+}(\tau_2-\tau_1)}_{\tau_2\geq\tau_1}&=\eval{G_{-+,\mathrm{LA}}(\tau_2-\tau_1)}_{\tau_2\geq\tau_1}\eval{G_{-+,\mathrm{LO}}(\tau_2-\tau_1)}_{\tau_2\geq\tau_1}\,,\\
\eval{G_{-+,\mathrm{LA}}(\tau_2-\tau_1)}_{\tau_2\geq\tau_1}&=\Tr\left\{\exp[i\sum_{n\notin \qty{\mathrm{LO}}}\omega_nb_n^{\dagger}b_n(\tau_2-\tau_1)]B_{-,\mathrm{LA}} \right. \notag\\
&\left. \qquad\times \exp[-i\sum_{n\notin \qty{\mathrm{LO}}}\omega_nb_n^{\dagger}b_n(\tau_2-\tau_1)]B_{+,\mathrm{LA}}\rho_{\mathrm{LA}} \right\}\,,\\
\eval{G_{-+,\mathrm{LO}}(\tau_2-\tau_1)}_{\tau_2\geq\tau_1}&=\Tr\qty[U_{\mathrm{eff, LO}}(\tau_2,\tau_1)^{-1}B_{-,\mathrm{LO}}U_{\mathrm{eff, LO}}(\tau_2,\tau_1)B_{+,\mathrm{LO}}\rho_{\mathrm{LO}}]\,,\label{eq:G_LO}\\
\eval{U_{\mathrm{eff, LO}}(\tau_2,\tau_1)}_{\tau_2\geq\tau_1}&=\exp\qty{-i\qty[\sum_{n\in\qty{\mathrm{LO}}}\omega_nb_n^{\dagger}b_n-\frac{i}{2}\sum_{n\in \qty{\mathrm{LO}}}\gamma_{\mathrm{LO}}^{(n)}b_n^{\dagger}b_n](\tau_2-\tau_1)}\,.
\end{align}
\subsubsection{LA correlation function}
The LA correlation function can also be written as 
\begin{equation}
G_{-+,\mathrm{LA}}(\tau_2-\tau_1)=\prod_{n\notin\qty{\mathrm{LO}}}\expval{D_n\qty(-\frac{g_n}{\omega_n}e^{i\omega_n(\tau_2-\tau_1)})D_n\qty(\frac{g_n}{\omega_n})}\,,
\end{equation}
where the expectation value is with respect to the thermal state of the $n$-th mode. The well known result, also used in~\cite{wigger2019phonon} is, irrespective whether $\tau_2\geq\tau_1$ or not, given by
\begin{align}
G_{-+,\mathrm{LA}}(\tau_2-\tau_1)&=\exp[\phi_{\mathrm{LA}}(\tau_2-\tau_1)]B_{\mathrm{LA}}^2\label{eq:C_LE}\,,\\
\phi_{\mathrm{LA}}(\tau_2-\tau_1)&=\int\limits_{0}^{\infty}\dd\omega\,\frac{J_{\mathrm{LA}}(\omega)}{\omega^2}\qty{\qty[n(\omega)+1]e^{-i\omega(\tau_2-\tau_1)}+n(\omega)e^{i\omega(\tau_2-\tau_1)}}\,,\\
J_{\mathrm{LA}}(\omega)&=\sum_{n\notin \qty{\mathrm{LO}}}|g_n|^2\delta(\omega_n-\omega)\,,\\
B_{\mathrm{LA}}&=\exp[-\frac{1}{2}\phi_{\mathrm{LA}}(0)]=\Tr\qty[B_{\pm,\mathrm{LA}}\rho_{\mathrm{LA}}]\,,
\end{align}
where we introduced the phonon spectral density $J_{\mathrm{LA}}$ of the LA modes and the thermal occupation of a phonon mode with frequency $\omega$
\begin{equation}
n(\omega)=\frac{1}{e^{\beta\hbar\omega}-1}\,,\quad \beta=(k_BT)^{-1}\,.
\end{equation}
\subsubsection{LO correlation function}
The LO correlation function from \eqref{eq:G_LO} can be evaluated by splitting it into a product of correlation functions for each mode
\begin{equation}
G_{-+,\mathrm{LO}}(\tau_2-\tau_1)=\prod\limits_{n\in\qty{\mathrm{LO}}}G_{n,\mathrm{LO}}(\tau_2-\tau_1)\,.
\end{equation}
For the $n$-th LO mode with frequency $\omega_n$, coupling constant $g_n$, and decay rate $\gamma_n=\gamma_{\mathrm{LO}}^{(n)}$, the correlation function takes the form
\begin{align}
G_{n,\mathrm{LO}}(\tau_2-\tau_1)&=\bra{0}\exp[ib_n^{\dagger}b_n\qty(\omega_n-\frac{i\gamma_n}{2})(\tau_2-\tau_1)]D_n\qty(-\frac{g_n}{\omega_n})\notag\\
&\qquad \times\exp[-ib_n^{\dagger}b_n\qty(\omega_n-\frac{i\gamma_n}{2})(\tau_2-\tau_1)]D_n\qty(\frac{g_n}{\omega_n})\ket{0}\notag\\
&=\bra{{\frac{g_n}{\omega_n}}}\exp[-ib_n^{\dagger}b_n\qty(\omega_n-\frac{i\gamma_n}{2})(\tau_2-\tau_1)]\ket{{\frac{g_n}{\omega_n}}}\notag\\
&=e^{-\frac{|g_n|^2}{\omega_n^2}}\sum_{k,l}\frac{(g_n^*/\omega_n)^k}{\sqrt{k!}}\bra{k}\exp[-ib_n^{\dagger}b_n\qty(\omega_n-\frac{i\gamma_n}{2})(\tau_2-\tau_1)]\ket{l}\frac{(g_n/\omega_n)^l}{\sqrt{l!}}\notag\\
&=e^{-\frac{|g_n|^2}{\omega_n^2}}\sum_{k}\frac{(|g_n|^2/\omega_n^2)^k}{k!}\exp[-ik\qty(\omega_n-\frac{i\gamma_n}{2})(\tau_2-\tau_1)]\notag\\
&=\exp\qty{-\frac{|g_n|^2}{\omega_n^2}\qty[1-e^{-i\qty(\omega_n-i\frac{\gamma_n}{2})(\tau_2-\tau_1)}]}\,.
\end{align}
Here, we have used the Fock basis representation of coherent states and that the LO modes are initially in a vacuum state. The complete LO correlation function is now a product of the separate correlation functions for each LO mode, yielding
\begin{align}\label{eq:C_LO}
\eval{G_{-+,\mathrm{LO}}(\tau_2-\tau_1)}_{\tau_2\geq\tau_1}&=\exp[\phi_{\mathrm{LO}}(\tau_2-\tau_1)]B_{\mathrm{LO}}^2\,,\\
\eval{\phi_{\mathrm{LO}}(\tau_2-\tau_1)}_{\tau_2\geq\tau_1}&=\int\limits_{0}^{\infty}\dd\omega\,\frac{J_{\mathrm{LO}}(\omega)}{\omega^2}\exp\qty{-i\qty[\omega-i\frac{\gamma(\omega)}{2}](\tau_2-\tau_1)}\,,\\
J_{\mathrm{LO}}(\omega)&=\sum_{n\in \qty{\mathrm{LO}}}|g_n|^2\delta(\omega_n-\omega)\,,\label{eq:LO_specdens}\\
B_{\mathrm{LO}}&=\exp[-\frac{1}{2}\phi_{\mathrm{LO}}(0)]=\Tr\qty[B_{\pm,\mathrm{LO}}\rho_{\mathrm{LO}}]\,,\\
\gamma(\omega_n)&=\gamma_n=\gamma_{\mathrm{LO}}^{(n)}\,.
\end{align}
From the property $C(\tau_1,\tau_2)^*=C(\tau_2,\tau_1)$, which also holds for the LO and LA correlation functions separately, we can extend the result to the case $\tau_2<\tau_1$ via
\begin{align}
G_{-+,\mathrm{LO}}(\tau_2,\tau_1)&=\exp[\phi_{\mathrm{LO}}(\tau_2-\tau_1)]B_{\mathrm{LO}}^2\notag\,,\\
\phi_{\mathrm{LO}}(\tau_2-\tau_1)&=\int\limits_{0}^{\infty}\dd\omega\,\frac{J_{\mathrm{LO}}(\omega)}{\omega^2}\exp\qty[-i\omega(\tau_2-\tau_1)-\frac{\gamma(\omega)}{2}|\tau_2-\tau_1|]\,.
\end{align}
We see that the decay of the LO modes leads to an exponential damping in the LO correlation function. We can also model the same correlation function by using a modified Lorentzian spectral density for each LO phonon mode, as long as $\gamma(\omega_n)\ll\omega_n$ (which is always the case)
\begin{align}
\phi_{\mathrm{LO}}(\tau_2-\tau_1)&\approx\int\limits_{0}^{\infty}\dd\omega\,\frac{\tilde{J}_{\mathrm{LO}}(\omega)}{\omega^2}\exp\qty[-i\omega(\tau_2-\tau_1)]\notag\,,\\
\tilde{J}_{\mathrm{LO}}(\omega)&=\sum_{n\in\qty{\mathrm{LO}}} \frac{|g_n|^2\omega^2}{\omega_n^2}\sqrt{\frac{2}{\pi}}\frac{\gamma(\omega_n)}{\gamma(\omega_n)^2+(\omega-\omega_n)^2}\,.
\end{align}
This can be seen by substituting $\omega'=\omega-\omega_n$ and shifting the lower limit of the integration to $-\infty$. The exponential decay term then emerges as the Fourier transform of the Lorentzian factor in the spectral density. Thus, on this level, there is no difference between a decaying dispersion-less mode and a mode with a broadened spectral density, e.g., due to a non-vanishing dispersion.
\subsubsection{Full phonon correlation function}\label{sec:phonon_corr}
The full phonon correlation function is a product of the LO correlation function and the LA correlation function. Note that for the thermal expectation value of the multi-mode displacement operators we have $B=\Tr\qty[B_{\pm}\rho_{\mathrm{phon}}]=B_{\mathrm{LA}}B_{\mathrm{LO}}$, such that
\begin{equation}\label{eq:G_mp}
G_{-+}(\tau)=B^2\exp[\phi(\tau)]\,,
\end{equation}
with 
\begin{align}\label{eq:phi}
\phi(\tau)&=\underbrace{\int\limits_{0}^{\infty}\dd\omega\,\frac{J_{\mathrm{LA}}(\omega)}{\omega^2}\qty{\qty[n(\omega)+1]e^{-i\omega\tau}+n(\omega)e^{i\omega\tau}}}_{\phi_{\mathrm{LA}}(\tau)}\notag\\
&+\underbrace{\int\limits_{0}^{\infty}\dd\omega\,\frac{J_{\mathrm{LO}}(\omega)}{\omega^2}\exp\qty[-i\omega\tau-\frac{\gamma(\omega)}{2}|\tau|]}_{\phi_{\mathrm{LO}}(\tau)}\,.
\end{align}
Since we will assume dispersionless LO modes, the LO spectral density $J_{\mathrm{LO}}(\omega)$ will be a sum of $\delta$-functions. The LA spectral density $J_{\mathrm{LA}}(\omega)$ will be discussed in Sec.~\ref{sec:la_specdens}.
\subsubsection{Phonon correlation function $G_{+-}(\tau)$}
The phonon correlation function $G_{+-}(\tau)$, which appears in the correlation function $\tilde{C}$ in \eqref{eq:C_tilde_sol1}, is given by
\begin{equation}\label{eq:G_pm_def}
G_{+-}(\tau)=\Tr\qty[B_+U_{\mathrm{eff}}(\tau,0)B_-\rho_{\mathrm{phon}}U_{\mathrm{eff}}^{(-1)}(\tau,0)]\,.
\end{equation}
This correlation function has the same structure as $G_{-+}(\tau)$ in \eqref{eq:G_mp_def}, however the operators $B_{\pm}$ are interchanged. If we change the sign of all coupling constants $g_n$ in the independent boson Hamiltonian $H_{\mathrm{IB}}$ in \eqref{eq:H_IB}, we see from \eqref{eq:B_pm_def}, that this leads to an interchange of $B_+$ with $B_-$. Since changing the sign of the couplings $g_n$ does not change the form of $G_{-+}(\tau)$, which depends only on the spectral density and thus on $|g_n|$, we conclude that $G_{-+}(\tau)=G_{+-}(\tau)$.
\subsubsection{Full correlation functions - final result}
The correlation functions $C(\tau_2,\tau_1)$ and $\tilde{C}(\tau_2,\tau_1)$ are now given by
\begin{align}\label{eq:C_final}
C(\tau_2,\tau_1)&=C(\tau_2-\tau_1)=e^{-i\omega_{\mathrm{ZPL}}(\tau_2-\tau_1)-\frac{\gamma_{\mathrm{pd}}}{2}|\tau_2-\tau_1|}G_{-+}(\tau_2-\tau_1)\\
\tilde{C}(\tau_2,\tau_1)&=\tilde{C}(\tau_2-\tau_1)=e^{i\omega_{\mathrm{ZPL}}(\tau_2-\tau_1)-\frac{\gamma_{\mathrm{pd}}}{2}|\tau_2-\tau_1|}G_{-+}(\tau_2-\tau_1)\,,\label{eq:C_tilde_final}
\end{align} 
where $G_{-+}$ is given in \eqref{eq:G_mp} together with \eqref{eq:phi} . We thus see that both correlation functions are completely determined by the phonon spectral density, the decay rates of the LO phonons, the polaron shifted transition frequency, and the pure dephasing rate of the exciton.
\subsection{Modeling the LA phonon spectral density}\label{sec:la_specdens}
So far we have not considered a specific model for the LA spectral density. The derivation of the correlation function in Sec.~\ref{sec:correlation_functions} does not require any structure of the coupling constants $g_n$ and we have seen that the phonon quantity of interest for the coherent control experiment and the PL spectra is the spectral density $J_\mathrm{LA}(\omega)$, which we can model to fit the experiment best. Therefore, to keep the model as simple as possible, we employ a phenomenological spectral density of the form 
\begin{equation}\label{eq:LA_specdens}
J_{\mathrm{LA}}(\omega)=\alpha\omega^3\frac{\omega_c^2}{\omega^2+\omega_c^2}\Theta(\Omega_c-\omega)\Theta(\omega)\,,
\end{equation}
which is a standard super-ohmic spectral density with a Lorentz-Drude cutoff~\cite{breuer2002theory} and a sharp high-energy cutoff at a frequency $\Omega_c$ describing the maximum acoustic phonon energy. By this cutoff also the polaron shift in \eqref{eq:omega_ZPL} is well defined. The frequency $\omega_c$ essentially determines the width of the LA PSB in the PL spectra and $\alpha$ determines its height.

\section{Impact of the system parameters}
In this section we discuss the impact of the fitted system parameters in the theoretical results. Some of the intricacies of the fitting process, especially the influence of the spectrometer resolution and the spectral shape of the laser pulses, are highlighted.
\subsection{Spectrometer response}
To determine the spectral response $R(\omega,\Omega)$ of the spectrometer, we use a spectrally narrow light source (light from a green laser diode), as discussed after \eqref{eq:spectrum_laser}. The spectrum of this laser diode is proportional to $R(\omega,\Omega)$ with $\Omega$ being the optical carrier frequency, which is given by the maximum of the laser spectrum. In the experiment, two different diffraction gratings are used, as discussed in Sec.~\ref{setup}. We refer to the spectrometer when using the 300 lines/mm grating as ``300-grating" and when using the 1200 lines/mm grating as ``1200-grating".

\begin{figure}[h]
	\centering
	\includegraphics[width=\linewidth]{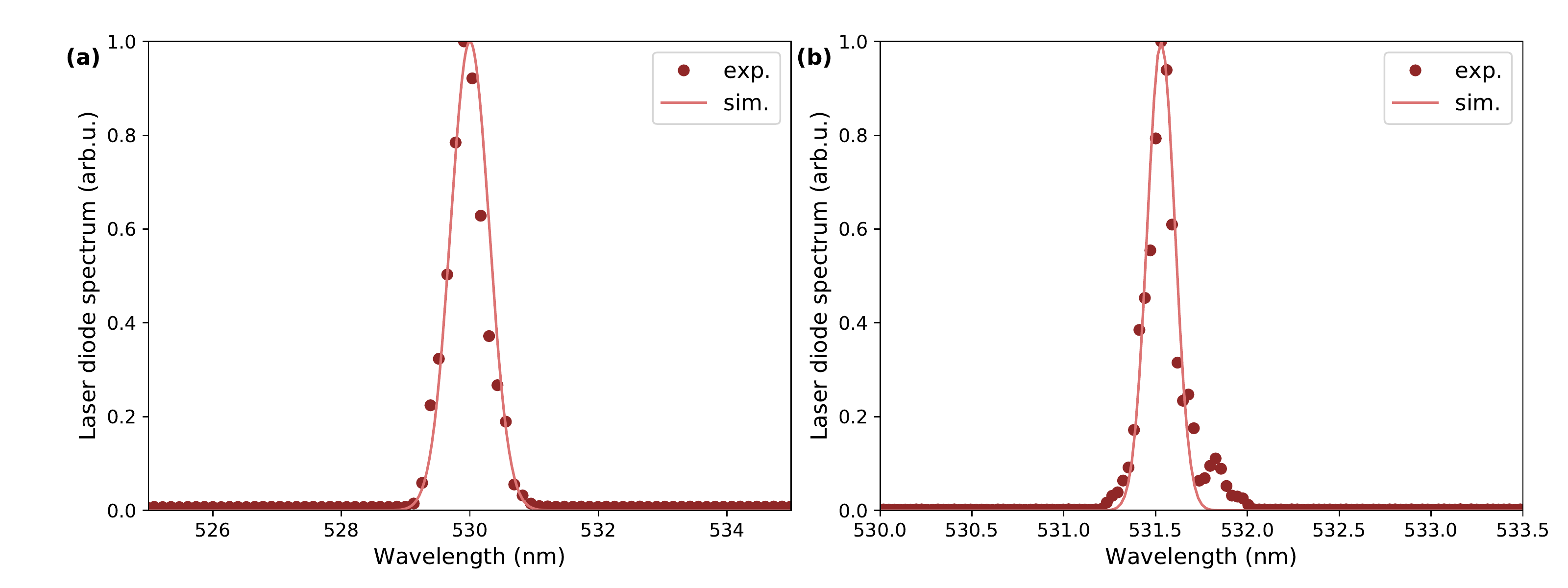}
	\caption{Measured spectra (dots) of a green laser diode using (a) the 300-grating and (b) the 1200-grating. The solid lines represent a Gaussian with standard deviation $\Delta\lambda=0.3$~nm (a) and $\Delta\lambda=0.075$~nm (b).}
	\label{fig:response}
\end{figure}

The spectra, depending on the wavelength of the detected light, are shown in Fig.~\ref{fig:response} as dots. Figure~\ref{fig:response}(a) shows the case of the 300-grating and Fig.~\ref{fig:response}(b) the case of the 1200-grating. The spectra are approximated by Gaussians (lines), with a standard deviation of $\Delta\lambda_{300}\approx0.3$~nm in the case of the 300-grating and $\Delta\lambda_{1200}\approx0.075$~nm in the case of the 1200-grating.\\
The spectrometer response in the spectral domain is thus given by
\begin{equation}
R(\omega,\Omega)=\exp\qty{-\frac{\qty[\lambda(\omega)-\Lambda(\Omega)]^2}{2(\Delta\lambda)^2}}
\end{equation}
with $\lambda(\omega)=2\pi c/\omega$ and analog for $\Lambda(\Omega)$. Since usually $\lambda\gg\Delta \lambda$, we can expand the exponent around $\lambda\approx \Lambda$, or equivalently $\omega\approx \Omega$ into a Taylor series yielding
\begin{equation}
\lambda(\omega)-\Lambda(\Omega)\approx -\lambda'(\omega)(\Omega-\omega)=\frac{\lambda(\omega)}{\omega}(\Omega-\omega)\,.
\end{equation}
Thus the spectrometer response can be approximated by the Gaussian
\begin{equation}
R(\omega,\Omega)\approx \exp\qty[-\frac{\qty(\omega-\Omega)^2}{2(\Delta\omega)^2}]\,,\quad \Delta\omega=\Delta\lambda \frac{\omega}{\lambda(\omega)}\,.\label{eq:R_fit}
\end{equation}
If we assume a constant resolution $\Delta \lambda$ in wavelength space, the resolution $\Delta \omega$ in the frequency domain depends on the detected frequency $\omega$.

\subsection{Modeling the laser pulses}
To find an accurate theoretical description for the laser pulse, we (i) employ \eqref{eq:spectrum_laser} for the detected laser spectrum using the spectrometer response from \eqref{eq:R_fit} (keeping in  mind that $\expval{E^{(+)}}(t)\sim\mathcal{E}(t)$ for linearly polarized light) and (ii) use \eqref{eq:laser_interference} for the laser interference, detected in the coherent control setup. To retrieve the most suitable model for the laser pulses, we simultaneously use both, the spectral (i) and the temporal domain (ii) to get a reasonable fit for the laser pulse. We also found that the PL visibility dynamics in the coherent control experiment critically depend on the spectral shape of the laser pulses, when exciting close to the ZPL on the LA sideband (see Sec.~\ref{sec:vis_LA} and the main text). This is discussed in more detail in Sec.~\ref{sec:LA_vis}. We achieve the best overall results by using an asymmetric Pearson-4 distribution~\cite{stuart1994kendall} for the Fourier transform $\mathcal{E}_s(\omega)$ of the laser field of a single pulse $\mathcal{E}_s(t)=\tilde{\mathcal{E}}(t)e^{-i\omega_l t}$ with
\begin{equation}\label{eq:pearson}
\mathcal{E}_s(\omega)=\int_{\mathbb{R}}\dd t\, e^{i\omega t}\mathcal{E}_s(t)\sim \qty[1+\qty(\frac{\omega-\omega_0}{\alpha_l})^2]^{-m_l}\exp[-\nu_l \arctan(\frac{\omega-\omega_0}{\alpha_l})]\,.
\end{equation}
This distribution converges to a Gaussian with standard deviation $1/\sigma_l$ for $\alpha_l=\sqrt{2m_l}/\sigma_l$ and $m_l\rightarrow\infty$. Another interesting limiting case is $m_l=1$ and $\nu_l=0$, where $\mathcal{E}_s(\omega)$ is a Lorentzian.  For $m_l$ close to unity, the parameter $\alpha_l$ is a measure for the width of the distribution. The parameter $\nu_l$ is a measure for the asymmetry of the spectrum. For $\nu_l<0$ the spectrum has a longer tail to the right, i.e., a positive skewness. The location parameter $\omega_0$ is a measure for the position of the maximum. The true maximum however is given by $\omega_0-\alpha_l\nu_l/(2m_l)$.\\
\begin{figure}[h]
	\centering
	\includegraphics[width=\linewidth]{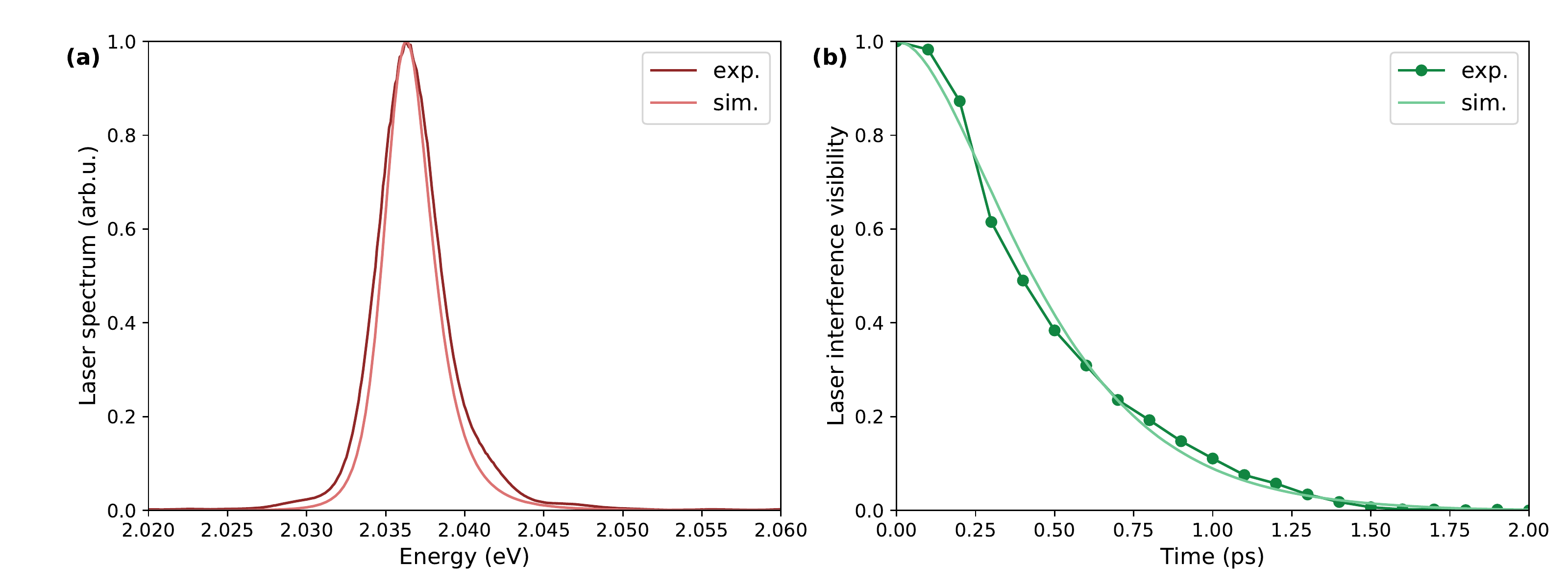}
	\caption{Characterization of the laser pulse. (a) Measurement (dark) and simulation (bright) of the laser spectrum. The spectrum is modeled via \eqref{eq:pearson} with $\alpha_l=\sqrt{2m_l}/\sigma_l$, $\omega_0=\omega_l+\alpha_l\nu_l/(2m_l)$, $\sigma_l=350$~fs, $m_l=1.2$ and $\nu_l=-0.5\sqrt{m_l}$. The spectrum is measured with the 1200-grating, such that the resolution for the spectrometer response in \eqref{eq:R_fit} is $\Delta\lambda_{1200}=0.075$nm. (b) Measurement (dark dots) and simulation (bright line) of the laser interference visibility, i.e., the pulse autocorrelation.}
	\label{fig:laser}
\end{figure}
We define the carrier frequency of the laser $\omega_l$ as the maximum of the distribution, as this frequency is easy to identify in the experiment. For the fit of the spectrum and the laser autocorrelation, we use \eqref{eq:pearson} together with $\alpha_l=\sqrt{2m_l}/\sigma_l$ and $\omega_0=\omega_l+\alpha_l\nu_l/(2m_l)$. A fit, leading to an overall good agreement with the experiment is achieved for the parameters $\sigma_l=350$~fs, $m_l=1.2$ and $\nu_l=-0.5\sqrt{m_l}$. This choice of parameters leads to a value of $\alpha_l\approx (230~\text{fs})^{-1}$, which we refer to as a pulse of 230~fs duration in the main text. In Fig.~\ref{fig:laser}(a) and (b) the simulated (bright) and detected (dark) spectrum and autocorrelation of the laser pulse are shown, respectively. For the detection of the spectrum, the 1200-grating is used and in the simulation we use the respective response in \eqref{eq:R_fit} with $\Delta\lambda=\Delta\lambda_{1200}=0.075$~nm.

\subsection{Fitting PL spectra of the emitter}
\begin{figure}[h]
	\centering
	\includegraphics[width=0.5\linewidth]{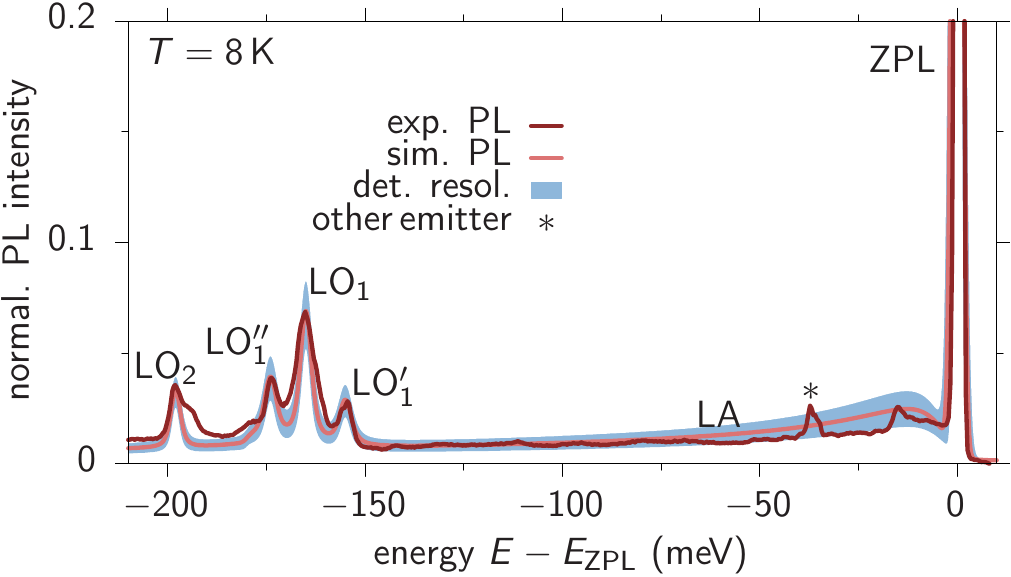}
	\caption{Light emission at $T=8$~K. Measured (dark red) and simulated (bright red) PL intensity detected with the 300-grating. The blue band shows simulations of the PL for a spectrometer resolution between $\Delta\lambda=0.2$~nm and $0.4$~nm.}
	\label{fig:spectrumsi8k}
\end{figure}

By fitting the detected PL spectra of the emitter for temperatures of 8~K and 80~K we can determine the spectral density of the acoustic phonons and the coupling strengths and lifetimes of the dispersionless optical modes. These parameters have to be chosen such that also the PL visibility dynamics in the coherent control experiment are consistently reproduced.\\
The PL at 80~K is shown in the main text in Fig.~1. In addition we show the measured (dark red) and simulated (bright red) PL at $T=8$~K in Fig.~\ref{fig:spectrumsi8k}. In both cases, a constant background was subtracted from the experimental data. The PL spectra for both temperatures are well reproduced by the theory using the same set of parameters for the phonons, demonstrating consistency of the approach. The spectra are detected with the 300-grating, such that we model it via \eqref{eq:PL} together with \eqref{eq:R_fit} and $\Delta\lambda=\Delta\lambda_{300}=0.3$~nm. All phonon parameters are summarized in Tab.~\ref{table:LA} and Tab.~\ref{table:LO}. Note, that the spectral resolution of the spectrometer for a detection at $\hbar\Omega\approx 2$~eV is $\hbar\Delta\omega=\Delta\lambda\hbar\Omega^2/(2\pi c)\approx1$~meV. Therefore, in the PL spectra we can disregard the Gaussian spectral jitter on an energy scale of $\sim 10$~\textmu eV, which is discussed in the main text and in Sec.~\ref{sec:jitter_exp}.\\
To illustrate the effect of the spectrometer resolution on the PL, in Fig.~\ref{fig:spectrumsi8k} we additionally show simulations of the PL spectrum for resolutions between $\Delta\lambda=0.2$~nm and $\Delta\lambda=0.4$~nm as a blue band. For a spectrometer with worse resolution (0.4~nm, upper edge of the blue band), especially the LA PSB gains intensity relative to the ZPL compared to the case of better resolution (0.2~nm, lower edge of blue band). This can be understood intuitively by considering the limit $\Delta\lambda\rightarrow\infty$, i.e., a spectrometer that cannot destinguish between different wavelengths. In that case the spectrum should be constant, i.e., the PSBs and the ZPL would have the same height and cannot be distinguished. In the other extreme case of $\Delta\lambda\rightarrow 0$, i.e., an ideal spectral resolution, we would be able to definitely distinguish between PSB and ZPL.\\
The height of the LA sideband is mainly determined by the parameter $\alpha$ in \eqref{eq:LA_specdens}, which is directly connected to the deformation potentials when considering a more microscopic approach to the modeling of the LA spectral density~\cite{wigger2019phonon}. This implies that one cannot accurately determine deformation potential constants from PL spectra, if the spectrometer resolution is not well known.

\newpage
\subsection{Gaussian spectral jitter}
\begin{figure}[h]
	\centering
	\includegraphics[width=0.55\linewidth]{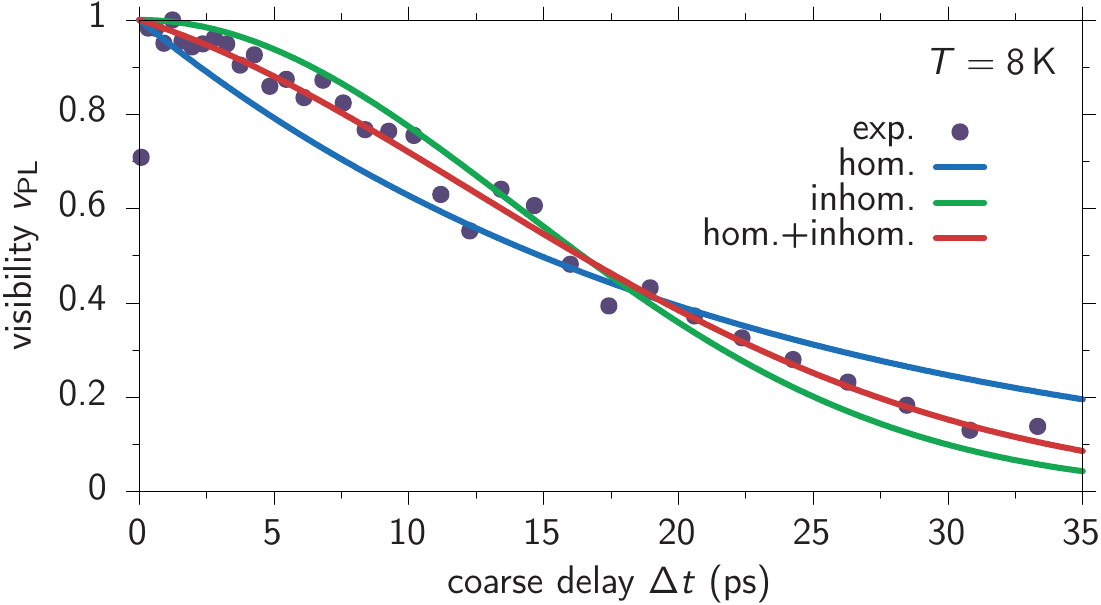}
	\caption{Coherent control visibility for resonant ZPL excitation at $T=8$~K. The data (dots) are the same as in Fig. 2 in the main text. The lines are fits to the data using \eqref{eq:fit}. The blue line is the best fit for the case $\sigma_\mathrm{j}=0$, i.e., no Gaussian jitter. The green line is the best fit for the case $\gamma_{\mathrm{pd}}=0$, i.e., no pure dephasing. The red line includes both pure dephasing and Gaussian jitter.}
	\label{fig:zplsi}
\end{figure}

In the context of Fig.~2 in the main text, the impact of Gaussian spectral jitter is discussed. In Fig.~\ref{fig:zplsi} we show the same data (dots), together with three fits (lines). All fits are simplified versions of the formula presented in \eqref{eq:Gauss_effect}. Since the excitation is in resonance with the ZPL, we can as a first approximation neglect the influence of phonons and since the dephasing occurs on a much longer timescale than the laser autocorrelation, we can approximate \eqref{eq:Gauss_effect} by
\begin{equation}\label{eq:fit}
|I_0(\Delta t)|\approx \exp\left[-\frac{(\Delta t)^2\sigma_\mathrm{j}^2}{2}-\frac{\gamma_{\mathrm{pd}}}{2}|\Delta t|\right]\,.
\end{equation}
The simulation in the main text of course fully includes the effect of phonons and of the finite width of the laser pulse. The blue line in Fig.~\ref{fig:zplsi} now represents the best fit, when $\sigma_\mathrm{j}=0$ is chosen, the green line represents the best fit for $\gamma_{\mathrm{pd}}=0$, and the red line is the best fit for non-vanishing $\sigma_\mathrm{j}$ and $\gamma_{\mathrm{pd}}$. We clearly see that the fit with Gaussian jitter and pure dephasing (red line) describes the data most accurately with $\hbar\sigma_\mathrm{j}\approx 36$~\textmu eV and $2/\gamma_{\mathrm{pd}}\approx55$~ps. The presence of Gaussian jitter is also supported by the discussion in Sec.~\ref{sec:jitter_exp} and the fitted value of $\sigma_\mathrm{j}$ agrees with the values determined there.

\subsection{Summary of the parameters and relevant formulas}
\begin{table}[h]
	\caption{Parameters for modeling the laser pulses together with \eqref{eq:pearson}, yielding the laser spectrum and laser interference dynamics presented in Fig.~\ref{fig:laser}.}
	\centering
	\begin{tabular}{c|c|c|c|c|c}
		$\alpha_l$ & $\omega_0$ & $\nu_l$ & $\sigma_l$ & $m_l$ & $\omega_l$ \\
		\hline
		$\sqrt{2m_l}/\sigma_l$ & $\displaystyle\omega_l+\frac{\alpha_l\nu_l}{2m_l}$ & $-0.5\sqrt{m_l}$ & 350~fs & $1.2$ & variable
	\end{tabular}\label{table:laser}
\end{table}
The spectrometer response $R(\omega,\Omega)$ is modeled by \eqref{eq:R_fit} with a resolution of $\Delta\lambda_{1200}=0.075$~nm in the case of the 1200-grating and a resolution of $\Delta\lambda_{300}=0.3$~nm in the case of the 300-grating. It is used in \eqref{eq:PL} to calculate the PL of the emitter and in \eqref{eq:spectrum_laser} to calculate the laser spectrum. Whenever a measured PL is displayed together with a simulated one, the detection was performed with the 300-grating. The laser spectrum was always detected with the 1200-grating.\\
The Fourier transform of the laser field of a single pulse is modeled by \eqref{eq:pearson}. The parameters are given in Tab.~\ref{table:laser}. With this, we can determine the laser spectrum using  \eqref{eq:spectrum_laser}, the laser interference using \eqref{eq:laser_interference}, and most importantly the expression $\tilde{I}_0(\Delta t)$ in \eqref{eq:I_0_abs}, which yields the normalized coherent control visibility and the PLE depending on the carrier frequency $\omega_l$ for $\Delta t=0$.\\
\begin{table}[h]
	\caption{Parameters for modeling the LA spectral density via \eqref{eq:LA_specdens}.}
	\centering
	\begin{tabular}{c|c|c}
		$\alpha/\hbar^2$ & $\hbar\omega_c$ & $\hbar\Omega_c$\\
		\hline 
		$1.5\cdot 10^{-3}$~meV$^{-2}$& 12~meV&180~meV
	\end{tabular}\label{table:LA}
\end{table}
\begin{table}[h]
	\caption{Parameters for modeling the LO spectral density via \eqref{eq:LO_specdens}.}
	\centering
	\begin{tabular}{c|c|c|c|c}
		$n$&LO$_1$&LO$_2$&LO$_1^\prime$&LO$_1^{\prime\prime}$\\
		\hline
		$\hbar\omega_n$ &165~meV &198~meV &155~meV &174~meV\\
		\hline 
		$\hbar g_n$ &55~meV &40~meV &32~meV &42~meV \\
		\hline
		$1/\gamma_n$ &300~fs &400~fs &250~fs &250~fs
	\end{tabular}\label{table:LO}
\end{table}\\
The LA spectral density is modeled by \eqref{eq:LA_specdens} with parameters given in Tab.~\ref{table:LA}. The LO spectral density is modeled by four dispersionless modes (sum of four delta-functions, see \eqref{eq:LO_specdens}) with frequencies $\omega_n$ and coupling constants $g_n$ given in Tab.~\ref{table:LO}. The lifetimes $1/\gamma_n=1/\gamma(\omega_n)$ are also presented there. These parameters are used in \eqref{eq:G_mp} together with \eqref{eq:phi} to calculate the phonon correlation function $G_{-+}$, which is relevant to all optical signals retrieved from the emitter.\\
The ZPL frequency $\omega_{\mathrm{ZPL}}$ and the pure dephasing rates $\gamma_{\mathrm{pd}}$ for 8~K and 80~K are given in Tab.~\ref{table:TLS}. The spectral width $\sigma_\mathrm{j}$ of the slow Gaussian jitter, which modifies all visibility calculations according to \eqref{eq:Gauss_effect} is also presented there, together with the parameters for the probability distribution describing the impact of random telegraph noise on the visibility dynamics in Fig.~4 in the main text. This probability distribution is modeled by a sum of two-delta functions 
\begin{equation}\label{eq:rtn}
A\delta\qty(\omega_{\mathrm{ZPL}}-\omega_{\mathrm{ZPL}}^{(0)}+\frac{\Delta E_{\mathrm{ZPL}}}{2\hbar})+B\delta\qty(\omega_{\mathrm{ZPL}}-\omega_{\mathrm{ZPL}}^{(0)}-\frac{\Delta E_{\mathrm{ZPL}}}{2\hbar})
\end{equation}
and incorporated into the calculation of the visibility dynamics according to the discussion in Sec.~\ref{sec:jitter_theory}. It is not necessary to use a normalized probability distribution, as the visibility is automatically normalized. Thus, $A$ and $B$ give the relative strength of the two frequency components of the random telegraph noise, which are separated by an energy of $\Delta E_{\mathrm{ZPL}}$.\\
All parameters given here are considered throughout the main text, leading to a consistent description of the experiment, including PL, PLE, and PL visibility signals.
\begin{table}[h!]
	\caption{All parameters of the emitter (two-level system), including those used for the modeling of Gaussian jitter and random telegraph noise.}
	\centering
	\begin{tabular}{c|c|c|c|c|c|c|c}
		$\hbar\omega_{\mathrm{ZPL}}$&$2/\gamma_{\mathrm{pd}}$ (8K)&$2/\gamma_{\mathrm{pd}}$ (80K)& $\hbar\sigma_\mathrm{j}$ & $\hbar\omega_{\mathrm{ZPL}}^{(0)}$ & $\Delta E_{\mathrm{ZPL}}$ & $A$ & $B$\\
		\hline
		2.0363~eV&55~ps&3.5~ps&35.9~\textmu eV&2.0363~eV&0.7~meV&0.9&1
	\end{tabular}\label{table:TLS}
\end{table}
\newpage
\section{Simulations for different pulse shapes}
\subsection{Coherent control visibility dynamics}\label{sec:LA_vis}
As mentioned in Sec.~\ref{sec:vis_LA}, the coherent control visibility for the case of slightly detuned excitation around the ZPL addressing the acoustic PSB, crucially depends on the form of the laser spectrum. Here, we present three auxiliary simulations to Fig. 6 in the main text.
\subsubsection{Asymmetric Pearson-4 laser spectrum}
\begin{figure}[h!]
	\centering
	\includegraphics[width=0.45\linewidth]{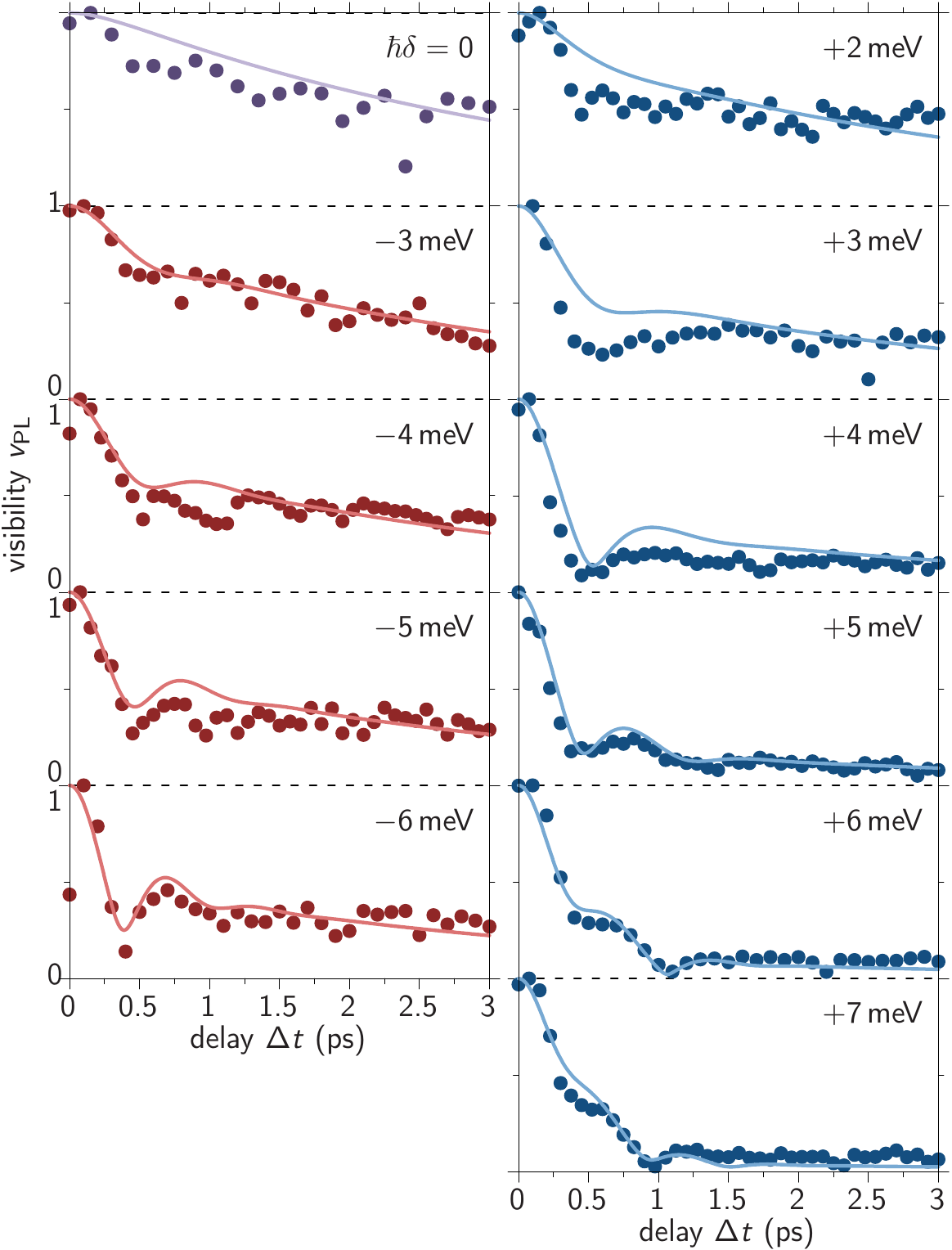}
	\caption{Coherent control visibility for excitation close to the ZPL, as in Fig. 6 of the main text. Experimental data as dots, simulation as lines. The case of resonant excitation (violet) and negative detuning (red) is shown in the left column, the case of positive detuning (blue) in the right.}
	\label{fig:pid80ksi}
\end{figure}

Figure~\ref{fig:pid80ksi} shows the same situation as Fig. 6 in the main text, however plotted on a slightly longer timescale and for more values of the detuning $\delta=\omega_l-\omega_{\mathrm{ZPL}}$. In the main text, the data was shown up to 2~ps to emphasize the non-trivial phonon induced dephasing dynamics occurring in the first picosecond of the delay. The experimental data is shown as dots, the simulation as lines. The left column shows the case of resonant excitation (violet) and negative detunings (red). The right column shows the case of positive detunings (blue). Simulation and experimental data match remarkably well, also for the cases not shown in the main text, i.e., $\hbar\delta=+2$, $\pm3$, $+7$~meV. By plotting data and simulation on a longer timescale of $3$~ps, we highlight that the chosen value of the pure dephasing rate $\gamma_\mathrm{pd}$ accurately describes the long-time dephasing behavior.
\subsubsection{Symmetric Pearson-4 laser spectrum}
\begin{figure}[h!]
	\centering
	\includegraphics[width=0.9\linewidth]{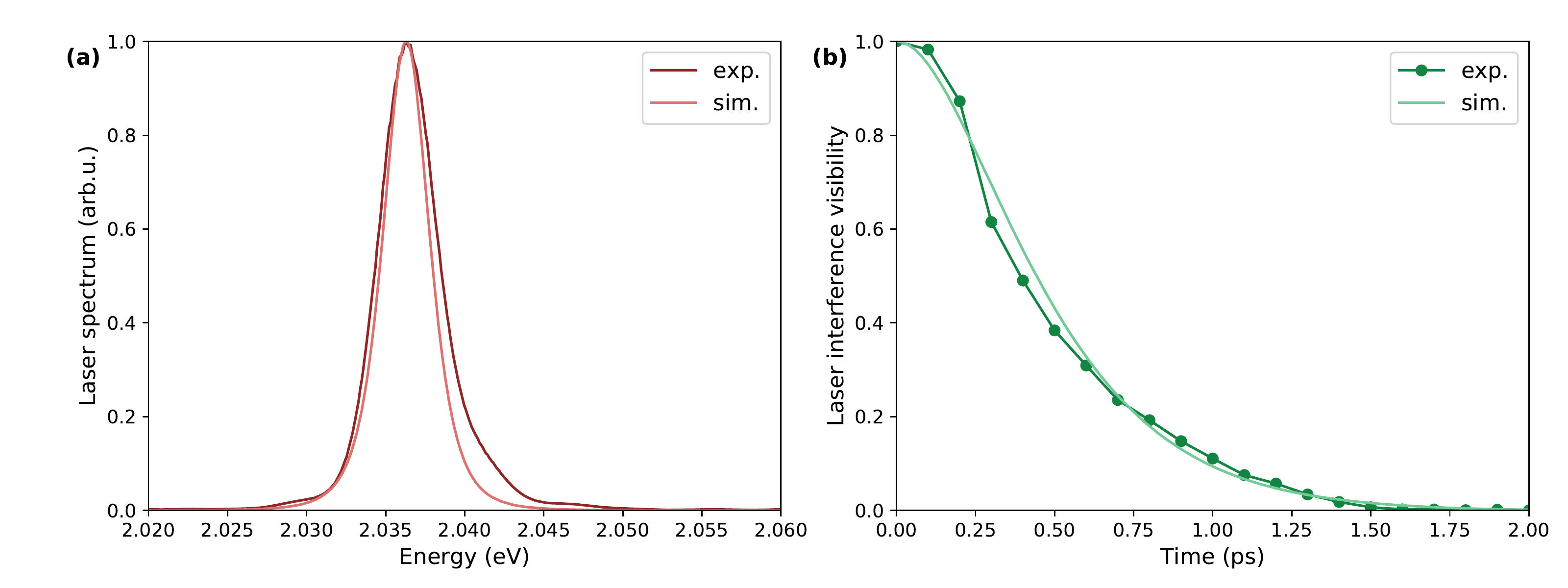}
	\caption{Characterization of the laser pulses with a symmetric spectrum. (a) Measurement (dark) and simulation (bright) of the laser spectrum. The spectrum is modeled via \eqref{eq:pearson} with $\alpha_l=\sqrt{2m_l}/\sigma_l$, $\omega_0=\omega_l+\alpha_l\nu_l/(2m_l)$, $\sigma_l=350$~fs, $m_l=1.2$, and $\nu_l=0$. The spectrum is measured with the 1200-grating, such that the resolution for the spectrometer response from \eqref{eq:R_fit} is $\Delta\lambda_{1200}=0.075$~nm. (b) Measurement (dark dotted line) and simulation (bright line) of the laser interference visibility, i.e., the pulse autocorrelation.}
	\label{fig:laser_symm}
\end{figure}

Now we consider the case of a symmetric spectrum of the Pearson-4 type from \eqref{eq:pearson} with the same parameters as in the main text (see Tab.~\ref{table:laser}), but with $\nu_l=0$.  In Fig.~\ref{fig:laser_symm}(a) and (b) the simulated (bright) and detected (dark) spectrum and autocorrelation of the laser pulse are shown, respectively. For the measurement of the spectrum, the 1200-grating is used, such that in the simulation we use the response in \eqref{eq:R_fit} with $\Delta\lambda=\Delta\lambda_{1200}=0.075$~nm. Compared to the case of an asymmetric spectrum considered in the main text (see Fig.~\ref{fig:laser}), this laser pulse model yields a worse fit to the measured spectrum. The laser interference however still agrees well with the experiment.
\begin{figure}[h!]
	\centering
	\includegraphics[width=0.45\linewidth]{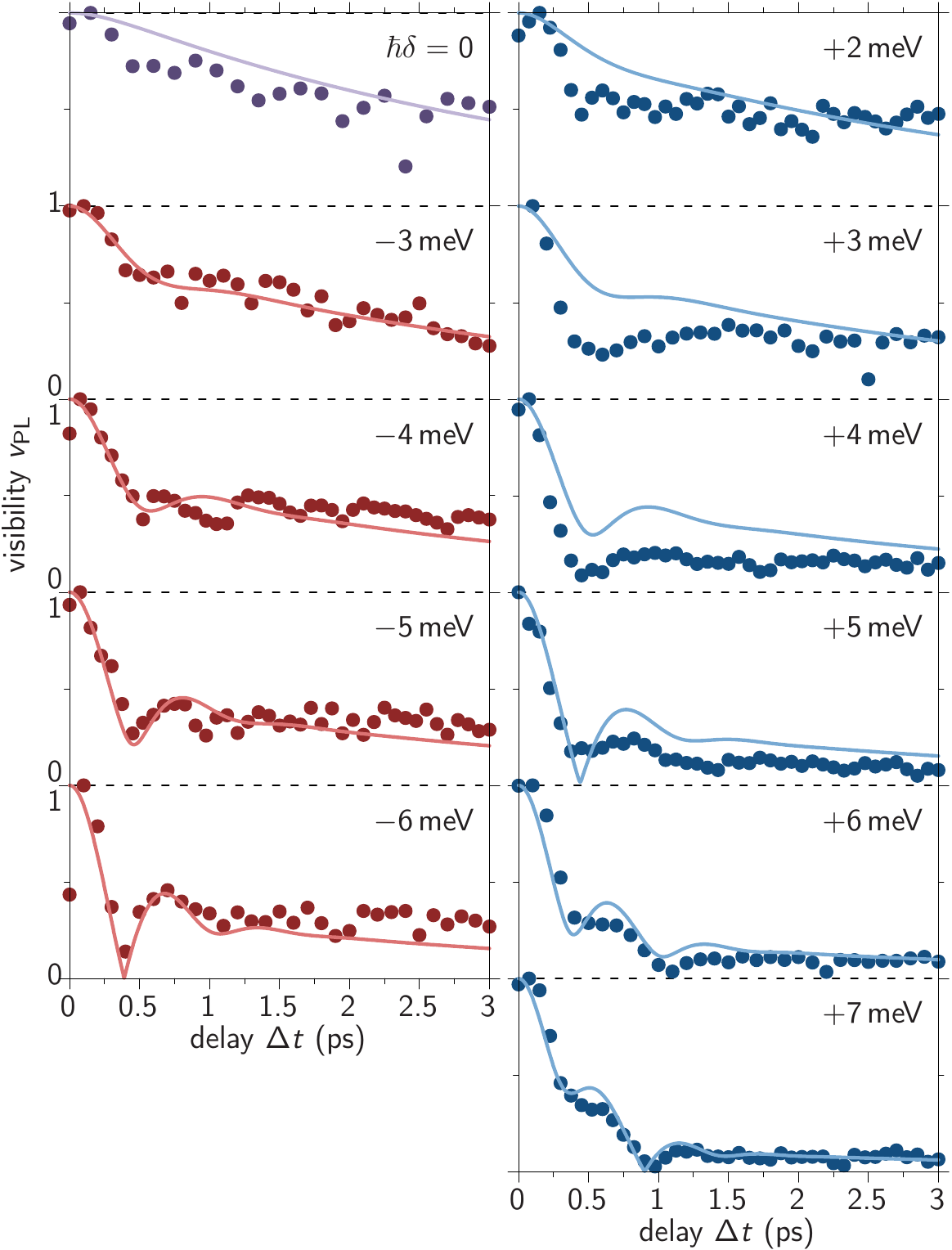}
	\caption{Coherent control visibility for excitation close to the ZPL, as in Fig. 6 of the main text, but with a symmetric ($\nu_l=0$) Pearson-4 pulse. Experimental data as dots, simulation as lines. The case of resonant excitation (violet) and negative detuning (red) is shown in the left column, the case of positive detuning (blue) in the right.}
	\label{fig:pid80ksipearsonsymm}
\end{figure}

Figure~\ref{fig:pid80ksipearsonsymm} shows the same data as Fig.~\ref{fig:pid80ksi}, but the simulations include the symmetric Pearson-4 pulse. We see that the simulation in the case of negative detuning (left) tends to lie below the experimental data at $\Delta t=3$~ps, while it tends to lie above the data for positive detuning (right). When comparing the PID drop of the signal, i.e., the final PL visibility value at $\Delta t=3$~ps, for detunings of the same absolute value, we find almost the same values. For the asymmetric Pearson-4 pulse in Fig.~\ref{fig:pid80ksi} the difference between $\pm\delta$ is larger. The reason for this is the asymmetry of the laser spectrum used in Fig.~\ref{fig:pid80ksi}. As discussed in Sec.~\ref{sec:vis_LA}, the value of the visibility for large times is determined by the overlap of the laser with the ZPL and the height of the LA PSB at the position of the maximum of the laser spectrum. In the case of an asymmetric laser spectrum with more spectral weight towards higher frequencies (see Fig.~\ref{fig:laser}), the overlap of the laser with the ZPL is enhanced for negative detunings resulting in larger visibility values at $\Delta t=3$~ps. When the asymmetry is lifted by reducing the high frequency flank of the spectrum the overlap with the ZPL reduces and the visibility values at $\Delta t=3$~ps shrink. The effect is less pronounced for positive detunings, because the lower frequency flank of the laser spectrum remains nearly the same when making the pulse symmetric (see Fig.~\ref{fig:laser}(a)).\\
Still in the case of a completely symmetric laser spectrum, the visibility dynamics for $+\delta$ and $-\delta$ behave differently. For positive detunings, phonon emission plays a more dominant role than phonon absorption and vice versa for negative detuning. Phonon absorption however can only take place if a thermal occupation $n_\mathrm{ph}$ is present and the process scales with $\sim n_\mathrm{ph}$. Phonon emission on the other hand includes spontaneous emission as well as induced emission and the full process scales with $n_\mathrm{ph}+1$. Therefore an asymmetry, which can best be seen when comparing the complete dynamics in Fig.~\ref{fig:pid80ksipearsonsymm}, e.g., for the case of $\hbar\delta=\pm 5$~meV, remains even in the case of a symmetric laser spectrum.

\subsubsection{Gaussian laser spectrum}
\begin{figure}[h]
	\centering
	\includegraphics[width=0.9\linewidth]{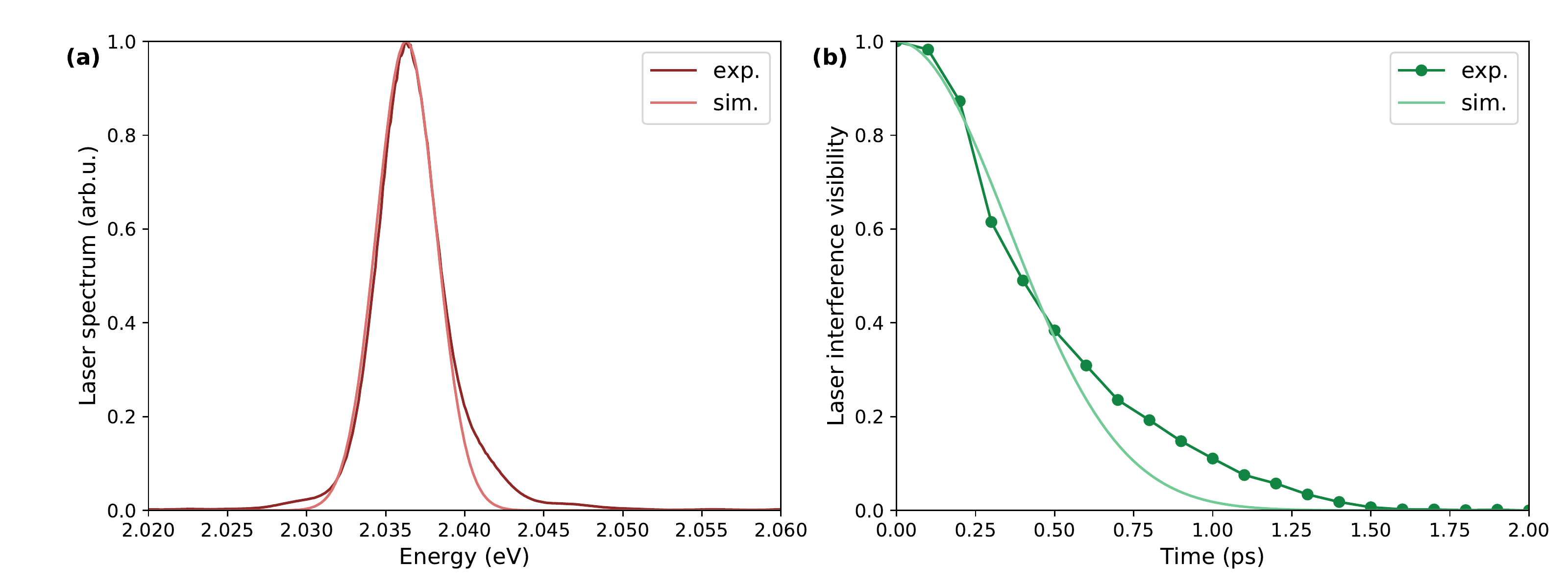}
	\caption{Characterization of the laser pulses with a Gaussian spectrum. (a) Measurement (dark) and simulation (bright) of the laser spectrum. The spectrum is modeled via \eqref{eq:laser_gauss} with $\alpha_l=(250~\text{fs})^{-1}$. The spectrum is measured with the 1200-grating, such that the resolution for the spectrometer response from \eqref{eq:R_fit} is $\Delta\lambda_{1200}=0.075$~nm. (b) Measurement (dark dotted line) and simulation (bright line) of the laser interference visibility, i.e., the pulse autocorrelation.}
	\label{fig:laser_gauss}
\end{figure}

A commonly used laser pulse spectrum to simulate the optical excitation of quantum system is given by a Gaussian with
\begin{equation}\label{eq:laser_gauss}
\mathcal{E}_s(\omega)\sim \exp[-\frac{(\omega-\omega_l)^2}{2\alpha_l^2}]\,.
\end{equation}
To test its performance in our situation we keep the temporal and spectral width comparable to the case considered in the main text and choose $\alpha_l=(250~\text{fs})^{-1}$. In Fig.~\ref{fig:laser_gauss}(a) and (b) we show the simulated (bright) and detected (dark) spectrum and autocorrelation of the laser pulse, respectively. For the detection of the spectrum, the 1200-grating is used, such that in the simulation we consider the response in \eqref{eq:R_fit} with $\Delta\lambda=\Delta\lambda_{1200}=0.075$~nm. Neither the spectrum, nor the autocorrelation of the laser are well reproduced by this model, compared to the case of an asymmetric Pearson-4 spectrum in Fig.~\ref{fig:laser}.
\newpage
\begin{figure}[h]
	\centering
	\includegraphics[width=0.45\linewidth]{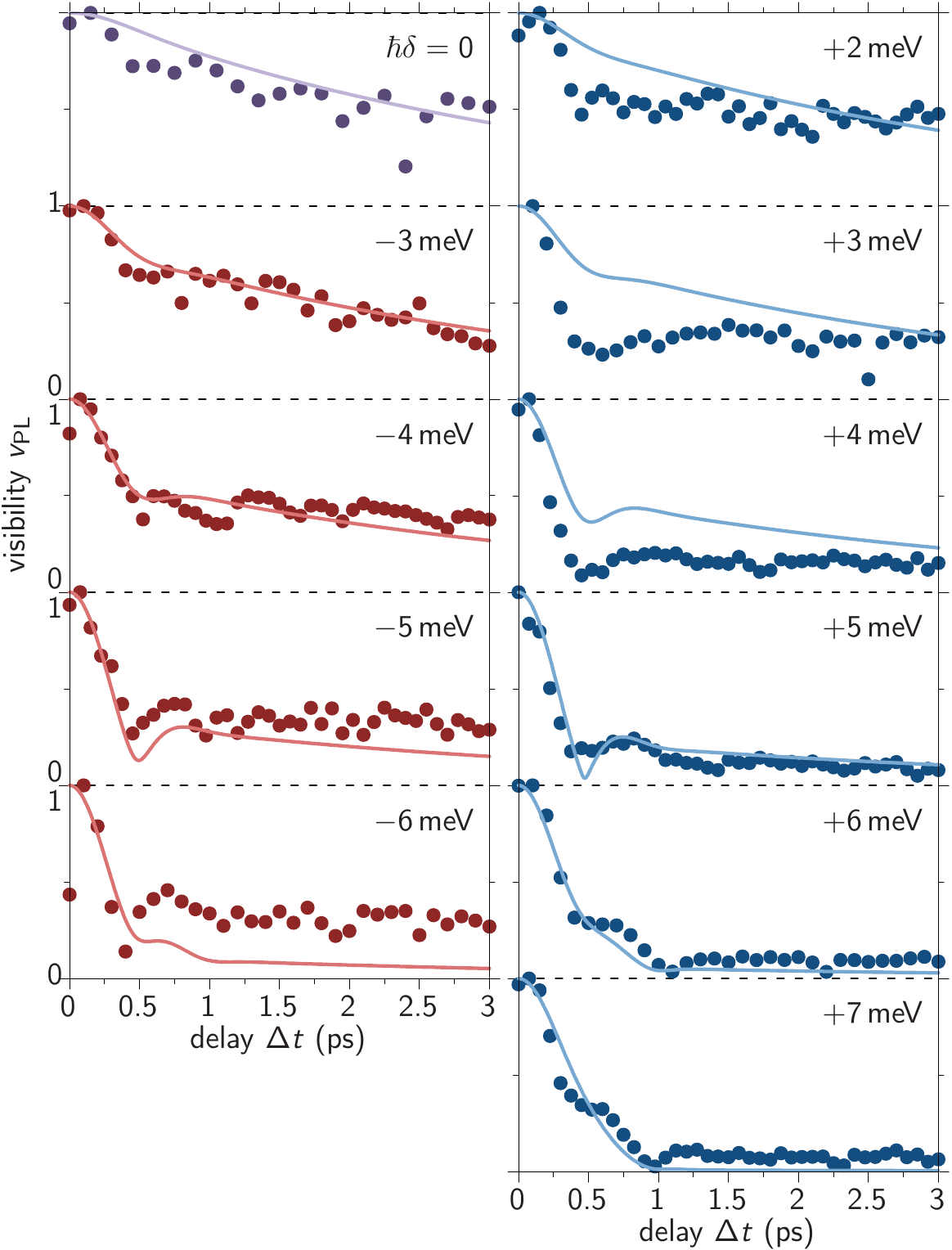}
	\caption{Coherent control visibility for excitation close to the ZPL, as in Fig. 6 of the main text, but with a Gaussian pulse (see \eqref{eq:laser_gauss} and Fig.~\ref{fig:laser_gauss}). Experimental data as dots, simulation as lines. The case of resonant excitation (violet) and negative detuning (red) is shown in the left column, the case of positive detuning (blue) in the right.}
	\label{fig:pid80ksigauss}
\end{figure}
Figure~\ref{fig:pid80ksigauss} again shows the same data as Fig.~\ref{fig:pid80ksi}, but using the Gaussian laser spectrum. While the simulations with the Pearson-4 spectra in Fig.~\ref{fig:pid80ksi} and Fig.~\ref{fig:pid80ksipearsonsymm} differed only slightly on a quantitative level, the visibility dynamics with the Gaussian pulse show a completely different qualitative behavior. Some of the cases, e.g., $\hbar\delta=-3, -4$~meV fit well with the experimental data, while others, e.g., $\hbar\delta=-6$~meV do not fit at all. The value of the visibility at $\Delta t=3$ps drops much faster than in the experiment, when going from $\hbar\delta=\pm3$~meV to $\hbar=\pm 6$~meV. This can be explained by the form of the Gaussian spectrum and the form of the Pearson-4 spectrum. The Gaussian spectrum drops $\sim\exp(-x^2)$ when moving from the maximum outwards. The Pearson-4 spectrum for our choice of parameters, specifically $m_l=1.2$, is close to a Lorentzian spectrum, which has more spectral weight far from the maximum. Consequently, the overlap with the ZPL is still significant for large absolute detunings in the case of the Pearson-4 spectrum and vanishes more quickly with the detuning for the Gaussian pulse.

\subsection{Impact of LAs and laser asymmetry on the PLE}
\begin{figure}[h!]
	\centering
	\includegraphics[width=0.45\linewidth]{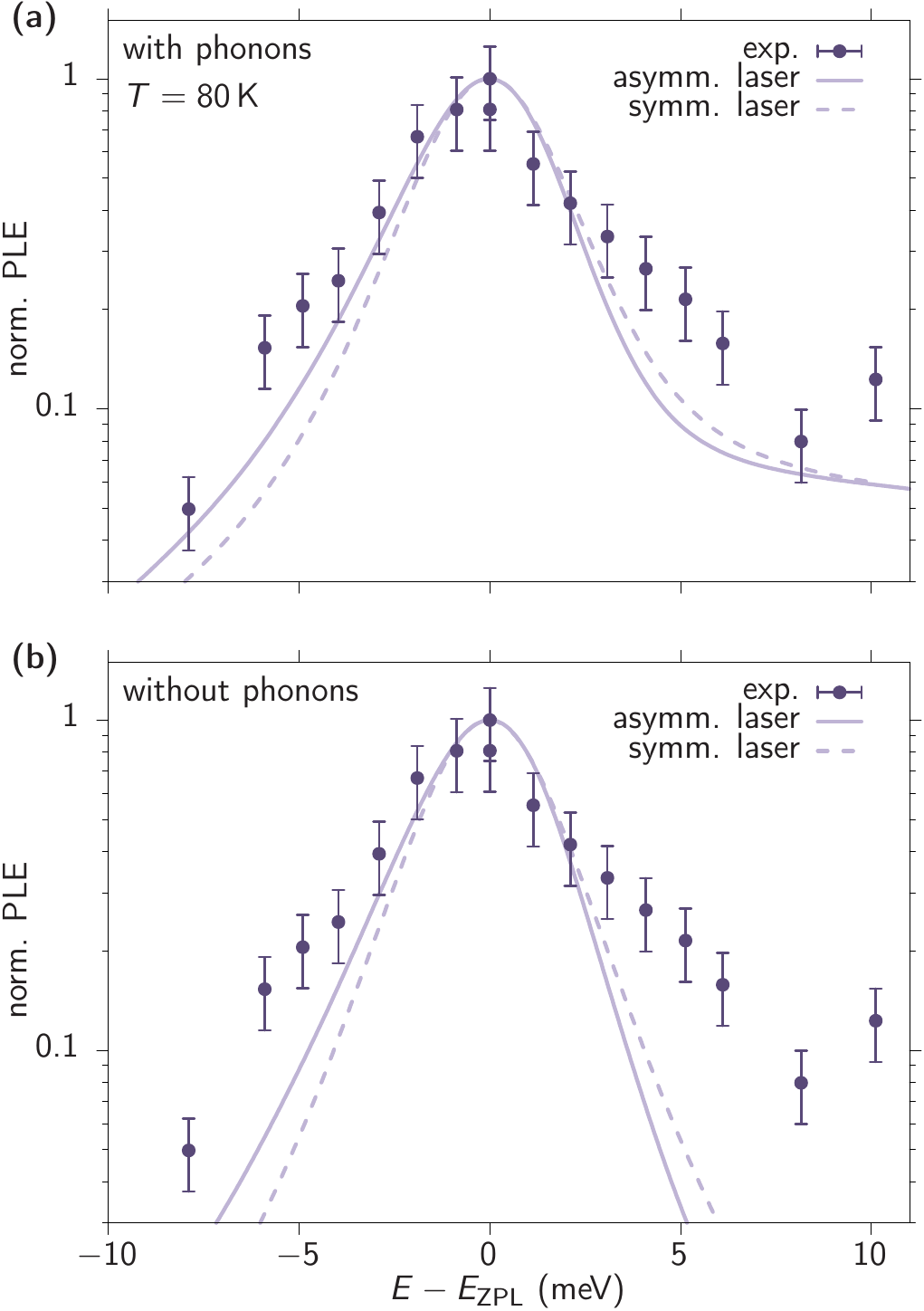}
	\caption{PLE spectrum at $T=80K$ as in Fig. 6 of the main text. Experimental data as dots, simulations as lines. (a) Simulation with and (b) without phonon coupling. Solid lines with asymmetric and dashed lines with symmetric laser spectra. Laser spectra are modeled by Pearson-4 distributions with parameters given in Tab.~\ref{table:laser}, but with $\nu_l=0$ for the symmetric cases.}
	\label{fig:ple80ksi}
\end{figure}
In this section we discuss the impact of the laser pulse spectrum on the PLE signal shown in Fig. 6(a) in the main text. Figure~\ref{fig:ple80ksi} shows the PLE spectrum of the emitter at $T=80$~K with the experimental data as dots and the simulations as lines. (a) shows simulations including the coupling to acoustic phonons and no phonons are considered in (b). The general impact of the phonon coupling is discussed in the main text. Here, we focus on the additional impact of the spectral shape of the laser. The solid lines present the asymmetric Pearson-4 pulse with the parameters in Tab.~\ref{table:laser} and the dashed lines the symmetric pulse with $\nu_l=0$.\\
In Fig.~\ref{fig:ple80ksi}(a) the simulated PLE with LA phonons is asymmetric between positive and negative detuning $E-E_{\mathrm{ZPL}}$. For negative detunings, the PLE vanishes, whereas it reaches a plateau for positive detunings. The reason is again the asymmetry between absorption and emission of phonons. When going further from the ZPL in the direction of negative detunings, phonons with the energy $\approx|E-E_{\mathrm{ZPL}}|$ need to be absorbed in order to yield a PLE signal. The thermal occupation decreases for increasing phonon energy, such that the PLE signal decreases as well for increasing negative detuning. For positive detuning, spontaneous emission, as well as induced emission of phonons can generate a PLE signal. While the probability for induced emission decreases when increasing the detuning, spontaneous emission processes can always take place, leading to the plateau for positive detuning.\\
If we now remove any influence of the LA phonons in Fig.~\ref{fig:ple80ksi}(b), e.g., by setting $\alpha=0$ in \eqref{eq:LA_specdens}, the absorption spectrum of the emitter close to the ZPL is just a Lorentzian. In that case the PLE, being the convolution of the mirrored laser spectrum with the absorption spectrum (see \eqref{eq:PLE}), is just a broadened version of the laser spectrum, mirrored with respect to $E-E_{\mathrm{ZPL}}=0$. Therefore an asymmetry of the PLE remains when removing the influence of LA phonons (solid line), since the laser spectrum is asymmetric (see Fig.~\ref{fig:laser}). When additionally making the pulse spectrum symmetric, as depicted by the dashed line in Fig.~\ref{fig:ple80ksi}(b), also the PLE spectrum becomes symmetric.\\
Thus we see that an asymmetry of the PLE in general stems from two influences, namely (i) the asymmetry between phonon emission and absorption and (ii) a possible asymmetry of the laser spectrum. Both have to be considered accurately to reproduce the measurement properly.

%